\PassOptionsToPackage{sort&compress}{natbib}
\documentclass[preprint,3p,12pt]{elsarticle}
\usepackage{mathrsfs}
\usepackage{amsmath}
\usepackage{stmaryrd}
\usepackage{bbding}
\usepackage{dcolumn}
\usepackage{graphicx}
\usepackage{amsfonts}
\usepackage{amssymb}
\usepackage{psfrag}
\usepackage{wrapfig}
\usepackage{makeidx}
\usepackage{tabularx}
\usepackage{bm}
\usepackage{epsf}
\usepackage{epsfig}
\usepackage{setspace}
\usepackage{epstopdf}
\usepackage{subcaption}
\usepackage{color}
\usepackage{multirow}
\usepackage{comment}
\usepackage{booktabs}
\usepackage[ruled,longend]{algorithm2e}
\usepackage{float}
\usepackage{placeins}
\usepackage{pdflscape}
\usepackage{url}
\usepackage{tikz}
\usepackage{hyperref}
\usetikzlibrary{arrows.meta,positioning,calc}
\DontPrintSemicolon
\SetKwInput{KwInput}{Input}
\SetKwInput{KwOutput}{Output}
\SetKwInput{KwData}{Data}

\newcommand{\Kxi}{K_{\xi}}

\begin{document}
	\def\floatpagepagefraction{1}
	\def\textpagefraction{.001}
	
	\title{A GPGPU-Oriented Full Phase-Space Parallel Unified Gas-Kinetic Scheme with Velocity-Block Pipelining}
	
	\author[XJTU,SKL,PKL]{Zhiwen Zhuang}
	\author[HKUST]{Yixiao Wang}
	\author[XJTU,SKL,PKL]{Xinhang Guo}
	\author[XJTU,SKL,PKL]{Xing Ji\corref{cor1}}
	\ead{jixing@xjtu.edu.cn}
	\author[XJTU,SKL,PKL]{Xian Wang}
	\author[HKUST]{Kun Xu}
	
	\address[XJTU]{School of Aerospace Engineering, Xi'an Jiaotong University, Xi'an 710049, Shaanxi, China}
	\address[SKL]{State Key Laboratory for Strength and Vibration of Mechanical Structures, Xi'an 710049, Shaanxi, China}
	\address[PKL]{Shaanxi Key Laboratory of Environment and Control for Flight Vehicle.Xi'an 710049,Shannxi,China}
	\address[HKUST]{The Hong Kong University of Science and Technology, Hong Kong, China}
	\cortext[cor1]{Corresponding author}
	
	\begin{abstract}
		The deterministic unified gas-kinetic scheme (UGKS) provides a multiscale framework for nonequilibrium gas dynamics, but its high-dimensional phase-space discretization leads to severe memory pressure and communication overhead, especially on large unstructured meshes. This paper presents a GPGPU-oriented UGKS with velocity-block pipelining and full phase-space MPI decomposition. In the proposed formulation, the discrete velocity space is partitioned into fixed-size velocity blocks for accelerator execution, while MPI ranks are organized into coupled physical-space and velocity-space communicators. As a result, each rank stores and advances only a local physical subdomain together with a contiguous subset of velocity blocks, and macroscopic moments are recovered through lightweight reductions over the velocity-space communicator. To improve concurrency and reduce exposed communication cost, a triple-buffered pipeline is further developed to overlap microscopic reconstruction, physical-halo exchange, nonequilibrium flux evaluation, and the first-stage distribution update during the local velocity-block sweep. The implementation targets SIMT-based GPGPU accelerators through a portable device-runtime abstraction. Numerical experiments demonstrate that the $P_v=8$ configuration achieves a $33.4$--$35.4\times$ strong-scaling speedup on 64 nodes, while an Orion-like capsule simulation reaches approximately $1.33\times10^{11}$ phase-space degrees of freedom on 4096 GPGPU accelerators. These results indicate that the proposed method preserves the original UGKS flux construction and two-stage time discretization, while substantially reducing microscopic storage per rank and improving the scalability of large unstructured phase-space simulations.
	\end{abstract}
	
	
	\maketitle
	
	\section{Introduction}\label{sec:introduction}
	Many gas-dynamic systems of practical interest are intrinsically multiscale: strong nonequilibrium effects, large density variations, and substantial local changes in the Knudsen number may coexist in a single configuration~\cite{wang2003predicting}. Such features are commonly encountered in hypersonic vehicles, rarefied external flows, and microscale transport devices. In these problems, continuum descriptions may lose validity in some regions, whereas conventional fixed-scale kinetic solvers can become prohibitively expensive in others.
	
	A broad class of kinetic and multiscale approaches has therefore been developed over the past decades. Early gas-kinetic schemes established a flux-construction framework in which macroscopic transport is recovered from the time-dependent kinetic evolution at cell interfaces~\cite{xu2001gas}. In parallel, the gas-kinetic unified algorithm (GKUA), developed from model Boltzmann equations, provided a practical framework for simulations spanning rarefied, transitional, and continuum regimes~\cite{li2014gas,hu2021gas}. For steady multiscale rarefied flows, the general synthetic iterative scheme (GSIS) and its extensions significantly improved convergence in the near-continuum regime while preserving the correct asymptotic behavior~\cite{su2021multiscale,zhu2021general,su2020can}. In another direction, generalized hydrodynamic models and nonlinear coupled constitutive relation (NCCR) models extended continuum-type constitutive descriptions toward local nonequilibrium regimes~\cite{myong1999thermodynamically,myong2001computational}.
	
	Among these approaches, the unified gas-kinetic scheme (UGKS) has become one of the representative frameworks for all-regime gas dynamics. Starting from the original formulation of Xu and Huang, UGKS couples particle transport and collision directly in the construction of the time-dependent interface flux. As a result, the same numerical formulation can recover flow physics from the rarefied regime to the hydrodynamic limit. The method was subsequently extended to multidimensional configurations and microflow simulations, establishing UGKS as a direct-modeling approach for multiscale transport~\cite{xu2010unified,huang2012unified,huang2013unified}. The discrete unified gas-kinetic scheme (DUGKS), introduced by Guo, Xu, and Wang, provides another important multiscale kinetic formulation based on coupled transport-collision flux evaluation and has been extended to unstructured meshes for computations over complex geometries~\cite{Guo2013DUGKS,Guo2015DUGKSCompressible,Zhu2016DUGKSUnstructured}. More recently, the unified gas-kinetic wave-particle (UGKWP) method enriched the UGKS family by combining deterministic wave evolution and stochastic particle transport within a unified multiscale framework, thereby improving efficiency across both rarefied and continuum regimes~\cite{Liu2020UGKWP1,Zhu2019UGKWP2,Liu2022WaveParticleReview}. Substantial efforts have also been devoted to improving UGKS itself, including implicit UGKS for steady and unsteady computations~\cite{Zhu2016ImplicitUGKSSteady,Zhu2019ImplicitUGKSUnsteady}, moving-mesh and velocity-space adaptation strategies~\cite{Chen2012MovingMeshUGKS}, and adaptive UGKS with continuous-discrete velocity-space switching for multiscale transport~\cite{Xiao2020AUGKS}.
	
	GPGPU computing refers to general-purpose computation on graphics processing units and, more broadly, to the use of GPU-like many-core processors as programmable accelerators for scientific workloads~\cite{Owens2008GPUComputing}. As high-fidelity simulations move toward larger scales and more complex configurations, high-performance computing has become indispensable for both continuum and kinetic flow solvers. Within the broader CFD community, GPU acceleration has developed along several complementary routes. Early studies demonstrated the feasibility of mapping Euler and hypersonic-flow solvers to graphics hardware, showing that GPU devices can be used effectively for structured-grid aerodynamic calculations~\cite{BrandvikPullan2008GPUCFD,Elsen2008HypersonicGPU}. Subsequent work extended GPU acceleration to unstructured-grid finite-volume solvers, high-order discontinuous Galerkin and flux-reconstruction formulations, and multi-level GPU implementations for incompressible flow, where the main issues are exposing fine-grained parallelism and reducing memory traffic~\cite{Corrigan2011UnstructuredGPU,Klockner2009DGGPUs,JacobsenSenocak2013MultiLevelGPU,Witherden2014PyFR,Vermeire2017GPUHighOrder}. More recent open frameworks and production-oriented solvers, including ZEFR, NekRS, and STREAmS, further illustrate the move toward reusable high-order, spectral-element, and multi-GPU infrastructures for turbulent and compressible-flow simulations~\cite{Romero2020ZEFR,Fischer2022NekRS,Bernardini2021STREAmS}. GPU-CFD reviews and recent heterogeneous CFD frameworks further emphasize portability through domain-specific languages, automatic code generation, and accelerator-oriented solver infrastructure~\cite{Niemeyer2014GPUCFD,Lusher2021OpenSBLI,Duffy2012,Piscaglia2023}. At the same time, high-fidelity CFD studies based on compact finite-difference schemes, high-order overset-grid technology, and turbomachinery DNS have shown that scalable domain decomposition, robust interface treatment, and complex-geometry capability are central to production-scale simulations~\cite{KimSandberg2012CompactParallel,DeuseSandberg2020Overset,SandbergMichelassi2019HighFidelityTurbomachinery}. Within gas-kinetic methods, GPU-related developments have expanded from multi-GPU high-order gas-kinetic solvers for compressible turbulence to GPU-enabled simulations of wind-turbine wakes~\cite{Wang2023MultiGPUHGKS,Huo2025GPUWindTurbine}. Related high-order gas-kinetic studies involving Ji and co-workers have also explored compact fourth-order reconstruction and GPU-oriented multigrid acceleration for continuum CFD applications~\cite{Ji2018CompactFourthGKS,Liu2026GPUCGKS}. These developments indicate that performance improvement increasingly depends not only on raw hardware capability, but also on algorithmic organization in terms of parallel granularity, data locality, portability, and communication efficiency.
	
	For kinetic methods, the challenge is more fundamental because the computational state is defined in phase space rather than in physical space alone. Consequently, deterministic methods such as the discrete velocity method (DVM), UGKS, and DUGKS generally require much larger memory capacity, incur higher communication overhead, and exhibit more irregular data access than conventional continuum solvers, especially for three-dimensional nonequilibrium flows. Considerable efforts have therefore been made to improve both algorithmic efficiency and hardware utilization. Early studies demonstrated the feasibility of solving model kinetic equations on many-core accelerators~\cite{Frezzotti2011}. Later work introduced GPU acceleration for iterative DVM solvers together with memory-reduction techniques~\cite{Zhu2019}, multi-level MPI/OpenMP implementations for large-scale rarefied-flow simulations~\cite{Ho2019}, and hybrid physical/velocity-space parallelization strategies for DUGKS on unstructured meshes~\cite{Zhang2022}. More recently, direct GPU implementations of DUGKS have been developed for low-speed and turbulent flow simulations~\cite{Liu2024,Karzhaubayev2024}, while memory-saving and communication-aware programming paradigms have been introduced for UGKS to reduce storage cost and improve large-scale parallel efficiency~\cite{Zhang2025}. Nevertheless, for unstructured phase-space discretizations on modern many-core accelerators, microscopic velocity-space evolution remains a dominant bottleneck. Further performance gains therefore require not only advances in physical modeling and asymptotic-preserving formulations, but also a more efficient execution organization for velocity-space evolution.
	
	This work develops a GPGPU-oriented velocity-block-pipelined UGKS for unstructured phase-space simulations. Unlike physical-domain-only parallelization, in which every MPI rank stores the complete discrete velocity space, the present implementation adopts a Full Phase-Space Parallel (FPP) decomposition. MPI ranks are arranged into physical-space and velocity-space communicators, so that each rank advances only its local physical subdomain and assigned velocity-block subset. Microscopic evolution is reorganized into a blockwise sweep in which update, reconstruction, and communication stages are overlapped. The velocity-block design improves data reuse and reduces intermediate memory traffic, while reductions over the velocity communicator assemble macroscopic moments across the distributed velocity space. The device layer is expressed through portable accelerator-runtime primitives, allowing the same algorithmic structure to target SIMT-based GPGPU accelerators. The resulting FPP-UGKS implementation preserves the UGKS flux construction while improving memory scalability and parallel efficiency for multiscale nonequilibrium-flow computations.
	
	\section{Method}
	\subsection{Unified gas-kinetic scheme}
	
	The gas evolution is modeled by the BGK--Shakhov kinetic equation
	\begin{equation}
		\frac{\partial f}{\partial t}
		+
		\mathbf{u}\cdot\nabla f
		=
		\frac{f^{+}-f}{\tau},
		\label{eq:bgk_shakhov}
	\end{equation}
	where $f=f(\mathbf{x},t,\mathbf{u},\boldsymbol{\xi})$ is the molecular velocity distribution function, $\mathbf{x}$ is the spatial coordinate, $t$ is time, $\mathbf{u}$ is the resolved translational molecular velocity, and $\boldsymbol{\xi}$ denotes the unresolved degrees of freedom. For a $D$-dimensional physical-space computation with $K$ molecular internal degrees of freedom, the dimension of $\boldsymbol{\xi}$ is
	\begin{equation}
		\Kxi = K+3-D.
		\label{eq:Kxi_def}
	\end{equation}
	Thus the total number of molecular degrees of freedom is $D+\Kxi=K+3$. The relaxation time is denoted by $\tau$, and $f^{+}$ is the Shakhov equilibrium state.
	
	To recover the correct Prandtl number, the equilibrium distribution is written in the Shakhov form
	\begin{equation}
		f^{+}
		=
		g\left(
		1+
		(1-\Pr)
		\frac{\mathbf{c}\cdot\mathbf{q}}{5pRT}
		\left(
		\frac{|\mathbf{c}|^{2}}{RT}-5
		\right)
		\right)
		=
		g+g^{+},
		\label{eq:shakhov_fp}
	\end{equation}
	where $\Pr$ is the Prandtl number, $p$ is the pressure, $R$ is the gas constant, $T$ is the temperature, $\mathbf{c}=\mathbf{u}-\mathbf{U}$ is the peculiar velocity, and $\mathbf{U}$ and $\mathbf{q}$ are the macroscopic velocity and heat flux, respectively. In Eq.~\eqref{eq:shakhov_fp}, $g^{+}$ denotes the Shakhov correction to the Maxwellian rather than an additional equilibrium distribution.
	
	The Maxwellian distribution is
	\begin{equation}
		g
		=
		\rho
		\left(\frac{\lambda}{\pi}\right)^{\frac{D+\Kxi}{2}}
		\exp\!\left(
		-\lambda\left(|\mathbf{u}-\mathbf{U}|^{2}+\xi^{2}\right)
		\right),
		\label{eq:maxwellian}
	\end{equation}
	where $\rho$ is the density and
	\begin{equation}
		\xi^{2}=\xi_{1}^{2}+\xi_{2}^{2}+\cdots+\xi_{\Kxi}^{2}.
		\label{eq:xi_square}
	\end{equation}
	The parameter $\lambda$ is related to temperature by
	\begin{equation}
		\lambda=\frac{m}{2k_{B}T}=\frac{1}{2RT},
		\label{eq:lambda_def}
	\end{equation}
	where $m$ is the molecular mass and $k_B$ is the Boltzmann constant.
	
	The BGK--Shakhov collision operator preserves the conservative moments,
	\begin{equation}
		\int (f^{+}-f)\,\boldsymbol{\psi}\,\mathrm{d}\mathbf{u}\,\mathrm{d}\boldsymbol{\xi}=0,
		\label{eq:compatibility}
	\end{equation}
	with the collision invariants
	\begin{equation}
		\boldsymbol{\psi}
		=
		\left(
		1,\,
		\mathbf{u},\,
		\frac{1}{2}\left(|\mathbf{u}|^{2}+\xi^{2}\right)
		\right)^{T},
		\label{eq:collision_invariant}
	\end{equation}
	and the internal-space measure
	\begin{equation}
		\mathrm{d}\boldsymbol{\xi}=\mathrm{d}\xi_{1}\,\mathrm{d}\xi_{2}\cdots \mathrm{d}\xi_{\Kxi}.
		\label{eq:dxi}
	\end{equation}
	
	The macroscopic conservative variables are obtained from the moments of the distribution function,
	\begin{equation}
		\mathbf{W}
		=
		\begin{pmatrix}
			\rho\\
			\rho\mathbf{U}\\
			\rho E
		\end{pmatrix}
		=
		\int \boldsymbol{\psi}\,f\,\mathrm{d}\mathbf{u}\,\mathrm{d}\boldsymbol{\xi},
		\label{eq:macro_W}
	\end{equation}
	while the heat flux is
	\begin{equation}
		\mathbf{q}
		=
		\frac{1}{2}
		\int
		(\mathbf{u}-\mathbf{U})
		\left(
		|\mathbf{u}-\mathbf{U}|^{2}+\xi^{2}
		\right)
		f\,\mathrm{d}\mathbf{u}\,\mathrm{d}\boldsymbol{\xi}.
		\label{eq:heat_flux}
	\end{equation}
	
	To avoid explicit discretization of the unresolved velocity space, two reduced distribution functions are introduced:
	\begin{equation}
		h(\mathbf{x},t,\mathbf{u})
		=
		\int f\,\mathrm{d}\boldsymbol{\xi},
		\qquad
		b(\mathbf{x},t,\mathbf{u})
		=
		\int \xi^{2} f\,\mathrm{d}\boldsymbol{\xi}.
		\label{eq:reduced_hb}
	\end{equation}
	In terms of $h$ and $b$, Eq.~\eqref{eq:heat_flux} becomes
	\begin{equation}
		\mathbf{q}
		=
		\frac{1}{2}
		\int
		(\mathbf{u}-\mathbf{U})
		\left(|\mathbf{u}-\mathbf{U}|^{2}h+b\right)
		\mathrm{d}\mathbf{u}.
		\label{eq:reduced_heat_flux}
	\end{equation}
	Multiplying Eq.~\eqref{eq:bgk_shakhov} by $1$ and $\xi^2$, respectively, and integrating over the unresolved degrees of freedom gives
	\begin{equation}
		\frac{\partial h}{\partial t}
		+
		\mathbf{u}\cdot\nabla h
		=
		\frac{h^{+}-h}{\tau},
		\qquad
		\frac{\partial b}{\partial t}
		+
		\mathbf{u}\cdot\nabla b
		=
		\frac{b^{+}-b}{\tau},
		\label{eq:reduced_bgk}
	\end{equation}
	where
	\begin{equation}
		h^{+}=H+H^{+},
		\qquad
		b^{+}=B+B^{+}.
		\label{eq:reduced_equilibrium}
	\end{equation}
	
	The reduced Maxwellian distributions are
	\begin{equation}
		H
		=
		\int g\,\mathrm{d}\boldsymbol{\xi}
		=
		\rho
		\left(\frac{\lambda}{\pi}\right)^{D/2}
		\exp\!\left(-\lambda|\mathbf{u}-\mathbf{U}|^{2}\right),
		\label{eq:H_def}
	\end{equation}
	and
	\begin{equation}
		B
		=
		\int \xi^{2} g\,\mathrm{d}\boldsymbol{\xi}
		=
		\frac{\Kxi}{2\lambda}H.
		\label{eq:B_def}
	\end{equation}
	The corresponding reduced Shakhov correction terms are
	\begin{equation}
		H^{+}
		=
		\int g^{+}\,\mathrm{d}\boldsymbol{\xi}
		=
		\frac{4(1-\Pr)\lambda^{2}}{5\rho}
		(\mathbf{u}-\mathbf{U})\cdot\mathbf{q}
		\left(
		2\lambda|\mathbf{u}-\mathbf{U}|^{2}-2-D
		\right)H,
		\label{eq:H_plus}
	\end{equation}
	and
	\begin{equation}
		B^{+}
		=
		\int \xi^{2} g^{+}\,\mathrm{d}\boldsymbol{\xi}
		=
		\frac{4(1-\Pr)\lambda^{2}}{5\rho}
		(\mathbf{u}-\mathbf{U})\cdot\mathbf{q}
		\,
		\frac{
			\left(2\lambda|\mathbf{u}-\mathbf{U}|^{2}-D\right)\Kxi-2K
		}{
			2\lambda
		}
		H.
		\label{eq:B_plus}
	\end{equation}
	These expressions are consistent with the implementation convention $\Kxi=K+3-D$; in a three-dimensional calculation, $\Kxi=K$.
	
	For brevity, the following finite-volume formulation is written for a generic distribution function $f$. In the actual computation, the same operations are applied separately to the reduced distributions $h$ and $b$.
	
	Let the physical domain $\Omega$ be partitioned into non-overlapping control volumes $\Omega_i$,
	\begin{equation}
		\Omega=\bigcup_i \Omega_i,
		\qquad
		\Omega_i\cap\Omega_j=\varnothing,
		\quad i\neq j.
		\label{eq:physical_partition}
	\end{equation}
	The molecular velocity space is discretized into velocity cells $\delta\mathbf{u}_k$. For a generic quantity $Q$, the velocity-space integration is approximated by
	\begin{equation}
		\int Q\,\mathrm{d}\mathbf{u}
		\approx
		\sum_k Q_k \omega_k,
		\label{eq:velocity_quadrature}
	\end{equation}
	where $Q_k$ is the cell-averaged value over $\delta\mathbf{u}_k$, and
	\begin{equation}
		\omega_k=\int_{\delta\mathbf{u}_k}\mathrm{d}\mathbf{u}
		\label{eq:omega_k}
	\end{equation}
	is the corresponding velocity-space volume.
	
	Integrating Eq.~\eqref{eq:bgk_shakhov}, or equivalently Eq.~\eqref{eq:reduced_bgk}, over a velocity cell $\delta\mathbf{u}_k$ gives the governing equation for the discrete distribution function,
	\begin{equation}
		\frac{\partial f_k}{\partial t}
		+
		\mathbf{u}_k\cdot\nabla f_k
		=
		\frac{f_k^{+}-f_k}{\tau},
		\label{eq:discrete_kinetic}
	\end{equation}
	where $f_k=f(\mathbf{x},t,\mathbf{u}_k)$.
	
	Integrating Eq.~\eqref{eq:discrete_kinetic} over the control volume $\Omega_i$ and over the time interval $[t^n,t^{n+1}]$ yields
	\begin{equation}
		f_{i,k}^{n+1}
		=
		f_{i,k}^{n}
		-
		\frac{1}{|\Omega_i|}
		\sum_{j\in N(i)} S_{ij}\,\mathcal{F}_{ij,k}
		+
		\int_{t^n}^{t^{n+1}}
		\frac{f_{i,k}^{+}-f_{i,k}}{\tau_i}\,\mathrm{d}t,
		\label{eq:fv_micro_update}
	\end{equation}
	where $f_{i,k}^{n}$ and $f_{i,k}^{n+1}$ are the cell-averaged microscopic distribution functions at time levels $t^n$ and $t^{n+1}$, respectively. Here, $N(i)$ is the set of face-neighboring cells of cell $i$, $S_{ij}$ is the area of the interface between cells $i$ and $j$, and $\mathcal{F}_{ij,k}$ is the time-integrated microscopic flux,
	\begin{equation}
		\mathcal{F}_{ij,k}
		=
		u_{k,n}
		\int_{t^n}^{t^{n+1}} f_{ij,k}(t)\,\mathrm{d}t,
		\label{eq:micro_flux}
	\end{equation}
	with
	\begin{equation}
		u_{k,n}=\mathbf{u}_k\cdot\mathbf{n}_{ij}.
		\label{eq:ukn}
	\end{equation}
	Here $\mathbf{n}_{ij}$ is the interface unit normal, and $f_{ij,k}(t)$ is the interface distribution function.
	
	Taking moments of Eq.~\eqref{eq:fv_micro_update} and using the compatibility condition gives the finite-volume update of the macroscopic conservative variables,
	\begin{equation}
		\mathbf{W}_i^{n+1}
		=
		\mathbf{W}_i^{n}
		-
		\frac{1}{|\Omega_i|}
		\sum_{j\in N(i)} S_{ij}\,\mathbf{F}_{ij},
		\label{eq:macro_update}
	\end{equation}
	where the macroscopic flux is evaluated from the time-integrated microscopic fluxes as
	\begin{equation}
		\mathbf{F}_{ij}
		=
		\sum_k
		\omega_k
		\begin{pmatrix}
			\mathcal{H}_{ij,k}\\
			\mathbf{u}_k \mathcal{H}_{ij,k}\\
			\frac{1}{2}\left(|\mathbf{u}_k|^{2}\mathcal{H}_{ij,k}+\mathcal{B}_{ij,k}\right)
		\end{pmatrix}.
		\label{eq:macro_flux}
	\end{equation}
	Here, $\mathcal{H}_{ij,k}$ and $\mathcal{B}_{ij,k}$ are the time-integrated fluxes associated with the reduced distributions $h$ and $b$:
	\begin{equation}
		\mathcal{H}_{ij,k}
		=
		u_{k,n}\int_{t^n}^{t^{n+1}} h_{ij,k}(t)\,\mathrm{d}t,
		\qquad
		\mathcal{B}_{ij,k}
		=
		u_{k,n}\int_{t^n}^{t^{n+1}} b_{ij,k}(t)\,\mathrm{d}t.
		\label{eq:HB_flux}
	\end{equation}
	
	The essential ingredient of UGKS is the time evolution of the interface distribution function. Along the characteristic line, the integral solution of Eq.~\eqref{eq:discrete_kinetic} is
	\begin{equation}
		f_k(\mathbf{x}_0,t)
		=
		\frac{1}{\tau}
		\int_{0}^{t}
		f_k^{+}(\mathbf{x}',t')
		\exp\!\left(-\frac{t-t'}{\tau}\right)\,\mathrm{d}t'
		+
		\exp\!\left(-\frac{t}{\tau}\right)
		f_{0,k}(\mathbf{x}_0-\mathbf{u}_k t),
		\label{eq:integral_solution}
	\end{equation}
	where $f_{0,k}$ is the initial distribution function at the beginning of the time step.
	
	Without loss of generality, the interface center is taken as $\mathbf{x}_0=\mathbf{0}$. To achieve second-order accuracy, the initial distribution at the interface is reconstructed from the left and right sides,
	\begin{equation}
		f_{0,k}(\mathbf{x})
		=
		\begin{cases}
			f_{k}^{l}+\mathbf{x}\cdot\nabla f_{k}^{l}, & u_{k,n}>0,\\[4pt]
			f_{k}^{r}+\mathbf{x}\cdot\nabla f_{k}^{r}, & u_{k,n}<0,
		\end{cases}
		\label{eq:initial_reconstruction}
	\end{equation}
	where $f_k^{l}$ and $f_k^{r}$ are the reconstructed interface values from the left and right cells, respectively.
	
	The equilibrium state around the interface is approximated by a first-order expansion in space and time,
	\begin{equation}
		f_k^{+}(\mathbf{x},t)
		\approx
		g_{0,k}+g_{0,k}^{+}
		+
		\mathbf{x}\cdot\nabla g_{0,k}
		+
		t\,\partial_t g_{0,k},
		\label{eq:equilibrium_expansion}
	\end{equation}
	where $g_{0,k}$ is the interface Maxwellian determined from the conservative variables transported from both sides of the interface, and $g_{0,k}^{+}$ is the corresponding Shakhov correction.
	
	Substituting Eqs.~\eqref{eq:initial_reconstruction} and~\eqref{eq:equilibrium_expansion} into Eq.~\eqref{eq:integral_solution}, the interface distribution function can be expressed as
	\begin{equation}
		f_k(\mathbf{0},t)
		=
		\begin{cases}
			c_1 f_k^{l}
			+
			c_2\,\mathbf{u}_k\cdot\nabla f_k^{l}
			+
			c_3\left(g_{0,k}+g_{0,k}^{+}\right)
			+
			c_4\,\mathbf{u}_k\cdot\nabla g_{0,k}
			+
			c_5\,\partial_t g_{0,k},
			& u_{k,n}>0,\\[6pt]
			c_1 f_k^{r}
			+
			c_2\,\mathbf{u}_k\cdot\nabla f_k^{r}
			+
			c_3\left(g_{0,k}+g_{0,k}^{+}\right)
			+
			c_4\,\mathbf{u}_k\cdot\nabla g_{0,k}
			+
			c_5\,\partial_t g_{0,k},
			& u_{k,n}<0,
		\end{cases}
		\label{eq:interface_distribution}
	\end{equation}
	with
	\begin{equation}
		c_1=\exp\!\left(-\frac{t}{\tau}\right),
		\qquad
		c_2=-t\exp\!\left(-\frac{t}{\tau}\right),
		\qquad
		c_3=1-\exp\!\left(-\frac{t}{\tau}\right),
		\label{eq:c123}
	\end{equation}
	and
	\begin{equation}
		c_4=
		t\exp\!\left(-\frac{t}{\tau}\right)
		-
		\tau\left(1-\exp\!\left(-\frac{t}{\tau}\right)\right),
		\qquad
		c_5=
		t
		-
		\tau\left(1-\exp\!\left(-\frac{t}{\tau}\right)\right).
		\label{eq:c45}
	\end{equation}
	Equation~\eqref{eq:interface_distribution} gives a multiscale interface evolution in which free transport and collisional relaxation are coupled within one time step.
	
	Boundary fluxes are constructed through ghost-cell treatment. For artificial boundaries, such as inflow, outflow, and far-field boundaries, the ghost-cell conservative variables are specified by the corresponding macroscopic boundary relations. The microscopic distribution in the ghost cell is then taken as the Maxwellian associated with those conservative variables, and the interface distribution is selected according to the sign of the molecular normal velocity.
	
	For solid walls, a diffuse-reflection Maxwell boundary condition is employed. The wall distribution function is written as
	\begin{equation}
		f_{w,k}
		=
		\begin{cases}
			f_{\mathrm{in},k}
			+
			\mathbf{u}_k\cdot\nabla f_{\mathrm{in},k}\,t,
			& u_{k,n}>0,\\[4pt]
			g_{w,k},
			& u_{k,n}<0,
		\end{cases}
		\label{eq:wall_bc}
	\end{equation}
	where $u_{k,n}=\mathbf{u}_k\cdot\mathbf{n}_w$ is the molecular normal velocity with respect to the wall normal $\mathbf{n}_w$, $f_{\mathrm{in},k}$ is obtained by one-sided interpolation from the interior side, and $g_{w,k}$ is the wall Maxwellian,
	\begin{equation}
		g_{w}
		=
		\rho_w
		\left(\frac{m}{2\pi k_B T_w}\right)^{\frac{D+\Kxi}{2}}
		\exp\!\left(
		-\frac{m\left(|\mathbf{u}-\mathbf{U}_w|^2+\xi^2\right)}{2k_B T_w}
		\right),
		\label{eq:wall_maxwellian}
	\end{equation}
	with $T_w$ and $\mathbf{U}_w$ denoting the prescribed wall temperature and wall velocity. The wall density $\rho_w$ is determined from the non-penetration condition
	\begin{equation}
		\sum_k
		\omega_k
		u_{k,n}
		\int_{t^n}^{t^{n+1}} h_{w,k}(t)\,\mathrm{d}t
		=
		0.
		\label{eq:wall_density_condition}
	\end{equation}
	Once the wall microscopic flux is obtained, the corresponding macroscopic surface quantities are evaluated from the moments of the wall flux.
	
	The collision source term is discretized in time by the trapezoidal rule,
	\begin{equation}
		\int_{t^n}^{t^{n+1}}
		\frac{f_{i,k}^{+}-f_{i,k}}{\tau_i}\,\mathrm{d}t
		=
		\frac{\Delta t}{2}
		\left(
		\frac{f_{i,k}^{+,n+1}-f_{i,k}^{n+1}}{\tau_i^{n+1}}
		+
		\frac{f_{i,k}^{+,n}-f_{i,k}^{n}}{\tau_i^{n}}
		\right),
		\label{eq:trapezoidal_source}
	\end{equation}
	where $f_{i,k}^{+,n+1}$ and $\tau_i^{n+1}$ are determined from the updated macroscopic conservative variables $\mathbf{W}_i^{n+1}$.
	
	Substituting Eq.~\eqref{eq:trapezoidal_source} into Eq.~\eqref{eq:fv_micro_update}, the fully discrete microscopic update becomes
	\begin{equation}
		f_{i,k}^{n+1}
		=
		\left(
		1+\frac{\Delta t}{2\tau_i^{n+1}}
		\right)^{-1}
		\left(
		f_{i,k}^{n}
		-
		\frac{1}{|\Omega_i|}
		\sum_{j\in N(i)}S_{ij}\,\mathcal{F}_{ij,k}
		+
		\frac{\Delta t}{2}
		\left(
		\frac{f_{i,k}^{+,n+1}}{\tau_i^{n+1}}
		+
		\frac{f_{i,k}^{+,n}-f_{i,k}^{n}}{\tau_i^{n}}
		\right)
		\right).
		\label{eq:full_discrete_micro}
	\end{equation}
	
	A direct evaluation of Eq.~\eqref{eq:full_discrete_micro} requires flux and reconstruction-related quantities for all discrete velocity points to be available simultaneously. For large three-dimensional computations, this leads to substantial memory consumption. To reduce the memory footprint, the microscopic update is reformulated into two stages.
	
	In the first stage, an intermediate distribution function is defined as
	\begin{equation}
		\widetilde{f}_{i,k}^{\,n+1}
		=
		f_{i,k}^{n}
		-
		\frac{1}{|\Omega_i|}
		\sum_{j\in N(i)} S_{ij}\,\mathcal{F}_{ij,k}
		+
		\frac{\Delta t}{2\tau_i^{n}}
		\left(
		f_{i,k}^{+,n}-f_{i,k}^{n}
		\right).
		\label{eq:f_tilde}
	\end{equation}
	This step accounts for transport and the old-time part of the collision term. Since the update is carried out sequentially in velocity space, only the microscopic information associated with the currently processed velocity point or velocity block needs to be stored.
	
	After the macroscopic conservative variables have been updated to time level $t^{n+1}$, the new equilibrium state $f_{i,k}^{+,n+1}$ and the new relaxation time $\tau_i^{n+1}$ are obtained from the updated macroscopic solution. The second-stage update is
	\begin{equation}
		f_{i,k}^{n+1}
		=
		\frac{
			\widetilde{f}_{i,k}^{\,n+1}
			+
			\dfrac{\Delta t}{2\tau_i^{n+1}}\,f_{i,k}^{+,n+1}
		}{
			1+\dfrac{\Delta t}{2\tau_i^{n+1}}
		}.
		\label{eq:f_final}
	\end{equation}
	The two-stage formulation is algebraically equivalent to the trapezoidal collision discretization in Eq.~\eqref{eq:full_discrete_micro}, but it is more suitable for memory-saving implementation. Once $\widetilde{f}_{i,k}^{\,n+1}$ has been obtained, the storage associated with $f_{i,k}^{n}$ can be reused. Therefore, microscopic variables can be advanced sequentially in velocity space without storing full-velocity gradients and residuals at the same time. In the reduced formulation, the same procedure is applied separately to $h$ and $b$.
	
	\subsection{GPGPU-oriented full phase-space parallel UGKS with block pipelining}
	\label{subsec:gpgpu_full_phase_space_ugks}
	
	The two-stage microscopic update introduced above allows distribution functions to be advanced sequentially in velocity space. It follows the memory-saving principle that microscopic gradients and residuals need not be retained for the complete velocity space at the same time~\cite{Zhang2025}. For GPGPU architectures, however, a pointwise velocity sweep is inefficient: a single velocity point provides insufficient parallel granularity for device kernels, halo exchange at the single-velocity level exposes excessive communication latency, and the same geometric and macroscopic interface data are repeatedly loaded from global memory. The present method therefore reorganizes microscopic evolution into a velocity-block sweep and combines it with a full phase-space MPI decomposition.
	
	\subsubsection{Full phase-space MPI decomposition}
	
	Let the total number of MPI ranks be $P=P_xP_v$, where $P_x$ is the number of physical-space partitions and $P_v$ is the number of velocity-space partitions. The global rank is mapped to the pair
	\begin{equation}
		r=pP_v+q,
		\qquad
		0\le p<P_x,
		\quad
		0\le q<P_v,
		\label{eq:rank_mapping}
	\end{equation}
	where $p$ is the physical-space partition index and $q$ is the velocity-space partition index. The physical domain and the discrete velocity space are decomposed as
	\begin{equation}
		\Omega=\bigcup_{p=0}^{P_x-1}\Omega^{(p)},
		\qquad
		\mathcal{V}=\bigcup_{q=0}^{P_v-1}\mathcal{V}^{(q)}.
		\label{eq:phase_space_partition}
	\end{equation}
	Rank $(p,q)$ therefore stores and advances the local phase-space block
	\begin{equation}
		\mathcal{D}_{p,q}=\Omega^{(p)}\times\mathcal{V}^{(q)}.
		\label{eq:local_phase_space_block}
	\end{equation}
	The physical-space communicator $\mathcal{C}_x(q)$ contains ranks with the same velocity index $q$ and different physical indices $p$; it is used for cell-halo and microscopic-gradient exchange across physical subdomain interfaces. The velocity-space communicator $\mathcal{C}_v(p)$ contains ranks with the same physical index $p$ and different velocity indices $q$; it is used for reductions and broadcasts of macroscopic quantities whose evaluation requires moments over the full velocity space.
	
	In the implementation, $P_v$ is specified by the input parameter, and the global MPI size must be divisible by $P_v$. The physical mesh is partitioned independently within each physical-space communicator, while the velocity-space communicator couples ranks that own the same physical subdomain but different velocity subsets. This organization reduces the persistent microscopic storage per rank from full-velocity storage to local-velocity storage, at the cost of low-dimensional moment reductions over the velocity communicator.
	
	\subsubsection{Local velocity-block sweeping}
	
	The global discrete velocity space is indexed as
	\begin{equation}
		\mathcal{V}=\{\mathbf{u}_0,\mathbf{u}_1,\ldots,\mathbf{u}_{N_v-1}\}.
		\label{eq:velocity_space_set}
	\end{equation}
	It is padded, if necessary, and divided into fixed-size velocity blocks. Let
	\begin{equation}
		M=\left\lceil \frac{N_v}{B_v}\right\rceil,
		\qquad
		N_v^{\ast}=MB_v,
		\label{eq:padded_velocity_number}
	\end{equation}
	where $B_v$ is the number of discrete velocity points processed in one device block. The padded entries $N_v\le k<N_v^{\ast}$ are assigned zero quadrature weights, so the velocity quadrature is unchanged. The padded velocity space is then partitioned as
	\begin{equation}
		\mathcal{V}^{\ast}
		=
		\bigcup_{m=0}^{M-1}\mathcal{V}_m,
		\qquad
		\mathcal{V}_m
		=
		\{\mathbf{u}_k\mid mB_v\le k < (m+1)B_v\}.
		\label{eq:velocity_block_partition}
	\end{equation}
	The parameter $B_v$ is a hardware- and case-dependent blocking parameter rather than a physical parameter of the UGKS discretization. It controls the granularity of GPGPU parallelism, the amount of data reuse inside a thread group, the size of microscopic MPI messages, and the temporary storage required by one sweep. In practice, $B_v$ is selected according to the velocity-space resolution, mesh size, accelerator occupancy, available memory, and communication environment.
	
	The velocity-space MPI partition is applied to the sequence of velocity blocks rather than to individual velocity points:
	\begin{equation}
		\mathcal{V}^{(q)}
		=
		\bigcup_{m=L_q}^{L_q+M_q-1}\mathcal{V}_m,
		\qquad
		\sum_{q=0}^{P_v-1}M_q=M,
		\qquad
		|M_q-M_{q'}|\le 1.
		\label{eq:velocity_mpi_block_partition}
	\end{equation}
	Here $L_q$ is the first global velocity-block index owned by velocity rank $q$. Thus each velocity rank owns a contiguous set of complete velocity blocks. This matches the local memory layout used by the device kernels and avoids splitting a single $B_v$-point execution block across MPI ranks.
	
	\begin{figure}[htbp]
		\centering
		\includegraphics[width=0.98\linewidth]{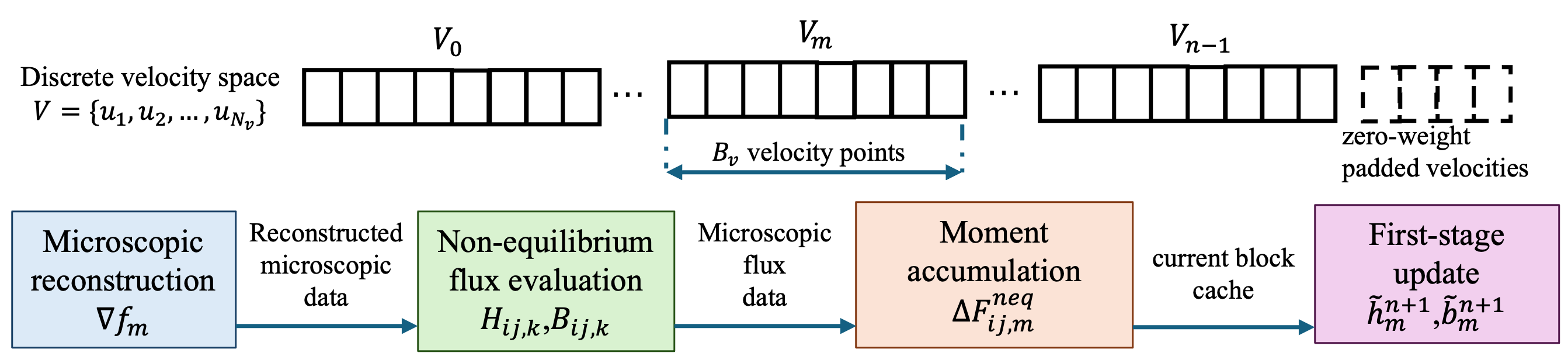}
		\caption{Schematic of the velocity-block sweeping strategy. The discrete velocity space is divided into blocks of size $B_v$. Each block is used as the basic unit for microscopic reconstruction, nonequilibrium flux evaluation, moment accumulation, and the first-stage microscopic update.}
		\label{fig:velocity_block_sweeping}
	\end{figure}
	
	With this organization, the macroscopic flux in Eq.~\eqref{eq:macro_flux} is accumulated in a blockwise and velocity-distributed form. On velocity rank $q$, the local contribution is
	\begin{equation}
		\mathbf{F}_{ij}^{(q)}
		=
		\delta_{q0}\mathbf{F}^{\mathrm{eq}}_{ij}
		+
		\sum_{m=L_q}^{L_q+M_q-1}\Delta\mathbf{F}^{\mathrm{neq}}_{ij,m},
		\label{eq:block_macro_flux_local}
	\end{equation}
		where $\delta_{q0}$ indicates that the equilibrium flux is owned only by the root velocity rank, chosen as $q=0$ in each communicator $\mathcal{C}_v(p)$. All non-root velocity ranks initialize the equilibrium part of their local face-flux accumulator to zero. The full macroscopic flux is
	\begin{equation}
		\mathbf{F}_{ij}
		=
		\sum_{q=0}^{P_v-1}\mathbf{F}_{ij}^{(q)}.
		\label{eq:block_macro_flux}
	\end{equation}
		Here, $\mathbf{F}^{\mathrm{eq}}_{ij}$ denotes the contribution determined by the interface equilibrium state and its derivatives. It depends only on macroscopic reconstructed quantities and is computed once on the root velocity rank before the velocity-block sweep. The interface equilibrium state, time-integration coefficients, and derivative coefficients needed by the nonequilibrium flux kernel are then broadcast from the root velocity rank to all other ranks in $\mathcal{C}_v(p)$. These broadcast data are read-only coefficients; the equilibrium flux accumulator itself is not added on the non-root velocity ranks. After the local block sweeps, a sum reduction over $\mathcal{C}_v(p)$ combines the local accumulators in Eq.~\eqref{eq:block_macro_flux_local}. Because $\mathbf{F}^{\mathrm{eq}}_{ij}$ appears only through $\delta_{q0}$, it is included exactly once, while the nonequilibrium contributions from all velocity blocks are included once through their owning velocity ranks. The nonequilibrium contribution from velocity block $\mathcal{V}_m$ is
	\begin{equation}
		\Delta\mathbf{F}^{\mathrm{neq}}_{ij,m}
		=
		\sum_{\mathbf{u}_k\in\mathcal{V}_m}
		\omega_k
		\begin{pmatrix}
			\mathcal{H}^{\mathrm{neq}}_{ij,k}\\
			\mathbf{u}_k\mathcal{H}^{\mathrm{neq}}_{ij,k}\\
			\dfrac{1}{2}\left(|\mathbf{u}_k|^2\mathcal{H}^{\mathrm{neq}}_{ij,k}+\mathcal{B}^{\mathrm{neq}}_{ij,k}\right)
		\end{pmatrix},
		\label{eq:block_noneq_flux}
	\end{equation}
		where $\mathcal{H}^{\mathrm{neq}}_{ij,k}$ and $\mathcal{B}^{\mathrm{neq}}_{ij,k}$ are the nonequilibrium parts of the time-integrated microscopic fluxes associated with the reduced distribution functions. After the contribution of one local velocity block has been accumulated into the velocity-rank local macroscopic interface-flux accumulator, the same block of microscopic fluxes is immediately consumed by Eq.~\eqref{eq:f_tilde} to update the intermediate distributions $\widetilde{h}^{\,n+1}$ and $\widetilde{b}^{\,n+1}$. Therefore, the microscopic flux cache is required only for the current velocity block, rather than for the complete velocity space.
	
	A finite velocity block is essential for matching the memory hierarchy of modern GPGPU accelerators. Similar to GPU-oriented CFD implementations, which must expose sufficient thread-level parallelism while controlling memory traffic and communication granularity~\cite{Corrigan2011UnstructuredGPU,Klockner2009DGGPUs,JacobsenSenocak2013MultiLevelGPU,Witherden2014PyFR}, the present block design increases the amount of work assigned to each device kernel without changing the velocity quadrature. For a given physical interface, the face normal, cell indices, geometric offsets, interface equilibrium state, time-integration coefficients, and macroscopic derivative coefficients are shared by all velocity points in the block. These quantities can be loaded once and reused by $B_v$ velocity threads. At the same time, the moment contribution in Eq.~\eqref{eq:block_noneq_flux} can be reduced locally within the thread group before being written to global memory. This reduces global-memory traffic and avoids excessive pointwise accumulation of macroscopic fluxes.
	
	\begin{algorithm}[htbp]
		\caption{Blockwise microscopic contribution and first-stage update.}
		\label{alg:block_microscopic_evolution}
		\KwInput{Velocity block $\mathcal{V}_m$; local and ghost-cell distributions $h_m^n,b_m^n$; gradients $\nabla h_m,\nabla b_m$; precomputed interface equilibrium data.}
		\KwOutput{Block contribution $\Delta\mathbf{F}_{ij,m}^{\mathrm{neq}}$ and intermediate distributions $\widetilde{h}_m^{\,n+1},\widetilde{b}_m^{\,n+1}$.}
		Initialize the block microscopic flux cache for $\mathcal{V}_m$\;
		\ForEach{interface $ij$ assigned to the GPGPU thread group}{
			Load interface geometry and precomputed macroscopic interface data once\;
			\ForEach{velocity point $\mathbf{u}_k\in\mathcal{V}_m$ in parallel}{
				Reconstruct the upwind values of $h_{ij,k}$ and $b_{ij,k}$ from the corresponding side of the interface\;
				Evaluate the time-integrated microscopic fluxes $\mathcal{H}_{ij,k}$ and $\mathcal{B}_{ij,k}$\;
				Form the weighted conservative-moment contribution for $\mathbf{u}_k$\;
			}
			Reduce the velocity-moment contributions inside the thread group\;
				Accumulate the reduced block contribution into the velocity-rank local face-flux accumulator\;
		}
		\ForEach{cell $i$ assigned to the GPGPU thread group}{
			Load cell-level macroscopic quantities once\;
			\ForEach{velocity point $\mathbf{u}_k\in\mathcal{V}_m$ in parallel}{
				Gather the microscopic fluxes through all faces of cell $i$\;
				Update $h_{i,k}^{n},b_{i,k}^{n}$ to $\widetilde{h}_{i,k}^{\,n+1},\widetilde{b}_{i,k}^{\,n+1}$ by the first-stage update\;
			}
		}
	\end{algorithm}
	
	\subsubsection{Triple-buffered microscopic reconstruction}
	
	In a multi-accelerator computation with the above phase-space decomposition, physical-halo exchange is performed inside each physical-space communicator $\mathcal{C}_x(q)$. Since ranks with different velocity indices own different velocity subsets, no microscopic distribution function is exchanged across $\mathcal{C}_v(p)$; the velocity-space communicator is used only for macroscopic reductions and broadcasts. For a local velocity block $\mathcal{V}_m\subset\mathcal{V}^{(q)}$, microscopic reconstruction near physical subdomain interfaces requires distribution functions in ghost cells, whereas nonequilibrium flux evaluation at parallel interfaces requires distribution-function gradients in ghost cells. Thus two types of physical-halo data are involved: the distribution values $f_m$ required before reconstruction and the gradients $\nabla f_m$ required before flux evaluation.
	
	A strictly sequential procedure would perform, for every velocity block, halo exchange of $f_m$, local reconstruction of $\nabla f_m$, halo exchange of $\nabla f_m$, and finally nonequilibrium flux calculation. Such an ordering exposes communication latency at every block. To avoid this serialization, the present method uses a triple-buffered velocity-block pipeline. During the sweep of block $\mathcal{V}_m$, three data lifetimes coexist:
	\begin{equation}
		\begin{aligned}
			&\text{current block:} && \nabla f_m \quad \text{is used for nonequilibrium flux evaluation},\\
			&\text{next block:} && \nabla f_{m+1} \quad \text{is packed and exchanged through MPI},\\
			&\text{future block:} && \nabla f_{m+2} \quad \text{is reconstructed on the device}.
		\end{aligned}
		\label{eq:three_lifetimes}
	\end{equation}
	Three rotating gradient buffers are therefore required to decouple these lifetimes. One buffer is read by the current flux kernel, one buffer is associated with in-flight halo communication, and one buffer is written by the reconstruction kernel. This design prevents overwrite conflicts and allows reconstruction, communication, and flux evaluation to proceed in a staggered manner.
	
	The microscopic communication package sent during the sweep of block $\mathcal{V}_m$ contains both the gradients needed by the next block and the distribution values needed by a future reconstruction. In abstract form,
	\begin{equation}
		\mathcal{P}_m
		=
		\left\{f_{m+3},\,\nabla f_{m+1}\right\}.
		\label{eq:mpi_packet_generic}
	\end{equation}
	For the present reduced formulation, this package becomes
	\begin{equation}
		\mathcal{P}_m
		=
		\left\{
		h_{m+3},\,b_{m+3},\,
		\nabla h_{m+1},\,\nabla b_{m+1}
		\right\}.
		\label{eq:mpi_packet_hb}
	\end{equation}
	The look-ahead distance in Eq.~\eqref{eq:mpi_packet_hb} follows from the data dependency of the pipeline: the gradient of block $m+1$ is needed for the next flux calculation, while the distribution values of block $m+3$ must be available before the reconstruction of that future block. At the beginning and end of a local sweep, unavailable look-ahead entries are treated as padded or empty entries.
	
	\begin{algorithm}[htbp]
		\caption{Triple-buffered scheduling for velocity-block sweeping.}
		\label{alg:triple_buffer_schedule}
			\KwInput{Current velocity-block index $m$; rotating gradient buffers; computation stream $S_c$; communication stream $S_p$; velocity-rank local face-flux accumulator.}
			\KwOutput{Scheduled execution of Algorithm~\ref{alg:block_flux_kernel}, first-stage update, next-block halo exchange, and future-block reconstruction.}
		Select the current, communication, and reconstruction buffers by cyclic rotation: $G^{\mathrm{cur}},G^{\mathrm{comm}},G^{\mathrm{rec}}$\;
		Wait until $G^{\mathrm{cur}}$ contains the unpacked ghost gradients $\nabla f_m$\;
			Launch Algorithm~\ref{alg:block_flux_kernel} for $\mathcal{V}_m$ on $S_c$ using $G^{\mathrm{cur}}$ and accumulate its block contribution locally\;
			Launch the first-stage microscopic update of $\mathcal{V}_m$ on $S_c$ after the current-block microscopic flux cache is produced\;
		Launch the reconstruction of $\nabla f_{m+2}$ into $G^{\mathrm{rec}}$ on $S_c$\;
		\If{$G^{\mathrm{comm}}$ contains the reconstructed local gradients $\nabla f_{m+1}$}{
			Pack the communication package $\mathcal{P}_m=\{f_{m+3},\nabla f_{m+1}\}$ on $S_p$\;
			Start the non-blocking halo exchange for $\mathcal{P}_m$\;
			Unpack the received ghost data after the MPI requests are completed\;
			Record a data-ready event for block $\mathcal{V}_{m+1}$\;
		}
		Before launching the flux evaluation of $\mathcal{V}_{m+1}$, wait only for its data-ready event\;
	\end{algorithm}
	
	\begin{figure}[htbp]
		\centering
		\includegraphics[width=0.94\linewidth]{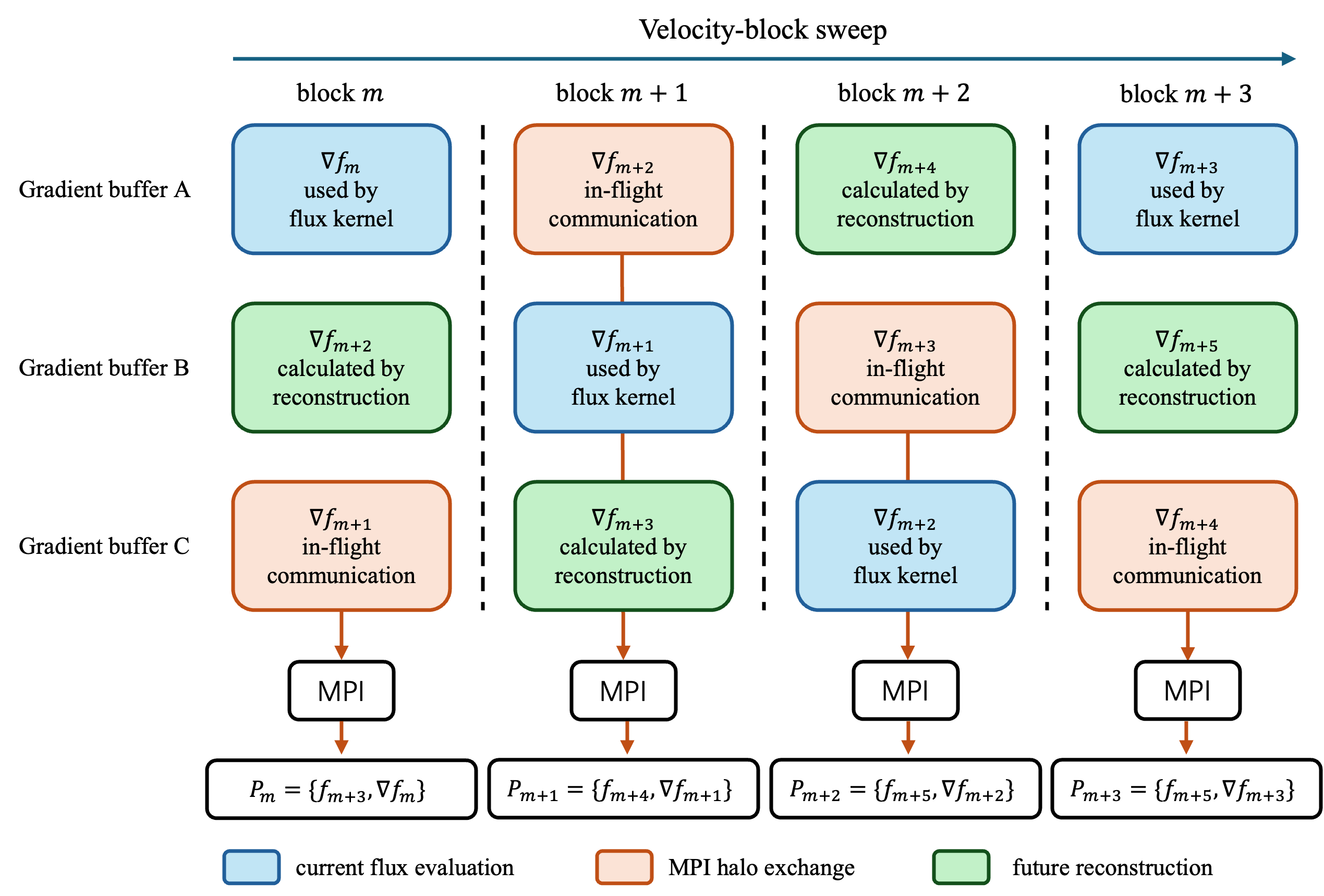}	
		\caption{Triple-buffered velocity-block pipeline. During the sweep of velocity block $m$, the current gradient buffer is consumed by nonequilibrium flux evaluation, another gradient buffer is associated with MPI halo exchange for block $m+1$, and the third buffer is written by the reconstruction of block $m+2$. The communication package contains $\{f_{m+3},\nabla f_{m+1}\}$.}
		\label{fig:triple_buffer_pipeline}
	\end{figure}
	
	Let $N_r$ be the number of reduced distribution functions. In the present $h/b$ formulation, $N_r=2$. For one MPI interface cell and one velocity block, the number of communicated scalar values is
	\begin{equation}
		N_{\mathrm{comm}}^{\mathrm{block}}
		=
		N_r(D+1)B_v,
		\label{eq:block_comm_size}
	\end{equation}
	where $N_rB_v$ corresponds to distribution values and $N_rDB_v$ corresponds to distribution gradients in $D$ physical dimensions. For the three-dimensional reduced formulation, this becomes
	\begin{equation}
		N_{\mathrm{comm}}^{\mathrm{block}}=8B_v.
		\label{eq:block_comm_size_3d}
	\end{equation}
	The communication granularity is therefore enlarged from a single velocity point to a velocity block. This reduces the relative impact of MPI latency while keeping the working set bounded by $B_v$.
	
	\subsubsection{Device-stream overlap and exposed communication cost}
	
	The velocity-block pipeline is executed by two asynchronous device-side streams. The computation stream performs macroscopic reconstruction, microscopic reconstruction, interface equilibrium precomputation, nonequilibrium flux evaluation, local velocity-moment reduction, and microscopic update. The communication stream performs halo-data packing, host-device staging when required, non-blocking MPI communication, and halo-data unpacking. The two streams are synchronized only at true data-dependency points.
	
	For example, the communication stream must wait until the reconstruction of $\nabla f_{m+1}$ has been completed before packing the gradient part of the communication package. Conversely, the computation stream must wait until the ghost-cell data for block $\mathcal{V}_{m+1}$ have been unpacked before launching the next flux evaluation. Between these dependency points, GPGPU computation and MPI communication proceed asynchronously. Thus, while the current velocity block is used for flux evaluation and intermediate update, halo exchange for the next block can progress concurrently. This overlap should be interpreted as a latency-reduction mechanism rather than complete communication hiding: if MPI waiting time or host-device staging time exceeds the available computation window, the remaining cost is still exposed in the timeline.
	
	\begin{algorithm}[htbp]
		\caption{Microscopic halo package exchange in the communication stream.}
		\label{alg:microscopic_halo_exchange}
		\KwInput{Send-side interface cells; receive-side ghost cells; distribution block $f_{m+3}$; gradient block $\nabla f_{m+1}$; communication stream $S_p$.}
		\KwOutput{Ghost-cell values of $f_{m+3}$ and $\nabla f_{m+1}$.}
		Wait on $S_p$ until the reconstruction event of $\nabla f_{m+1}$ is satisfied\;
		Pack $h_{m+3},b_{m+3},\nabla h_{m+1},\nabla b_{m+1}$ into the send buffer\;
		\eIf{device-aware MPI or RDMA is available}{
			Call non-blocking MPI routines with device send and receive buffers\;
		}{
			Asynchronously copy the packed send buffer from device memory to pinned host memory\;
			Call non-blocking MPI routines with pinned host buffers\;
			Asynchronously copy the received buffer from pinned host memory to device memory\;
		}
		Wait for the corresponding MPI requests before the received data are unpacked\;
		Unpack the received package into ghost-cell storage on $S_p$\;
		Record the data-ready event used by the computation stream\;
	\end{algorithm}
	
	\begin{figure}[htbp]
		\centering
		\includegraphics[width=0.94\linewidth]{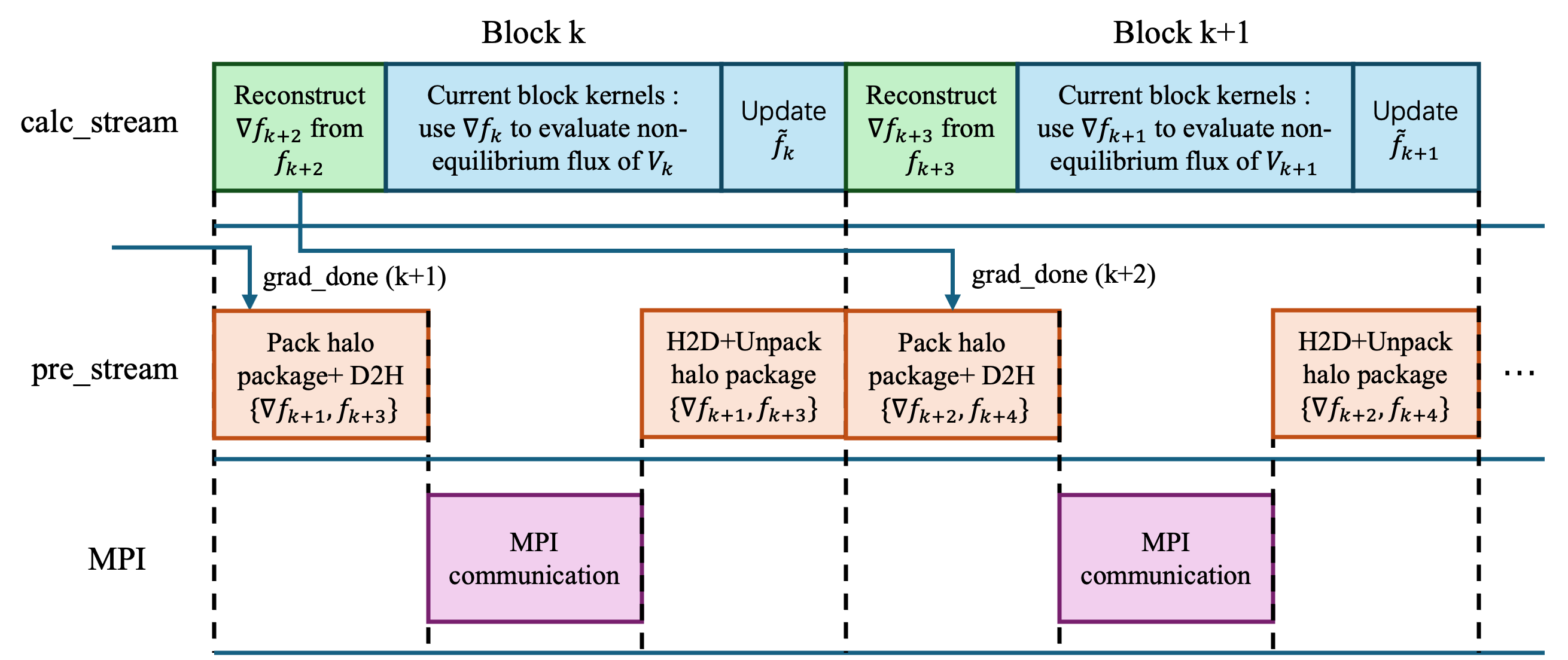}	
		\caption{Schematic of the two-stream execution model. The computation stream evaluates reconstruction, nonequilibrium fluxes, local reductions, and microscopic updates. The communication stream packs halo data, performs non-blocking MPI communication, and unpacks received data. Device events are used only at necessary dependency points, allowing part of the next-block communication to overlap with the computation of the current block.}
		\label{fig:device_mpi_overlap}
	\end{figure}
	
	The overlap is dependency-driven rather than unconditional. If the communication time is shorter than the available computation window of the current block, much of the MPI cost can be overlapped. If the communication time exceeds this window, the remaining part appears as exposed waiting time before the next block. In practical profiling, especially when host-device staging is required, MPI waiting time and device-host transfers may still dominate the end-to-end time of one velocity-block iteration. When device-aware MPI or RDMA is available, halo buffers can be transferred directly from device memory. Otherwise, pinned host memory is used as a staging buffer; in that case, asynchronous host-device transfers are still placed in the communication stream so that staging and network transfer can overlap with GPGPU computation as much as possible.
	
	\subsubsection{Block-level cache reuse and local moment reduction}
	
	The nonequilibrium interface flux calculation is well suited to velocity-block parallelization. A thread group is assigned to one physical interface and one velocity block. Interface-level quantities, such as geometric metrics, interface primitive variables, time-integration coefficients, and equilibrium derivative coefficients, are common to all velocity points in the block. These quantities are first loaded into registers or shared memory and then reused by the velocity threads. Each thread evaluates the contribution of one discrete velocity point, including the local coordinate transformation, upwind selection of the reconstructed distribution, gradient contribution, Shakhov correction, and microscopic flux construction.
	
		After the microscopic fluxes have been obtained, the same thread group performs a local reduction over the velocity block to compute the partial macroscopic moment in Eq.~\eqref{eq:block_noneq_flux}. Only the reduced block contribution is written back to the velocity-rank local face-flux accumulator. This organization increases arithmetic intensity because a relatively large amount of floating-point work is performed after loading the shared interface data, and it reduces global-memory pressure because the macroscopic moment is accumulated at the block level rather than at the velocity-point level.
	
	\begin{algorithm}[htbp]
		\caption{Block-level nonequilibrium interface flux kernel.}
		\label{alg:block_flux_kernel}
		\KwInput{Interface $ij$; velocity block $\mathcal{V}_m$; reconstructed distributions and gradients; interface equilibrium data.}
		\KwOutput{Microscopic flux cache for $\mathcal{V}_m$ and block-level macroscopic increment $\Delta\mathbf{F}^{\mathrm{neq}}_{ij,m}$.}
		Assign one GPGPU thread group to the pair $(ij,\mathcal{V}_m)$\;
		Load the face normal, geometric offsets, and macroscopic interface coefficients into the shared cache\;
		Initialize the local moment accumulator for $\Delta\mathbf{F}^{\mathrm{neq}}_{ij,m}$\;
		\ForEach{velocity point $\mathbf{u}_k\in\mathcal{V}_m$ handled by one thread}{
			Transform $\mathbf{u}_k$ and the distribution gradients to the local interface frame\;
			Select the left or right reconstructed distribution according to the sign of $\mathbf{u}_k\cdot\mathbf{n}_{ij}$\;
			Evaluate the free-transport term, gradient term, equilibrium derivative term, and Shakhov correction\;
			Store the time-integrated microscopic fluxes $\mathcal{H}_{ij,k}$ and $\mathcal{B}_{ij,k}$ in the current-block flux cache\;
			Add the weighted moment contribution to the thread-local accumulator\;
		}
		Reduce the thread-local accumulators inside the thread group\;
			Write only the reduced block contribution $\Delta\mathbf{F}^{\mathrm{neq}}_{ij,m}$ to the velocity-rank local face-flux accumulator\;
	\end{algorithm}
	
	A similar reuse pattern appears in the microscopic update. For a given physical cell and velocity block, the cell volume, macroscopic state, relaxation time, temperature, and heat flux are common to all velocity points in the block. These cell-level quantities are reused while the block of intermediate distributions is updated by Eq.~\eqref{eq:f_tilde}. Since the microscopic flux cache is consumed immediately after being produced, its lifetime is short and its active working set remains limited to the current velocity block.
	
	\begin{figure}[htbp]
		\centering
		\includegraphics[width=0.68\linewidth]{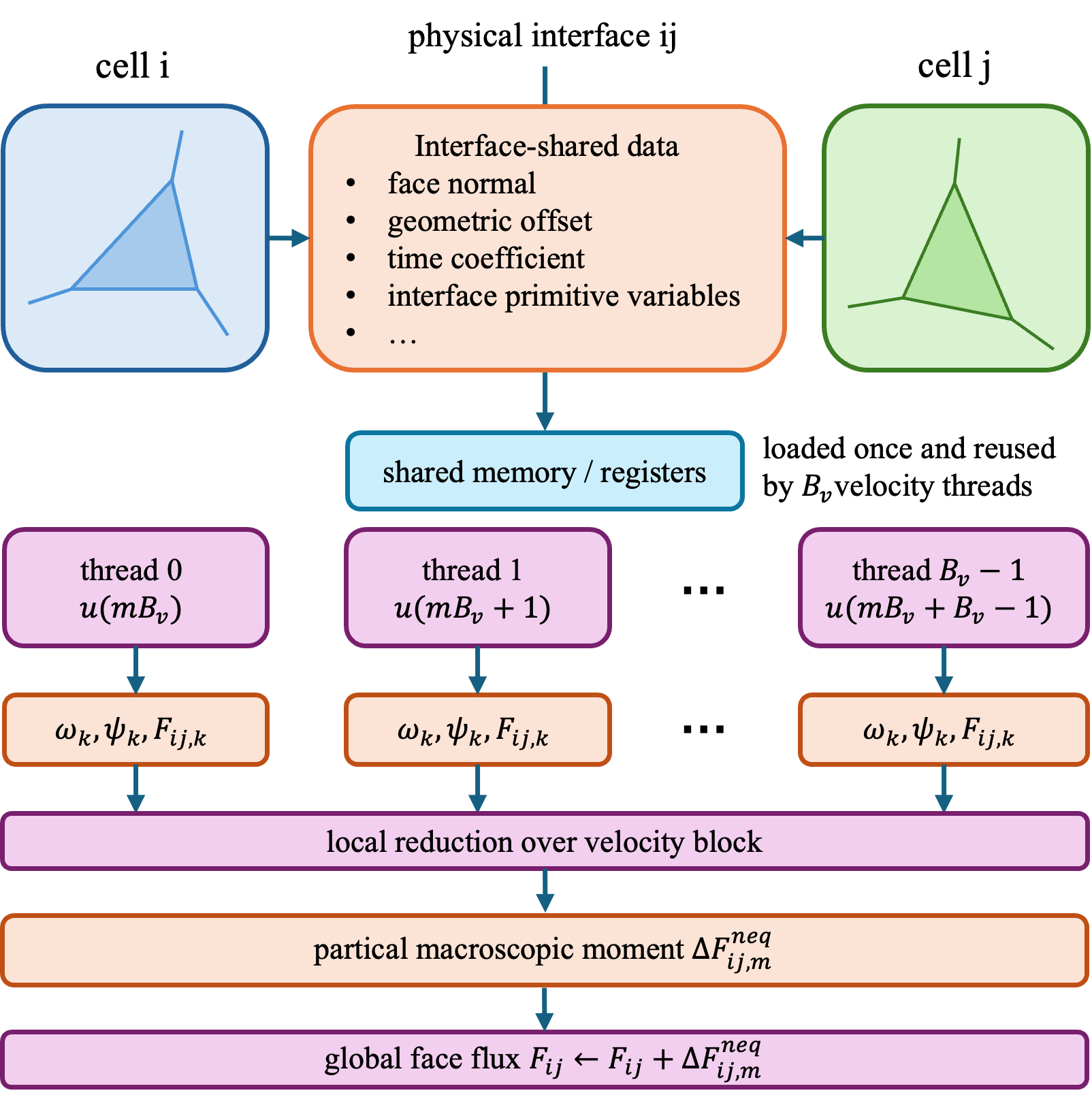}	
		\caption{Block-level data reuse and local moment reduction. For one physical interface, the geometric and macroscopic interface data are shared by all velocity threads in the block. Each thread computes one velocity-point contribution, and the partial macroscopic moment is reduced inside the thread group before a block-level contribution is written to global memory.}
		\label{fig:block_cache_reduction}
	\end{figure}
	
	\subsubsection{Memory footprint and suitability for GPGPU architectures}
	
	The full phase-space decomposition first reduces persistent microscopic storage by assigning only a local velocity subset to each velocity rank. Let $N_v^{(q)}=M_qB_v$ be the padded number of velocity points owned by velocity rank $q$, and let $N_c^{\mathrm{tot}}$ be the number of local cells including ghost cells. The persistent storage of the reduced distributions on rank $(p,q)$ scales as
	\begin{equation}
		\mathcal{O}\!
		\left(N_r N_c^{\mathrm{tot}} N_v^{(q)}\right),
		\label{eq:local_distribution_storage}
	\end{equation}
	instead of $\mathcal{O}(N_r N_c^{\mathrm{tot}}N_v^{\ast})$ for a physical-domain-only implementation. The velocity-block pipeline further reduces microscopic intermediate storage from the local velocity-space scale to the block scale. Let $N_f$ be the number of local faces. If all microscopic gradients were stored for the local velocity subset, the storage complexity would be
	\begin{equation}
		\mathcal{O}\!
		\left(N_r D N_c^{\mathrm{tot}} N_v^{(q)}\right).
		\label{eq:full_grad_storage}
	\end{equation}
	With the triple-buffered velocity-block pipeline, only three gradient buffers are retained, and the corresponding complexity becomes
	\begin{equation}
		\mathcal{O}\!
		\left(3N_r D N_c^{\mathrm{tot}} B_v\right).
		\label{eq:block_grad_storage}
	\end{equation}
	Similarly, the storage of microscopic interface fluxes is reduced from
	\begin{equation}
		\mathcal{O}\!
		\left(N_r N_f N_v^{(q)}\right)
		\label{eq:full_flux_storage}
	\end{equation}
	for full local velocity-space storage to
	\begin{equation}
		\mathcal{O}\!
		\left(N_r N_f B_v\right)
		\label{eq:block_flux_storage}
	\end{equation}
	for current-block storage. Since typically $N_v^{(q)}\gg B_v$, the reduction in temporary storage is substantial.
	
		This organization matches the high compute-to-memory-bandwidth ratio of modern GPGPU accelerators. The UGKS nonequilibrium flux evaluation involves local coordinate transformations, exponential Maxwellian evaluation, Shakhov correction, gradient expansion, time-integration coefficient combination, and velocity-moment accumulation. These operations provide abundant floating-point work once the required geometric and macroscopic data have been loaded. The velocity-block pipeline improves reuse of these data, confines temporary variables to a small working set, and reduces global-memory accumulation. With velocity-space MPI decomposition, each rank accumulates only a partial moment over its local velocity subset; heat fluxes, wall moments, and face fluxes are then completed by reductions over $\mathcal{C}_v(p)$. These reductions involve macroscopic arrays rather than full microscopic distribution arrays, so the dominant storage and physical-halo communication remain bounded by the local velocity-block working set.
		
		\subsubsection{Algebraic equivalence to the original UGKS discretization}
		
		The velocity-block sweep and the MPI velocity-space partition do not introduce a new physical approximation. They only reorganize the evaluation of the same discrete-velocity sums and the same microscopic update. To see this, consider any velocity moment appearing in the UGKS formulation,
		\begin{equation}
			\mathcal{M}
			=
			\sum_{k=0}^{N_v-1}
			\omega_k\,\boldsymbol{\phi}_k\,Q_k,
			\label{eq:original_velocity_moment}
		\end{equation}
		where $\boldsymbol{\phi}_k$ denotes the corresponding collision invariant or flux weight, and $Q_k$ may be a cell distribution, a wall distribution, a heat-flux integrand, or a time-integrated interface flux. After padding, $\omega_k=0$ for $N_v\le k<N_v^\ast$. Since the padded velocity blocks form a disjoint partition of the same velocity index set,
		\begin{equation}
			\mathcal{M}
			=
			\sum_{q=0}^{P_v-1}
			\sum_{m=L_q}^{L_q+M_q-1}
			\sum_{\mathbf{u}_k\in\mathcal{V}_m}
			\omega_k\,\boldsymbol{\phi}_k\,Q_k .
			\label{eq:partitioned_velocity_moment}
		\end{equation}
		Equation~\eqref{eq:partitioned_velocity_moment} is exactly the operation performed by the local block reductions followed by the MPI reduction over $\mathcal{C}_v(p)$. Therefore, the macroscopic conservative variables, heat fluxes, wall moments, and face fluxes obtained by the velocity-space partition are identical to those obtained by a direct full-velocity summation in exact arithmetic.
		
		For each fixed physical cell $i$ and velocity point $k$, the microscopic UGKS evolution is local in velocity once the macroscopic moments have been assembled. The interface distribution in Eq.~\eqref{eq:interface_distribution}, the microscopic fluxes in Eq.~\eqref{eq:HB_flux}, the first-stage update in Eq.~\eqref{eq:f_tilde}, and the second-stage update in Eq.~\eqref{eq:f_final} use the same reconstructed distribution, equilibrium state, relaxation time, heat flux, and time-integration coefficients as the original full-velocity implementation. Physical-subdomain halo exchange is still performed for the same velocity point $k$ inside $\mathcal{C}_x(q)$, so the left and right reconstructed states at each partition interface are unchanged. The velocity communicator is not used to exchange microscopic distributions; it only completes the moment sums in Eq.~\eqref{eq:partitioned_velocity_moment}.
		
		The equivalence follows by induction over time steps. Suppose that, at time level $t^n$, the block-pipelined phase-space implementation and the original full-velocity UGKS have the same $h_{i,k}^n$, $b_{i,k}^n$, and $\mathbf{W}_i^n$. The partitioned reductions then give the same heat flux and boundary moments, so the same Shakhov correction, wall density, interface equilibrium state, and equilibrium flux part are used. For every velocity point $k$, the block sweep evaluates the same time-integrated microscopic fluxes as the original scheme and applies the same first-stage update. Summing the block contributions by Eq.~\eqref{eq:partitioned_velocity_moment} gives the same macroscopic face flux, hence Eq.~\eqref{eq:macro_update} produces the same $\mathbf{W}_i^{n+1}$. The second-stage update then uses the same $\mathbf{W}_i^{n+1}$ and gives the same $h_{i,k}^{n+1}$ and $b_{i,k}^{n+1}$. Since the initial condition is identical, the complete discrete solution is unchanged at all time levels.
		
		Consequently, conservation, the coupled transport--collision flux construction, the trapezoidal collision discretization, and the multiscale numerical behavior of UGKS are preserved by the block sweep and by the MPI velocity-space decomposition. In floating-point arithmetic, different local and MPI reduction orders may lead only to round-off-level differences, as in standard parallel reductions; they do not represent a change of the UGKS discretization itself.
		
		The resulting time-marching procedure is summarized in Algorithm~\ref{alg:gpgpu_full_phase_space_ugks}.
	
	\begin{algorithm}[htbp]
		\caption{GPGPU-oriented full phase-space parallel UGKS with block pipelining for one time step.}
		\label{alg:gpgpu_full_phase_space_ugks}
		\KwInput{Local cell-averaged conservative variables $\mathbf{W}^n$ and local velocity-subset reduced distributions $h_q^n,b_q^n$.}
		\KwOutput{Updated conservative variables $\mathbf{W}^{n+1}$ and local velocity-subset reduced distributions $h_q^{n+1},b_q^{n+1}$.}
		Compute the local time step and partial heat flux; reduce the heat flux over $\mathcal{C}_v(p)$\;
		Apply boundary ghost-cell states\;
		Exchange macroscopic quantities and the leading microscopic halo data over $\mathcal{C}_x(q)$\;
		Reconstruct the macroscopic variables and the first microscopic velocity block\;
		Exchange macroscopic gradients over $\mathcal{C}_x(q)$ for the interface equilibrium construction\;
			On the root velocity rank $q=0$, precompute the interface equilibrium state, time-integration coefficients, equilibrium derivatives, and $\mathbf{F}^{\mathrm{eq}}_{ij}$\;
			Initialize the local face-flux accumulator with $\mathbf{F}^{\mathrm{eq}}_{ij}$ only on $q=0$ and with zero on $q\ne0$\;
			Broadcast the read-only interface coefficients over $\mathcal{C}_v(p)$ while keeping the equilibrium flux only in the $q=0$ accumulator\;
		Initialize the velocity-block pipeline by prefetching future distribution values and exchanging the first required gradient block\;
		\For{$m\leftarrow L_q$ \KwTo $L_q+M_q-1$}{
				Execute the triple-buffered scheduling in Algorithm~\ref{alg:triple_buffer_schedule}\;
			}
			Complete the physical boundary flux treatment in the local face-flux accumulator\;
			Reduce the local face-flux accumulators by summation over $\mathcal{C}_v(p)$ to assemble the full macroscopic interface flux\;
		Update the macroscopic conservative variables by Eq.~\eqref{eq:macro_update}\;
		\For{$m\leftarrow L_q$ \KwTo $L_q+M_q-1$}{
			Finish the second-stage microscopic update by Eq.~\eqref{eq:f_final} for all velocity points in the local block $\mathcal{V}_m$\;
		}
		\end{algorithm}
		
		In Algorithm~\ref{alg:gpgpu_full_phase_space_ugks}, the call to Algorithm~\ref{alg:triple_buffer_schedule} is a scheduling wrapper for the current velocity block: it launches the block-level flux kernel in Algorithm~\ref{alg:block_flux_kernel}, accumulates the returned nonequilibrium contribution into the local face-flux accumulator, and performs the first-stage microscopic update once. These operations are therefore not repeated elsewhere in the outer velocity-block loop.
		
		The proposed block-pipelined formulation does not alter the UGKS flux model or the two-stage time discretization. Instead, it reorganizes the execution order of microscopic nonequilibrium evolution so that GPGPU computation, local cache reuse, and MPI communication overlap can be exploited within a single velocity-space sweep.
	
	\FloatBarrier
	
	\section{Numerical results}\label{sec:numerical_results}
	
	\subsection{Three-dimensional lid-driven cavity flow}
	\label{subsec:cavity_flow}
	
	The three-dimensional lid-driven cavity flow is first used to assess the accuracy of the present block-pipelined UGKS implementation for rarefied internal flows. The gas is enclosed in a cubic cavity. The upper wall moves tangentially with velocity $u_w$, while the remaining walls are stationary. Diffuse-reflection boundary conditions are imposed on all solid walls. Two Knudsen numbers, $\mathrm{Kn}=0.1$ and $\mathrm{Kn}=1.0$, are considered to cover the transition and highly rarefied regimes. The physical space is discretized by an $80\times80\times80$ uniform mesh, and the velocity space uses a $21\times21\times21$ tensor-product Gauss--Hermite quadrature. In the following plots, velocity components are normalized by $u_w$, and temperature is nondimensionalized by the reference wall temperature.
	
	The discrete velocity set is generated from one-dimensional Gauss--Hermite points. Let $\xi_i$ and $\omega_i^{\mathrm{GH}}$ be the roots and weights of the physicists' Hermite polynomial $H_{N_{\mathrm{GH}}}$, with $N_{\mathrm{GH}}=21$. The one-dimensional nodes are scaled to a prescribed velocity range by
	\begin{equation}
		s=\frac{U_{\max}}{\max_i|\xi_i|},\qquad
		v_i=s\xi_i,\qquad U_{\max}=6.
		\label{eq:cavity_dvs_scaling}
	\end{equation}
	The standard Gauss--Hermite weights integrate functions in the form $\int e^{-\xi^2}\phi(\xi)\,d\xi$. Since the solver reads ordinary velocity-space quadrature weights, the one-dimensional weights written to the velocity-space file are
	\begin{equation}
		\omega_i=s\,\omega_i^{\mathrm{GH}}\exp(\xi_i^2),
		\label{eq:cavity_dvs_weight}
	\end{equation}
	where the factor $s$ is the Jacobian associated with the coordinate scaling. The three-dimensional velocity point and its quadrature weight are then given by
	\begin{equation}
		\mathbf{u}_{ijk}=(v_i,v_j,v_k),\qquad
		W_{ijk}=\omega_i\omega_j\omega_k,
		\label{eq:cavity_dvs_tensor_product}
	\end{equation}
	which gives $21^3=9261$ discrete velocity points.
	
\begin{figure}[htbp]
		\centering
		\begin{minipage}{0.48\linewidth}
			\centering
			\includegraphics[width=\linewidth,trim=30bp 180bp 80bp 150bp,clip]{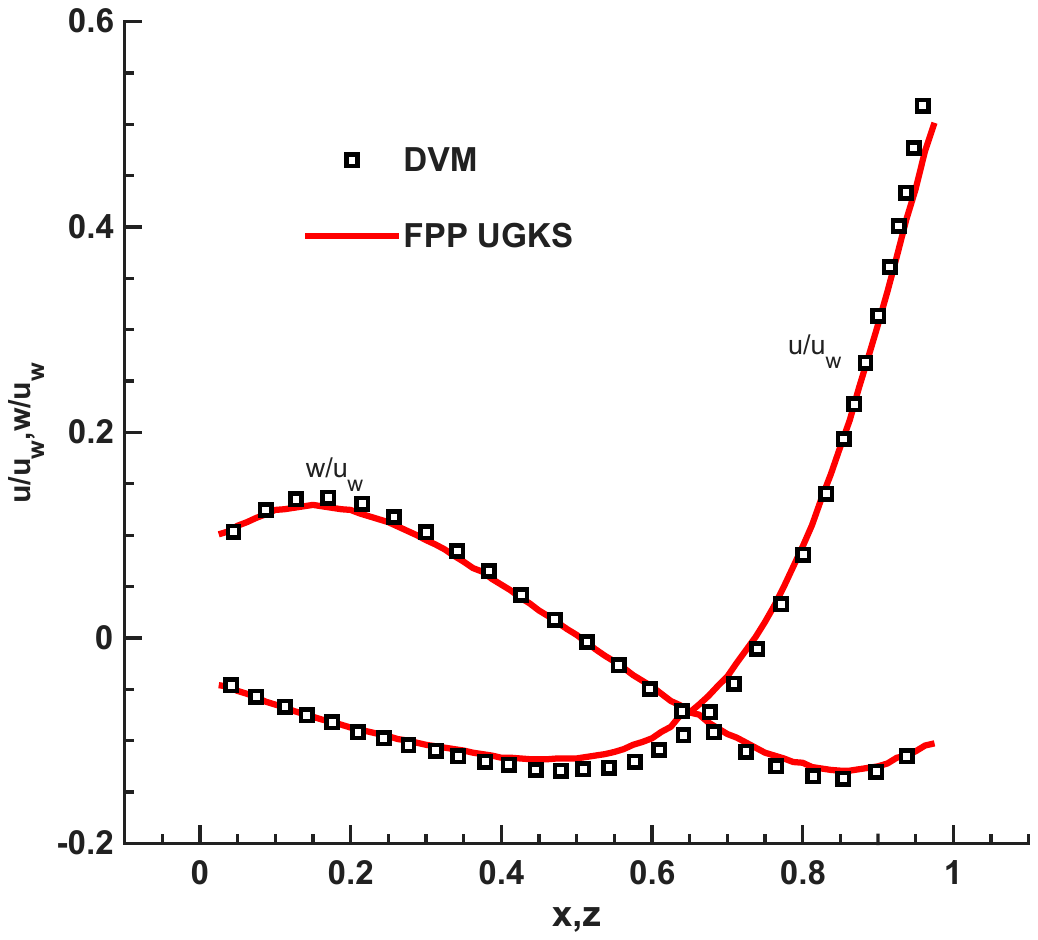}\\
			(a) $\mathrm{Kn}=0.1$
		\end{minipage}
		\hfill
		\begin{minipage}{0.48\linewidth}
			\centering
			\includegraphics[width=\linewidth,trim=30bp 180bp 80bp 150bp,clip]{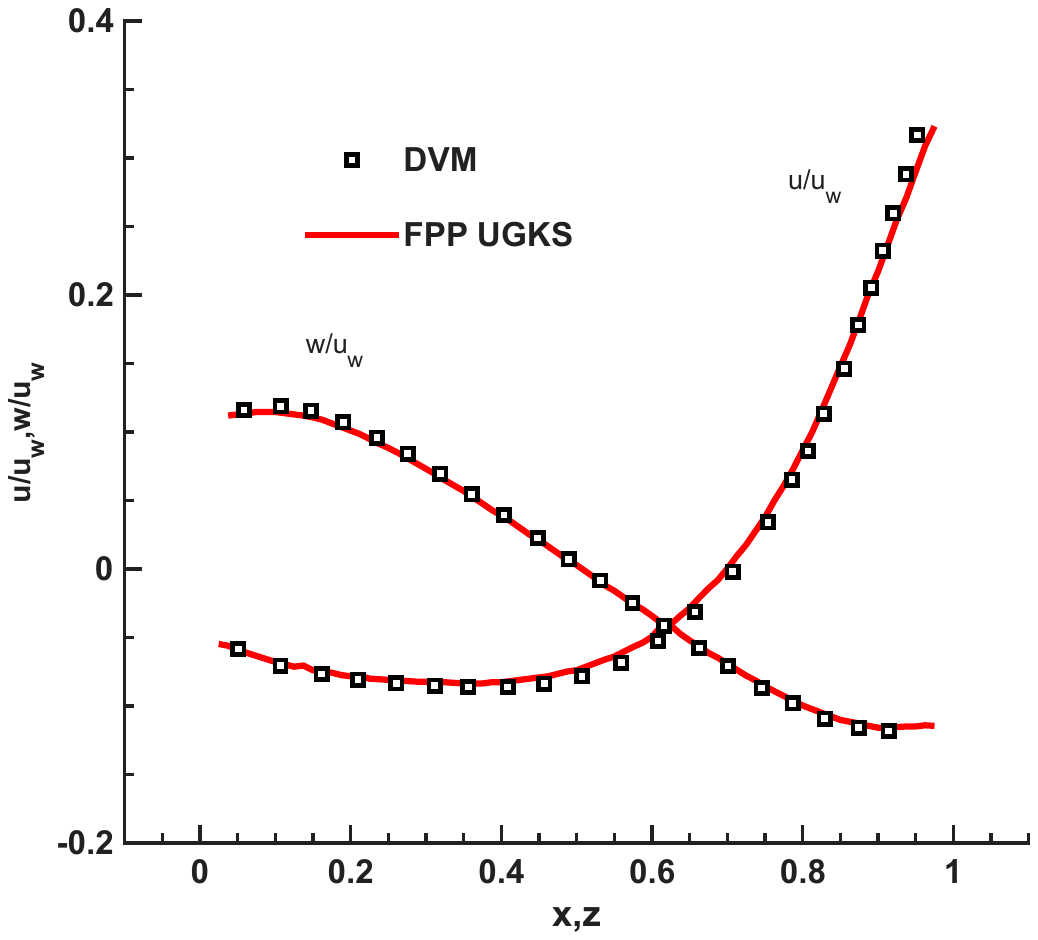}\\
			(b) $\mathrm{Kn}=1.0$
		\end{minipage}
		\caption{Centerline velocity profiles for the three-dimensional lid-driven cavity flow. Symbols denote DVM reference data, and solid lines denote the FPP-UGKS results.}
		\label{fig:cavity_velocity_profiles}
	\end{figure}

	Figure~\ref{fig:cavity_velocity_profiles} compares the centerline velocity profiles obtained by the present UGKS with DVM reference data. For both Knudsen numbers, the present results agree closely with the DVM data for the streamwise velocity $u/u_w$ and the transverse velocity $w/u_w$. At $\mathrm{Kn}=0.1$, the driven-wall motion produces a stronger streamwise velocity near the moving wall and a broader recirculating response in the cavity core. At $\mathrm{Kn}=1.0$, the velocity magnitude is reduced and the profiles become smoother because the flow is more rarefied. The agreement in both cases indicates that the velocity-block formulation and the phase-space decomposition preserve the original UGKS solution behavior.
	
\begin{figure}[htbp]
		\centering
		\begin{minipage}{0.49\linewidth}
			\centering
			\includegraphics[width=\linewidth]{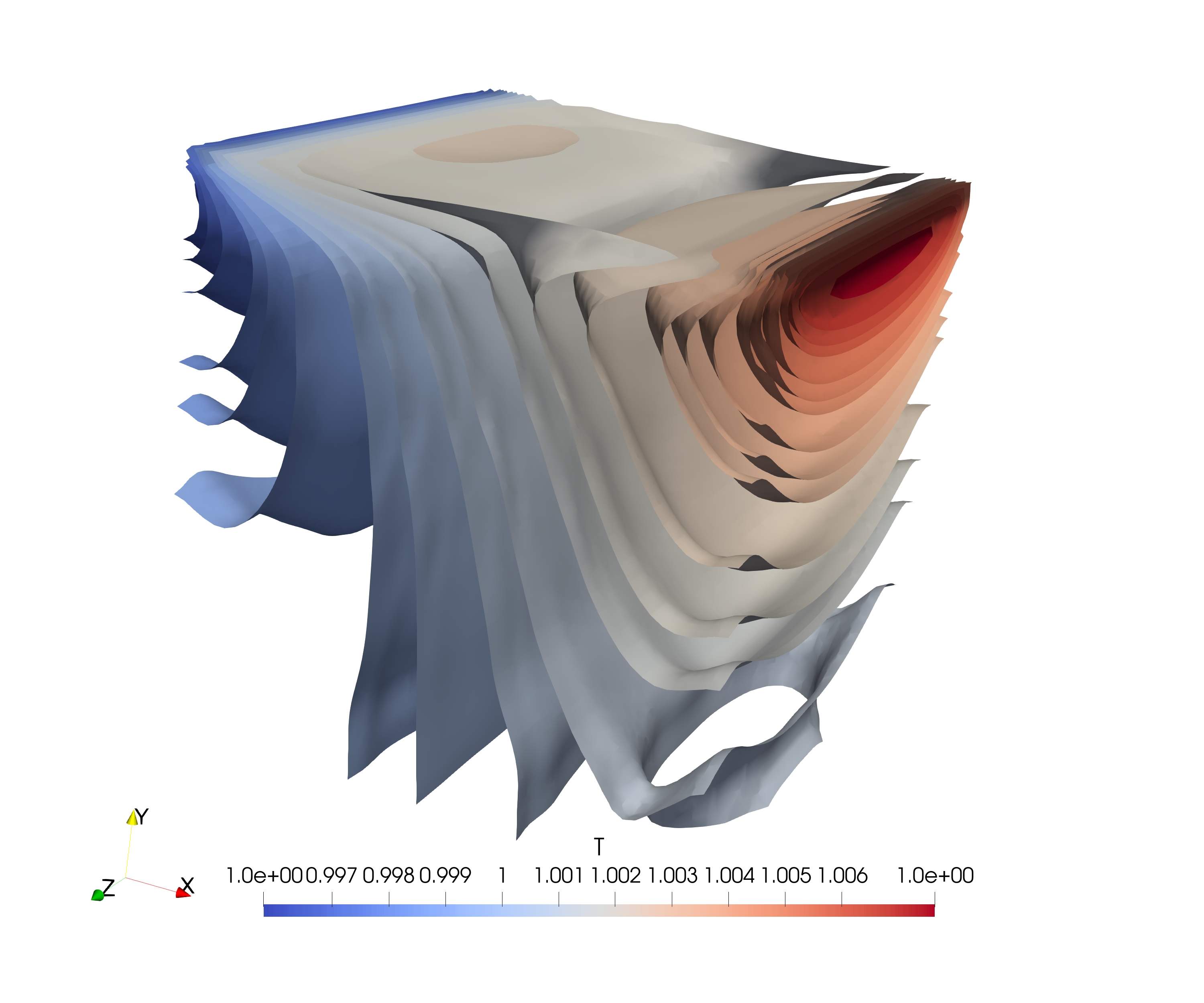}\\
			(a) $\mathrm{Kn}=0.1$
		\end{minipage}
		\hfill
		\begin{minipage}{0.49\linewidth}
			\centering
			\includegraphics[width=\linewidth]{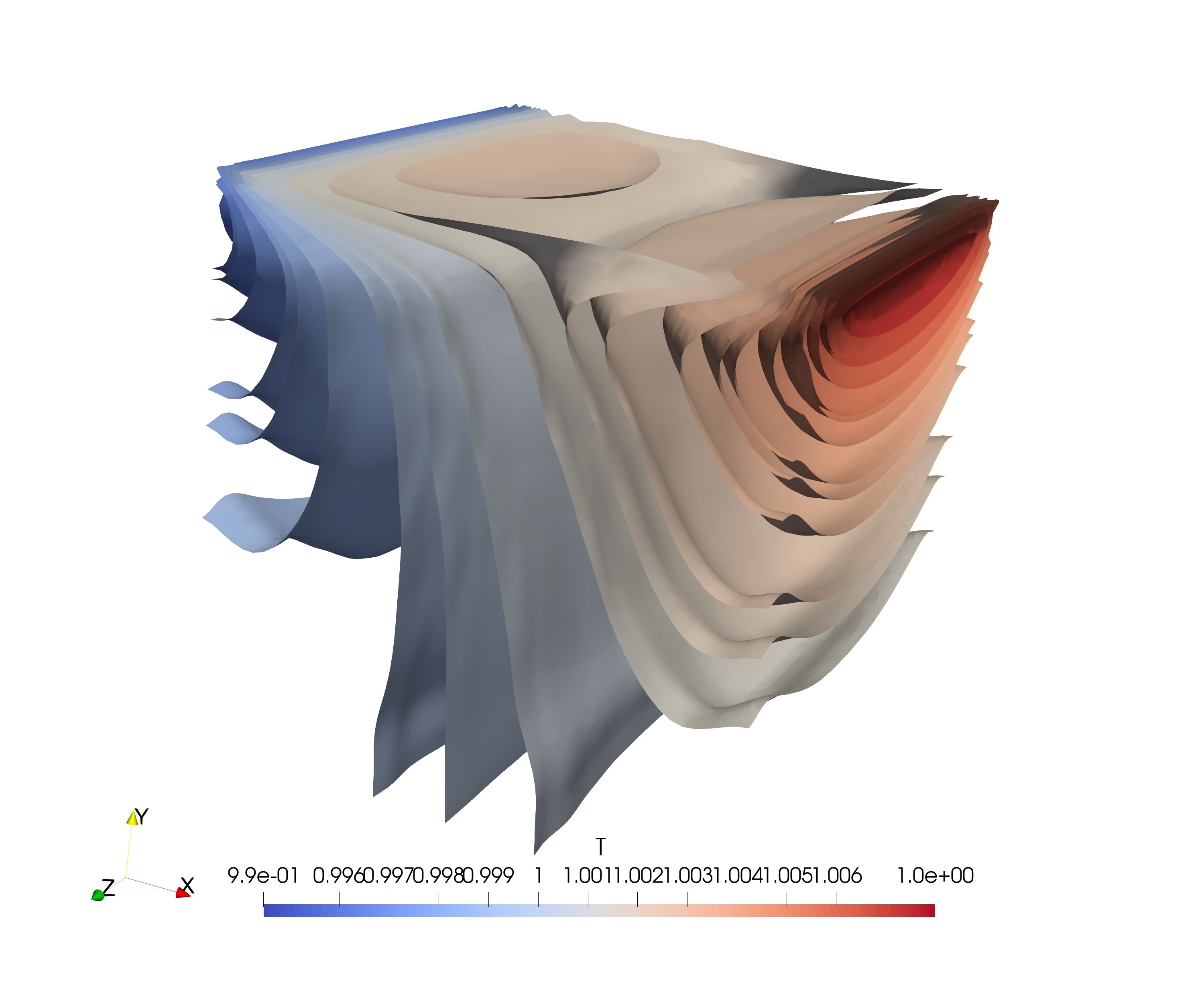}\\
			(b) $\mathrm{Kn}=1.0$
		\end{minipage}
		\caption{Temperature distributions of the three-dimensional lid-driven cavity flow.}
		\label{fig:cavity_temperature}
	\end{figure}

	Figure~\ref{fig:cavity_temperature} shows the corresponding temperature distributions. The temperature variation is moderate around the reference wall temperature, but the thermal nonequilibrium structure is clearly visible inside the cavity. A colder region appears near the upstream side of the recirculating flow, while a warmer region forms near the downstream side where the driven-wall motion compresses and redistributes the gas. The $\mathrm{Kn}=1.0$ case exhibits a more diffuse thermal structure than the $\mathrm{Kn}=0.1$ case, consistent with the stronger rarefaction effect and weaker momentum coupling across the cavity.
	
	\FloatBarrier
	
	\subsection{Scalability test}\label{subsec:speedup_test}
	
	The scalability of the proposed GPGPU-oriented FPP-UGKS is then evaluated using the same three-dimensional lid-driven cavity geometry. The hardware and software description in this subsection applies only to the scalability tests reported here. The tests were carried out on a heterogeneous supercomputing platform in China with a scale-up/scale-out architecture. Each node contains two 64-bit CISC-based server processors and eight SIMT-based GPGPU accelerators. Each processor has 64 cores at 2.4~GHz, a NUMA memory organization, eight-channel DDR5-6400 memory, and PCIe Gen5 connectivity, with host-to-device and device-to-host bandwidths of 64~GB/s. Each accelerator contains 320 SIMD units, 64~GB of HBM with a bandwidth of 1.8~TB/s, 768~KB of registers, 64~KB of LDS, and 8~MB of L2 cache. The peak performance of one accelerator is 37~TFLOPS in FP64, 74~TFLOPS in FP32, and 470~TFLOPS in FP16. The nodes are connected through a proprietary InfiniBand-like native RDMA interconnect with four 400~Gbps ports per node and are organized as a three-level Clos topology with two data planes. The software environment consists of Anolis OS 8.9, a GPGPU programming environment compatible with standard accelerator runtime APIs, GCC-8.5.0, BLIS-2.0, FFTW-3.3.10, OpenMPI-5.0.3, and a collective communications library compatible with GPU collective-communication APIs.
	
	In all tests, one MPI process is bound to one GPGPU accelerator; therefore, the MPI process number $n_p$ is also the number of accelerators used in the run. In the raw timing records, \texttt{mode} distinguishes strong and weak scaling, \texttt{nodes} is the number of nodes, \texttt{np} is the number of MPI processes, $N$ is the grid resolution in each coordinate direction, and \texttt{cells} is the total number of physical cells, i.e., $N^3$. The discrete velocity space is fixed to 15000 velocity points. Four velocity-block sizes, $B_v=\texttt{DVS\_PER\_BLOCK}=32,64,128,$ and $256$, are tested, and the device thread-block size is set to 256. Each timing result is the wall-clock time for 100 time steps. Four velocity-space MPI settings are compared: $P_v=1$, corresponding to a single velocity-space MPI partition; $P_v=N_{\mathrm{node}}$, for which the number of velocity partitions equals the number of nodes; and two fixed velocity partition numbers, $P_v=4$ and $P_v=8$. These settings correspond to the raw timing columns denoted by \texttt{v1\_algorithm\_s}, \texttt{vnode\_algorithm\_s}, \texttt{v4\_algorithm\_s}, and \texttt{v8\_algorithm\_s}, respectively. The raw timing data are listed in Appendix~\ref{app:raw_scalability_timings}.
	
	For the strong-scaling test, the physical mesh is fixed at $80^3$ cells. For each velocity-block size $B_v$, the baseline runtime is defined as the one-node run with one velocity-space partition, namely the entry \texttt{v1\_algorithm\_s} at $N_{\mathrm{node}}=1$. Since each node contains eight GPGPU accelerators, this baseline corresponds to eight MPI processes and eight accelerators. The strong-scaling speedup and parallel efficiency are computed as
	\begin{equation}
		S_n=\frac{T_{\mathrm{ref}}}{T_n},\qquad
		E_{\mathrm{strong}}
		=
		\frac{S_n}{N_{\mathrm{node},n}/N_{\mathrm{node},\mathrm{ref}}},
		\label{eq:strong_efficiency}
	\end{equation}
	where $T_{\mathrm{ref}}$ is the above one-node baseline runtime and $N_{\mathrm{node},\mathrm{ref}}=1$. All strong-scaling curves in a given $B_v$ panel are normalized by this same baseline, so configurations with the same measured runtime at the same node count have the same speedup and efficiency.
	
	For the weak-scaling test, the mesh size increases from $30^3$ to $120^3$ cells as the node count increases from 1 to 64. The one-node \texttt{v1\_algorithm\_s} runtime is again used as the reference time. Because the rounded weak-scaling meshes do not keep the number of cells per node exactly constant at every node count, the weak-scaling efficiency is weighted by the actual workload,
	\begin{equation}
		E_{\mathrm{weak}}
		=
		\frac{T_{\mathrm{ref}}}{T_n}
		\frac{N_{c,n}/N_{c,\mathrm{ref}}}{N_{\mathrm{node},n}/N_{\mathrm{node},\mathrm{ref}}},
		\label{eq:weighted_weak_efficiency}
	\end{equation}
	where $T_{\mathrm{ref}}$ is the one-node reference runtime, $T_n$ is the runtime at node count $n$, and $N_c$ is the physical-cell count.
	
	\begin{figure}[!htbp]
		\centering
		\includegraphics[width=0.98\linewidth]{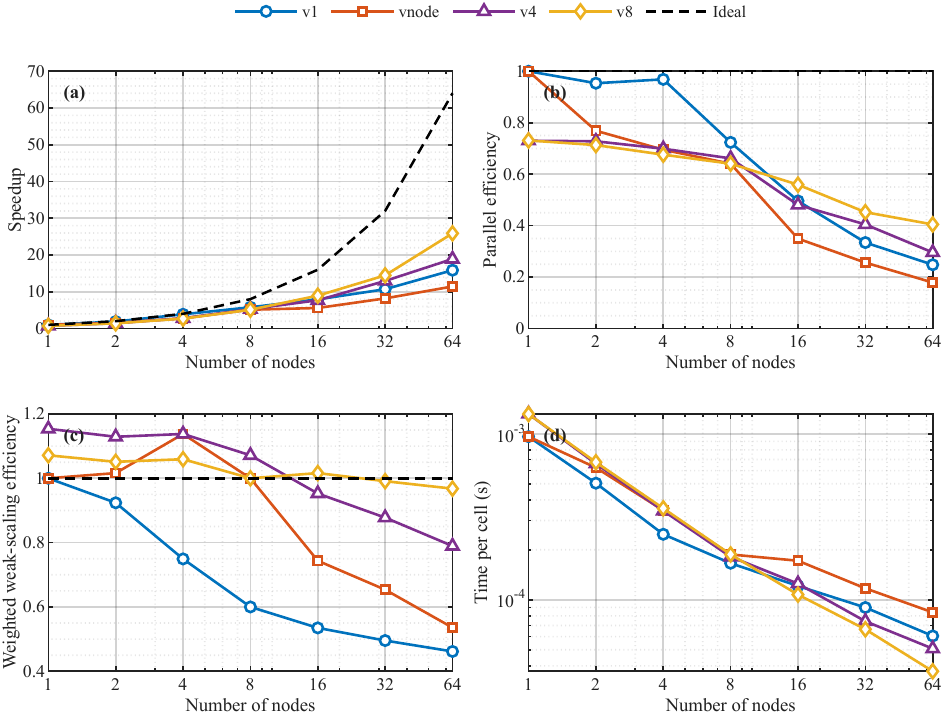}
		\caption{Strong and weak scalability for $B_v=32$. Panels show (a) strong-scaling speedup, (b) strong-scaling parallel efficiency, (c) workload-weighted weak-scaling efficiency, and (d) time per cell in the strong-scaling test.}
		\label{fig:scalability_bv32}
	\end{figure}
	
	\begin{figure}[!htbp]
		\centering
		\includegraphics[width=0.98\linewidth]{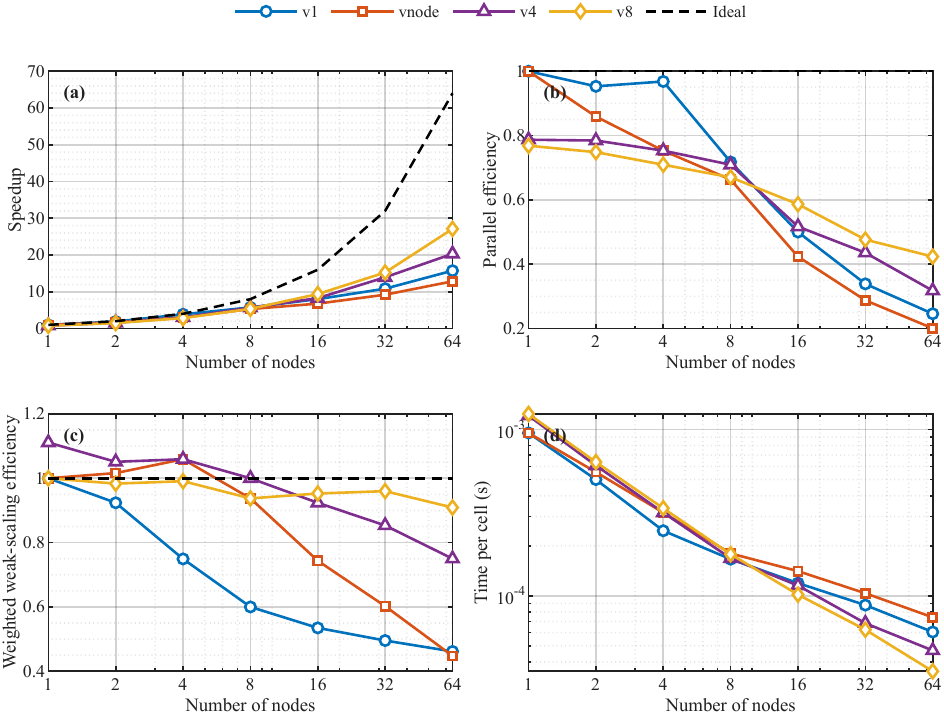}
		\caption{Strong and weak scalability for $B_v=64$. Panels show (a) strong-scaling speedup, (b) strong-scaling parallel efficiency, (c) workload-weighted weak-scaling efficiency, and (d) time per cell in the strong-scaling test.}
		\label{fig:scalability_bv64}
	\end{figure}
	
	\begin{figure}[!htbp]
		\centering
		\includegraphics[width=0.98\linewidth]{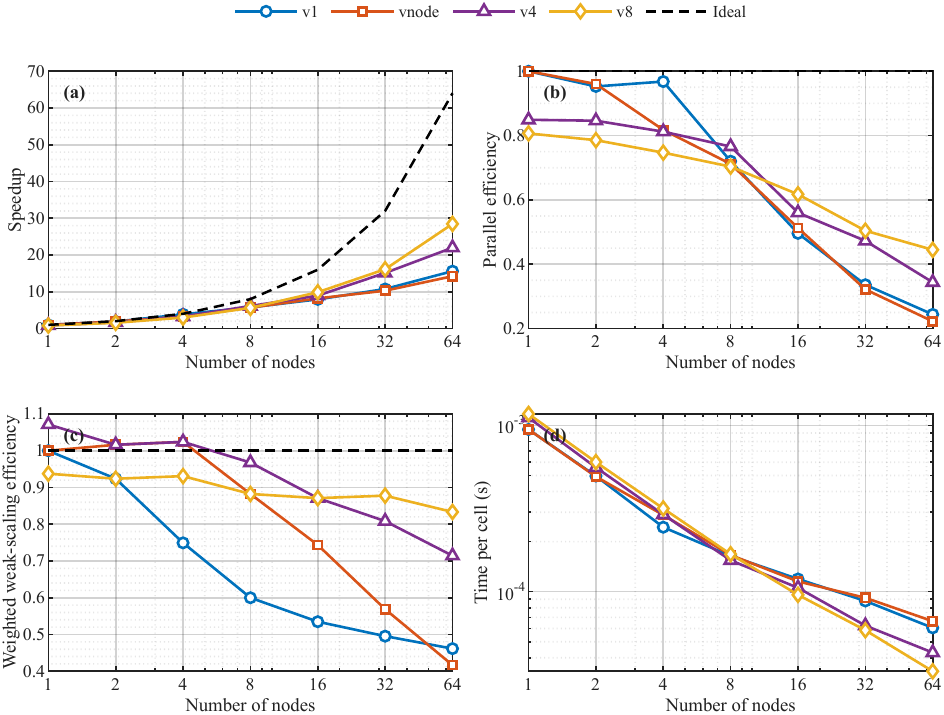}
		\caption{Strong and weak scalability for $B_v=128$. Panels show (a) strong-scaling speedup, (b) strong-scaling parallel efficiency, (c) workload-weighted weak-scaling efficiency, and (d) time per cell in the strong-scaling test.}
		\label{fig:scalability_bv128}
	\end{figure}
	
	\begin{figure}[!htbp]
		\centering
		\includegraphics[width=0.98\linewidth]{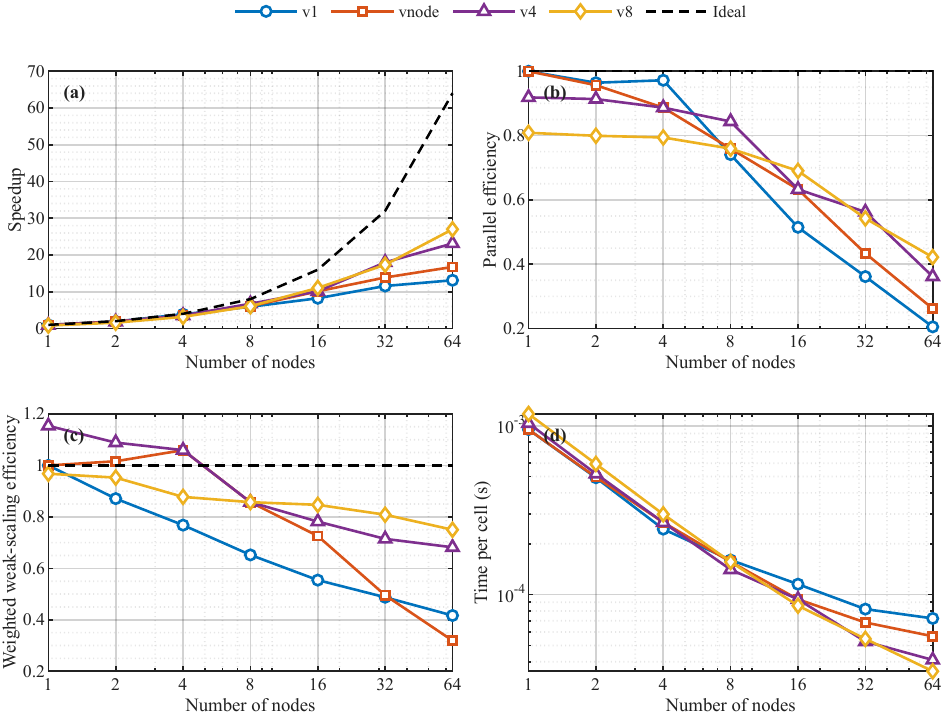}
		\caption{Strong and weak scalability for $B_v=256$. Panels show (a) strong-scaling speedup, (b) strong-scaling parallel efficiency, (c) workload-weighted weak-scaling efficiency, and (d) time per cell in the strong-scaling test.}
		\label{fig:scalability_bv256}
	\end{figure}
	
	The strong-scaling results in Figs.~\ref{fig:scalability_bv32}--\ref{fig:scalability_bv256} show that velocity-space decomposition becomes increasingly important at large node counts. With $P_v=1$, the full discrete velocity space is stored and advanced on every physical-space partition, and the speedup starts to saturate once the physical subdomain per accelerator becomes small. In contrast, the fixed velocity-space decompositions continue to reduce the local microscopic workload. At 64 nodes, $P_v=8$ gives the shortest runtime for all tested block sizes: 19~s for $B_v=32$, 18~s for $B_v=64$, 17~s for $B_v=128$, and 18~s for $B_v=256$. The corresponding strong-scaling speedups are approximately 35.4, 35.3, 35.3, and 33.4, respectively, and are consistently higher than those obtained with $P_v=1$, $P_v=N_{\mathrm{node}}$, and $P_v=4$.
	
	The weak-scaling results further indicate that a moderate fixed velocity-space partition gives the best balance between microscopic workload reduction and velocity-communicator overhead. The $P_v=8$ curves maintain the highest workload-weighted weak-scaling efficiency at large node counts for all tested block sizes. At 64 nodes, the $P_v=8$ runtimes are 31~s, 33~s, 36~s, and 40~s for $B_v=32,64,128,$ and $256$, respectively, whereas the physical-domain-only $P_v=1$ setting takes 65--72~s. The $P_v=N_{\mathrm{node}}$ setting can be competitive at small node counts, but its efficiency deteriorates when the velocity communicator becomes too fine. These results show that the proposed phase-space decomposition is most effective when the velocity partition is large enough to reduce local microscopic storage and work, but not so large that macroscopic moment reductions dominate the runtime.

	An x38 production case was also profiled with Nsight Systems over the CUDA profiler API capture range covering a 50-step time loop. The available capture records MPI rank 0, and the values reported here and in Appendix~\ref{app:x38_nsys_profiling} are rank-0 per-rank measurements. On the profiled rank, the dominant GPU work is the non-equilibrium interior-face flux kernel, followed by the DVS least-squares reconstruction and the distribution-function update. These three kernels account for 89.9\% of the accumulated GPU kernel time. The same capture reports 68.27~GB of host-to-device transfers, 68.28~GB of device-to-host transfers, and 62.85~GB of point-to-point MPI send volume on rank 0 over the 50 steps. These data give a quantitative view of the compute-side kernel balance, staged host-device copies, MPI payloads, and resident microscopic gradient-cache footprint under velocity-block pipelining.
	
	\FloatBarrier
	\subsection{Supersonic flow around a sphere}
	\label{subsec:supersonic_sphere}
	
	Supersonic flow around a sphere is further considered to assess the present method for an external rarefied flow with strong compression and thermal nonequilibrium. The hardware and software description in this paragraph applies to the sphere and X38-like vehicle calculations reported below. These two cases were carried out on a local GPU cluster using eight nodes and 64 NVIDIA Tesla V100-SXM2 accelerators in total. Each node contains two Intel Xeon Gold 5117 processors at 2.00~GHz, with 28 physical CPU cores and 56 hardware threads, and 250~GiB of main memory. Each node is equipped with eight Tesla V100-SXM2-16GB GPUs of compute capability 7.0; each GPU has 16~GiB of device memory, a maximum SM clock of 1530~MHz, and a 300~W power limit. The GPUs within one node are connected through NVLink, with six NVLink connections reported for each GPU. The nodes are connected by a Mellanox ConnectX-4 InfiniBand interconnect with an active 100~Gb/s link. The software environment consists of Ubuntu 20.04.6 LTS, Linux kernel 5.15, NVIDIA driver 550.163.01, CUDA 12.4, GCC 9.4.0, and OpenMPI 4.1.8. In these production runs, one MPI process is bound to one GPU, so that the sphere and X38-like vehicle simulations use 64 MPI processes on 64 V100 GPUs.
	
	The two hardware environments are used for different purposes. The scalability and Orion-like capsule tests use the large heterogeneous accelerator platform to assess multi-node scalability and large-case feasibility, whereas the sphere and X38-like vehicle runs use the local NVIDIA V100 cluster for validation cases and Nsight Systems profiling. Therefore, wall-clock timings are compared only within the same hardware environment.
	
	Two Knudsen numbers, $\mathrm{Kn}=0.031$ and $\mathrm{Kn}=0.121$, are computed at a freestream Mach number of $M_\infty=4.25$. The wall temperature is fixed at $T_w=302~\mathrm{K}$, and the freestream total temperature is $T_0=300~\mathrm{K}$. The physical space is discretized by an unstructured mesh with 138240 cells, and the velocity space contains 18802 discrete velocity points.
	
\begin{figure}[htbp]
		\centering
		\begin{minipage}{0.48\linewidth}
			\centering
			\includegraphics[width=\linewidth]{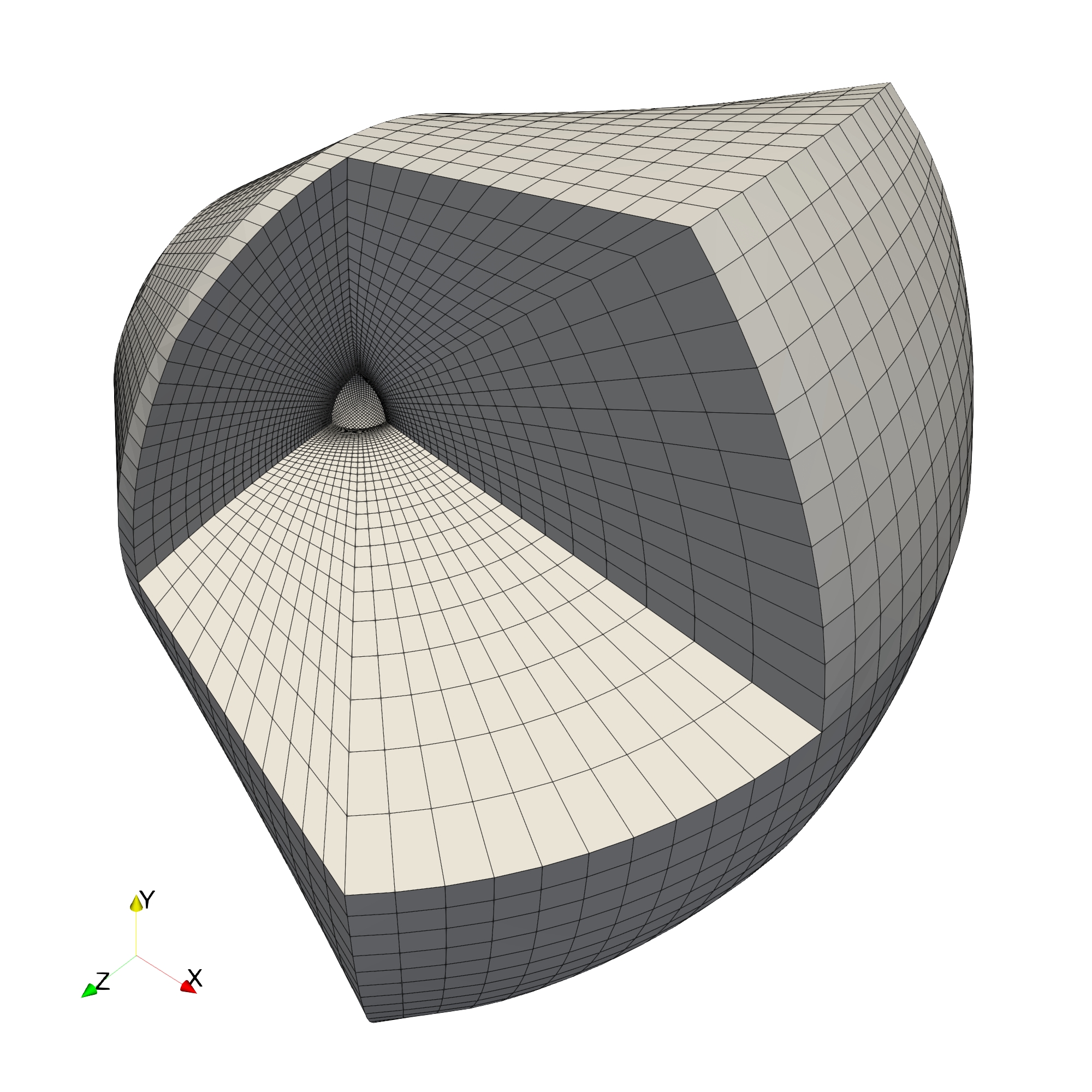}\\
			(a) Physical-space mesh
		\end{minipage}
		\hfill
		\begin{minipage}{0.48\linewidth}
			\centering
			\includegraphics[width=\linewidth]{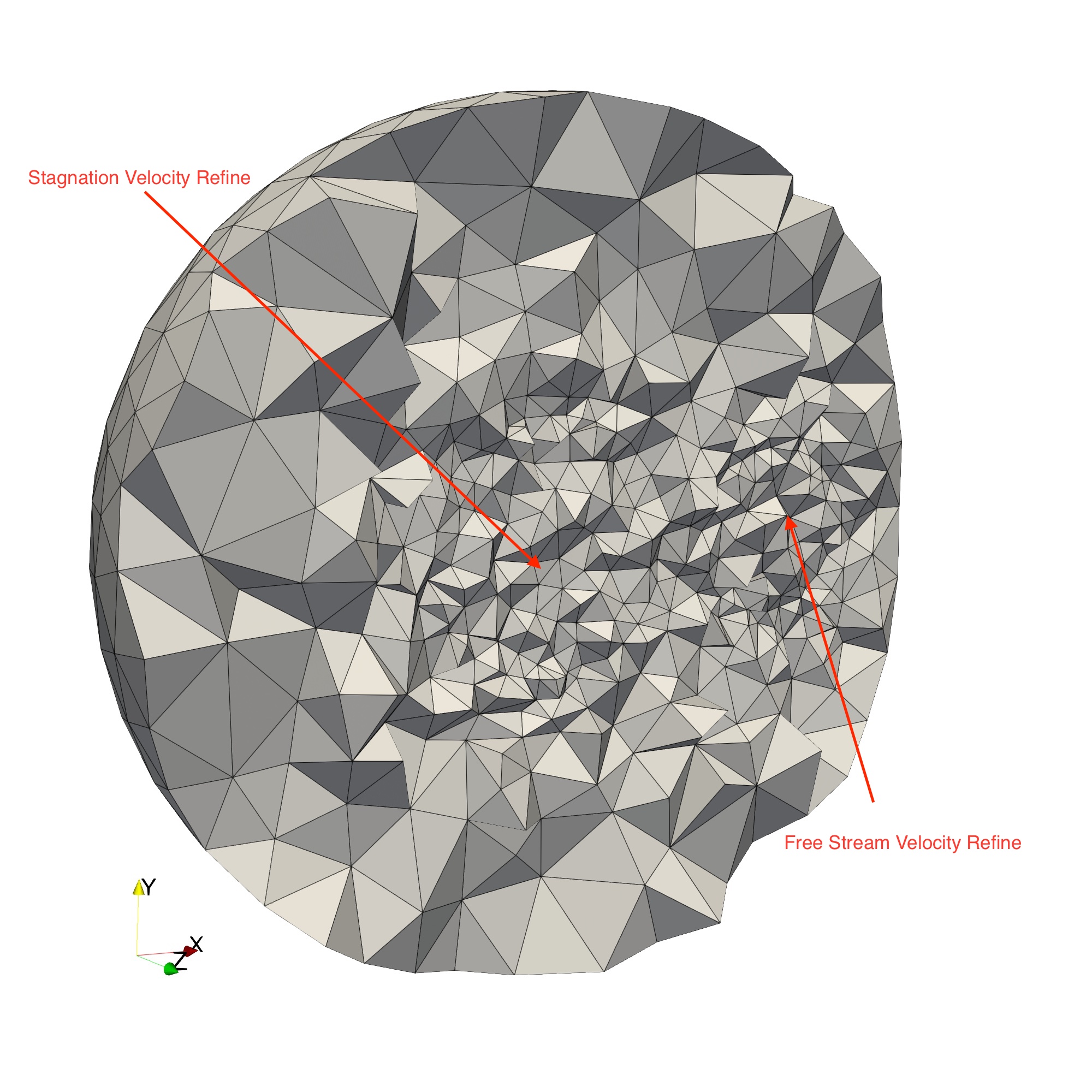}\\
			(b) Velocity-space mesh
		\end{minipage}
		\caption{Physical-space and velocity-space meshes for the supersonic sphere-flow calculation.}
		\label{fig:sphere_meshes}
	\end{figure}

	The physical-space and velocity-space discretizations used in this test are shown in Fig.~\ref{fig:sphere_meshes}. The physical mesh is body-fitted around the sphere and is locally refined near the wall and in the downstream flow region, so that the strong gradients in the shock layer and wake can be represented. The velocity mesh is nonuniform and refined around the freestream-velocity and stagnation-velocity regions. This allocation concentrates velocity points in the parts of velocity space that dominate the incoming molecular beam and the strongly compressed near-stagnation distribution, while keeping the total number of velocity points manageable for the full phase-space calculation.
	
\begin{figure}[htbp]
		\centering
		\begin{minipage}{0.48\linewidth}
			\centering
			\includegraphics[width=\linewidth]{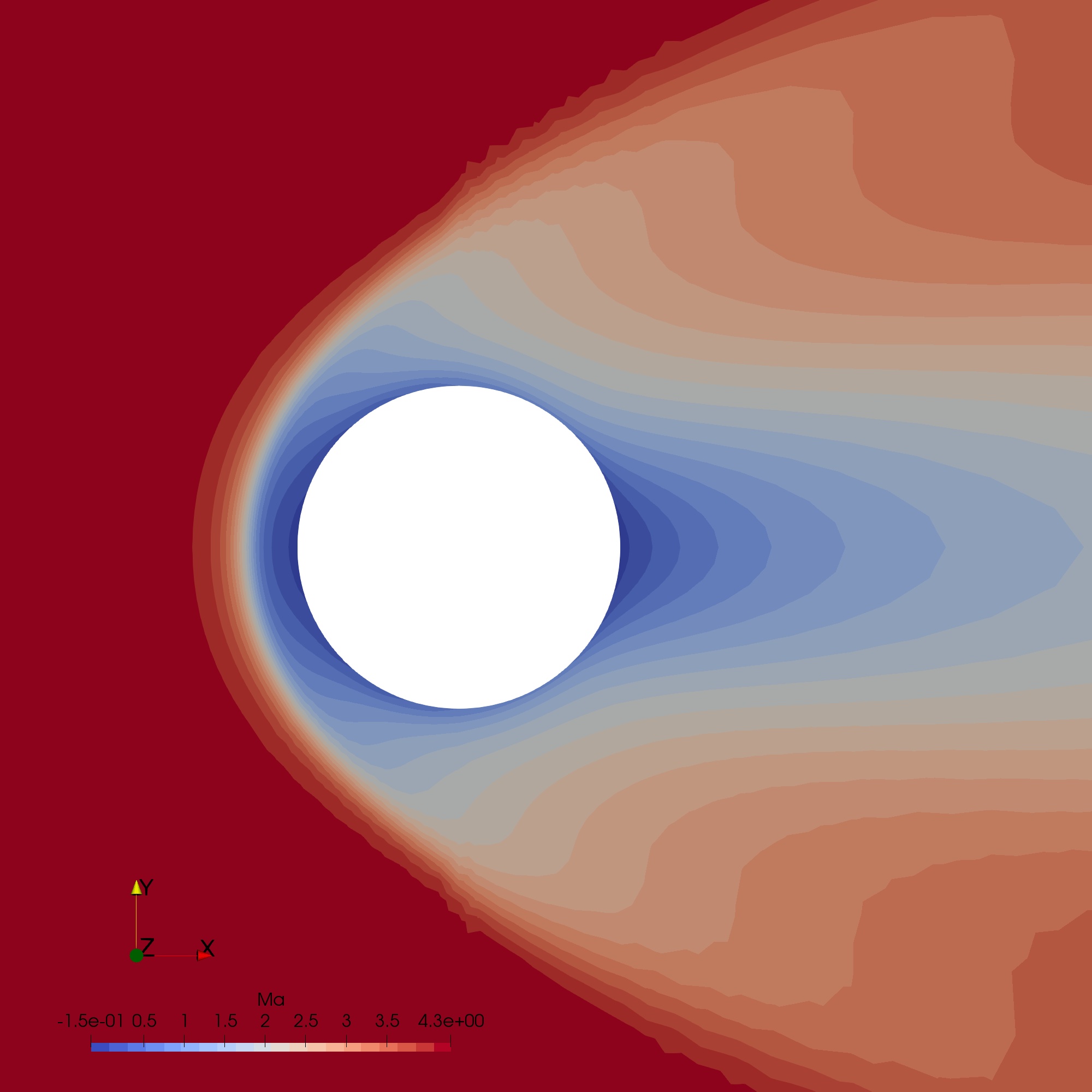}\\
			(a) $\mathrm{Kn}=0.031$
		\end{minipage}
		\hfill
		\begin{minipage}{0.48\linewidth}
			\centering
			\includegraphics[width=\linewidth]{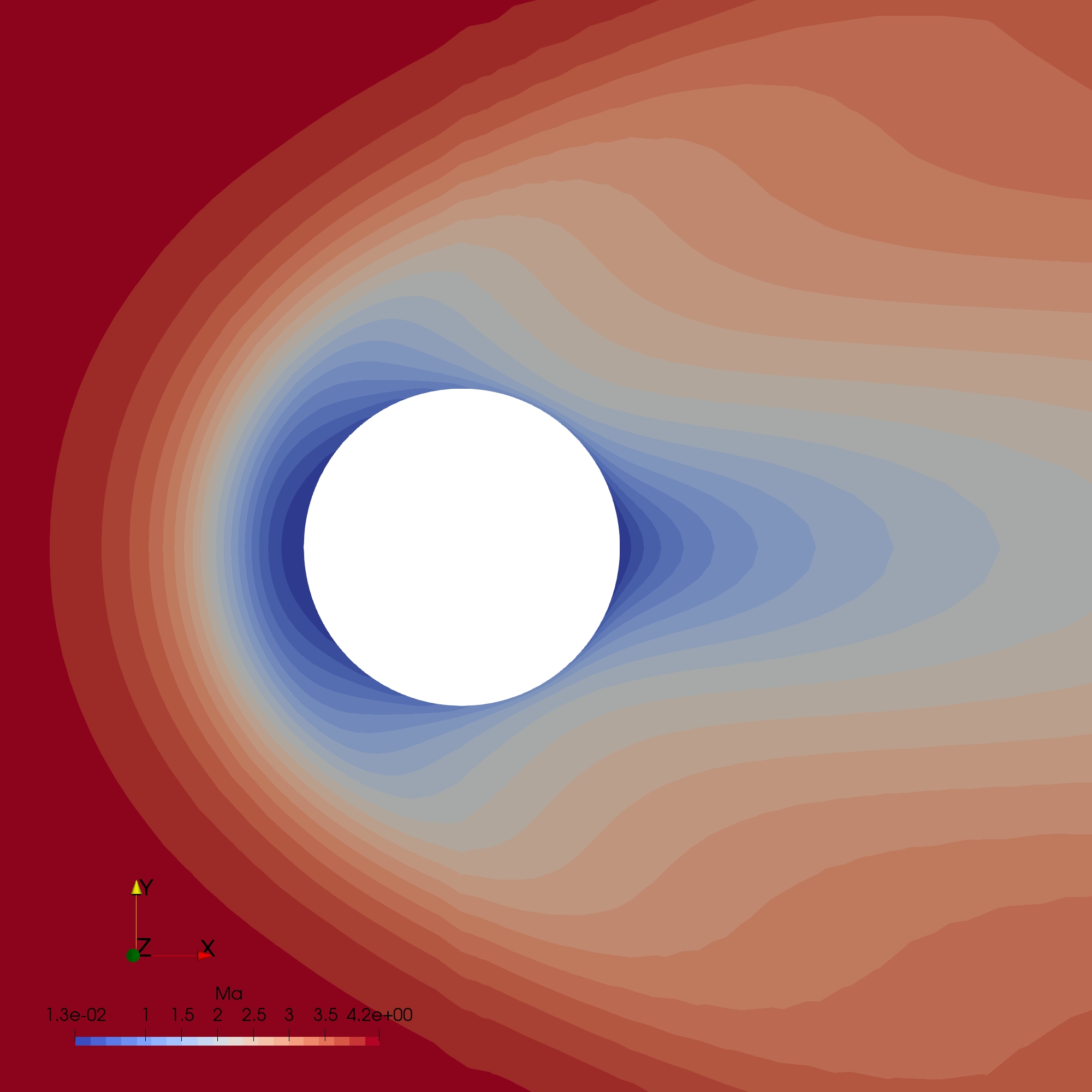}\\
			(b) $\mathrm{Kn}=0.121$
		\end{minipage}
		\caption{Mach-number distributions for supersonic flow around a sphere at $M_\infty=4.25$.}
		\label{fig:sphere_mach_contours}
	\end{figure}

\begin{figure}[htbp]
		\centering
		\begin{minipage}{0.48\linewidth}
			\centering
			\includegraphics[width=\linewidth]{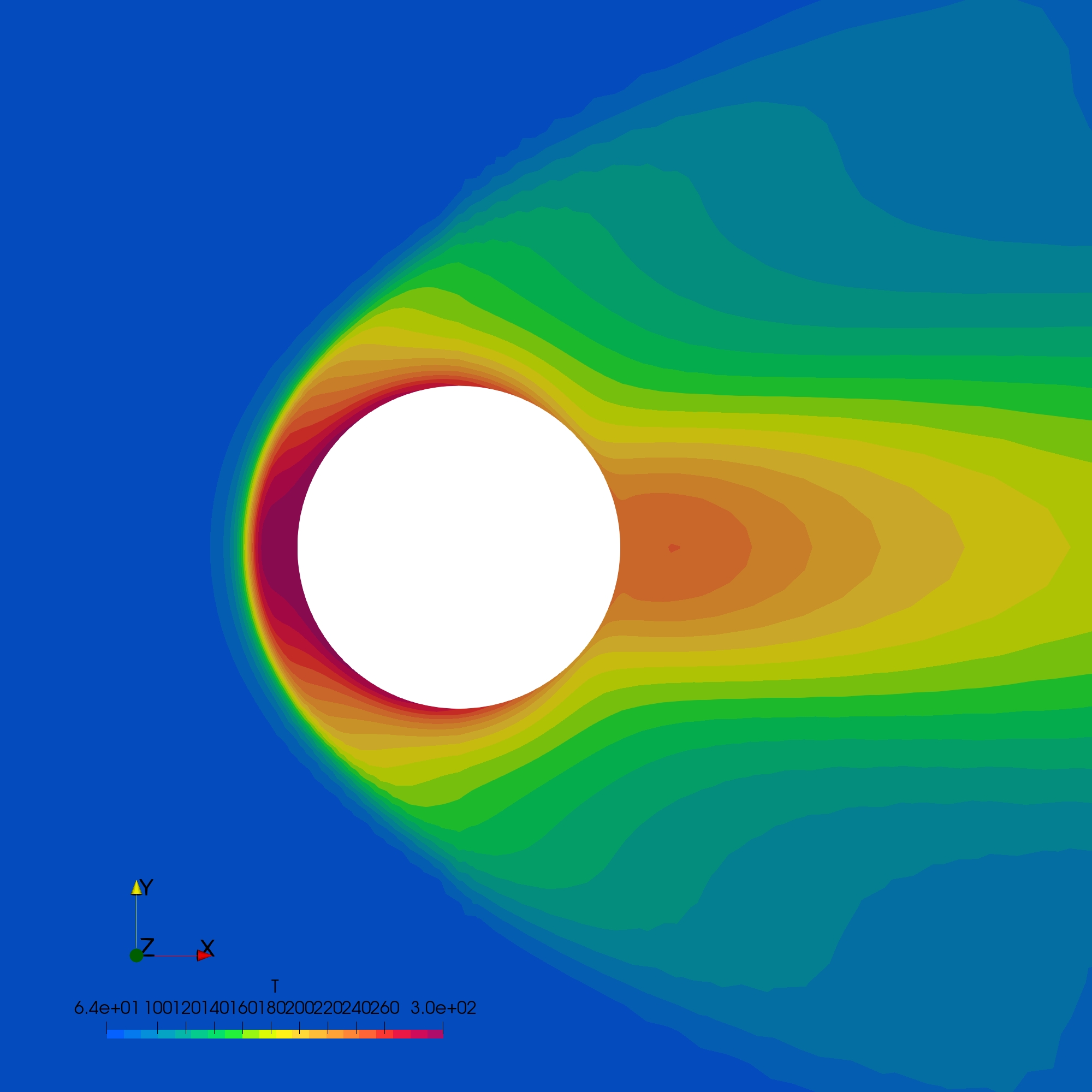}\\
			(a) $\mathrm{Kn}=0.031$
		\end{minipage}
		\hfill
		\begin{minipage}{0.48\linewidth}
			\centering
			\includegraphics[width=\linewidth]{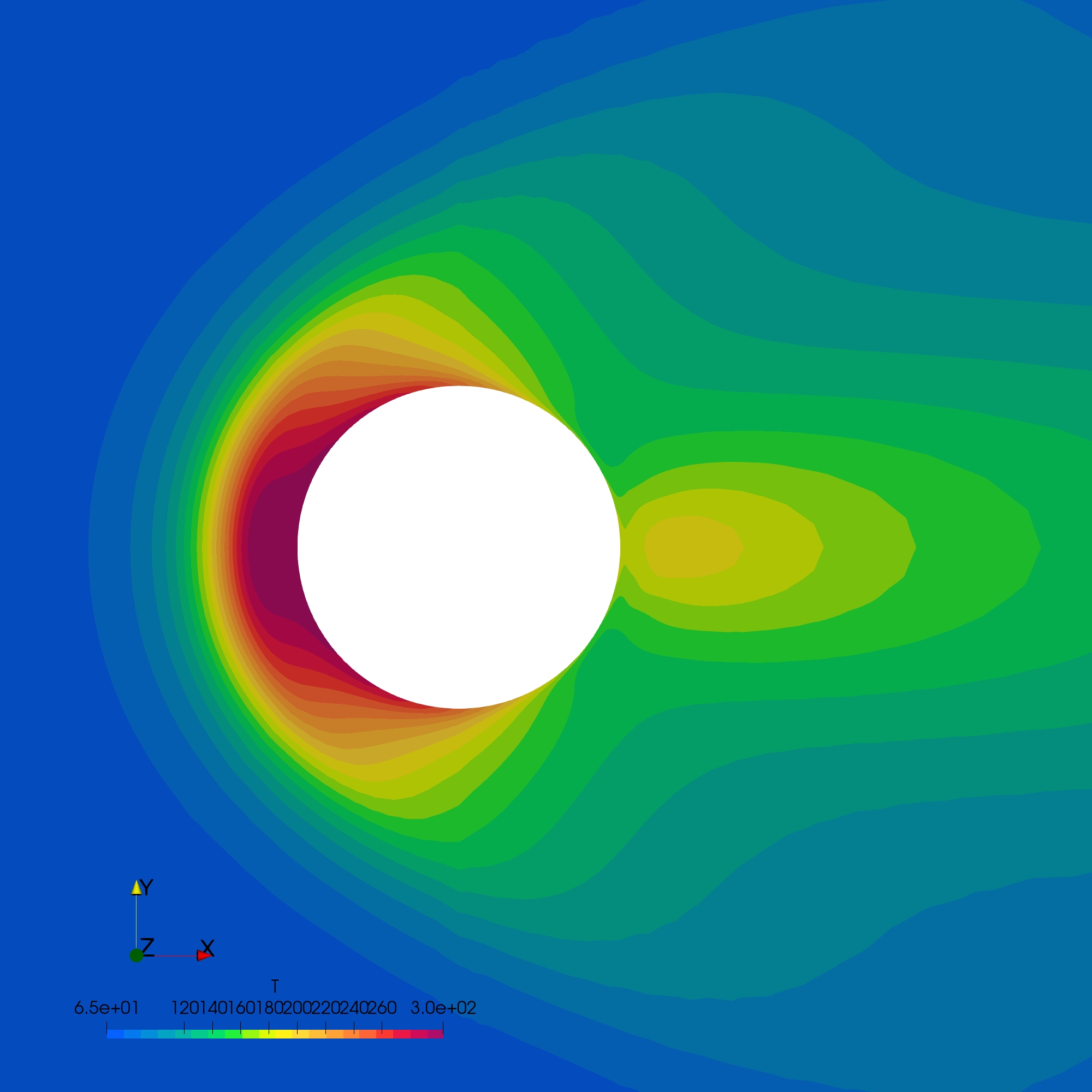}\\
			(b) $\mathrm{Kn}=0.121$
		\end{minipage}
		\caption{Temperature distributions for supersonic flow around a sphere at $M_\infty=4.25$.}
		\label{fig:sphere_temperature_contours}
	\end{figure}

	Figures~\ref{fig:sphere_mach_contours} and~\ref{fig:sphere_temperature_contours} show the computed Mach-number and temperature distributions on the symmetry plane. In both cases, a detached bow shock forms upstream of the sphere, followed by a low-speed stagnation region and a wake behind the body. When the Knudsen number is increased from 0.031 to 0.121, the shock layer and wake become more diffuse, and the near-wall gradients are smoothed by rarefaction effects. The temperature field exhibits strong aerodynamic heating around the forebody and a thermal wake downstream, while the higher-Knudsen-number case shows a broader nonequilibrium region.
	
	\FloatBarrier
	
\begin{figure}[htbp]
		\centering
		\begin{minipage}{0.48\linewidth}
			\centering
			\includegraphics[width=\linewidth]{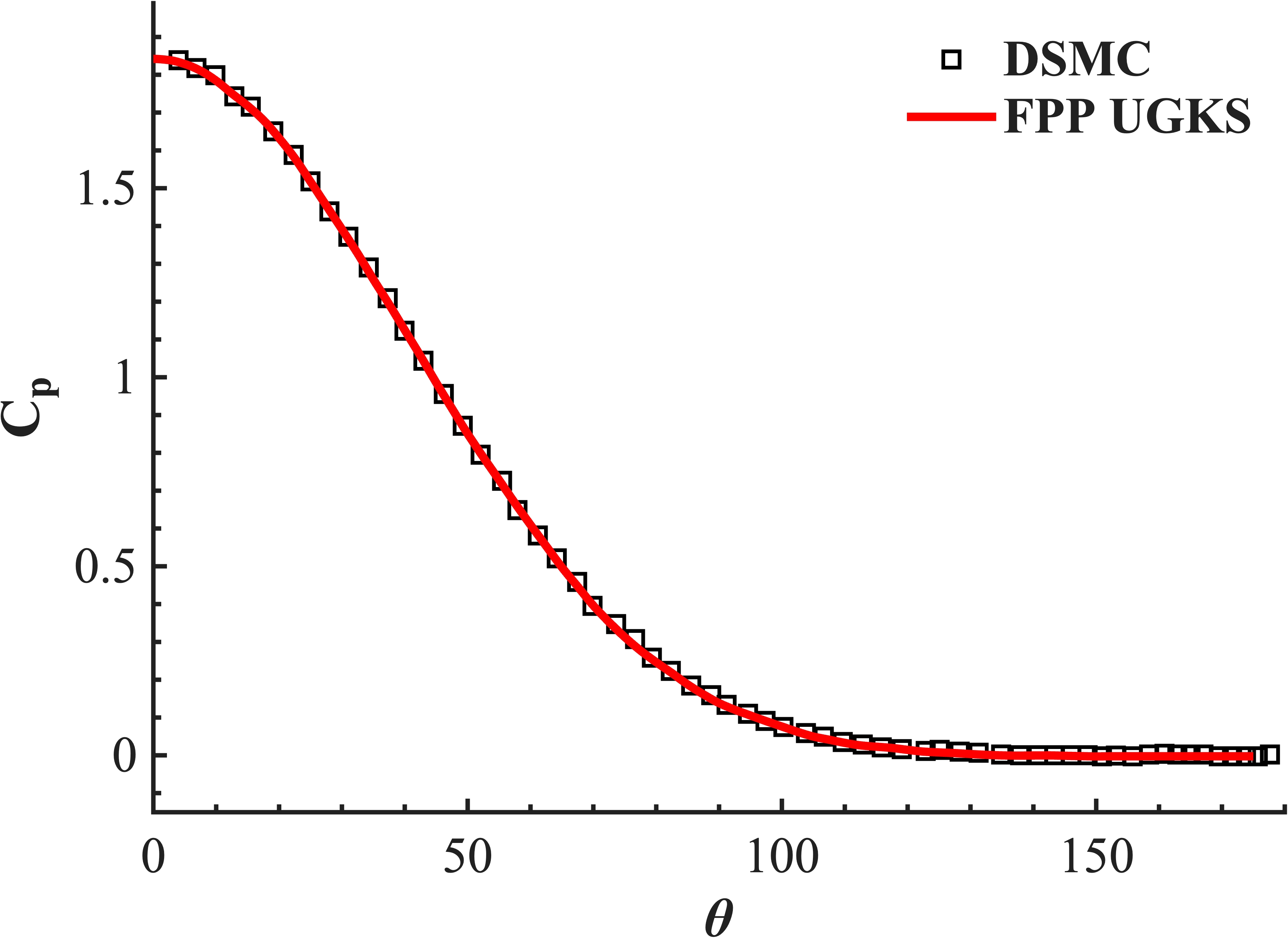}\\
			(a) $\mathrm{Kn}=0.031$
		\end{minipage}
		\hfill
		\begin{minipage}{0.48\linewidth}
			\centering
			\includegraphics[width=\linewidth]{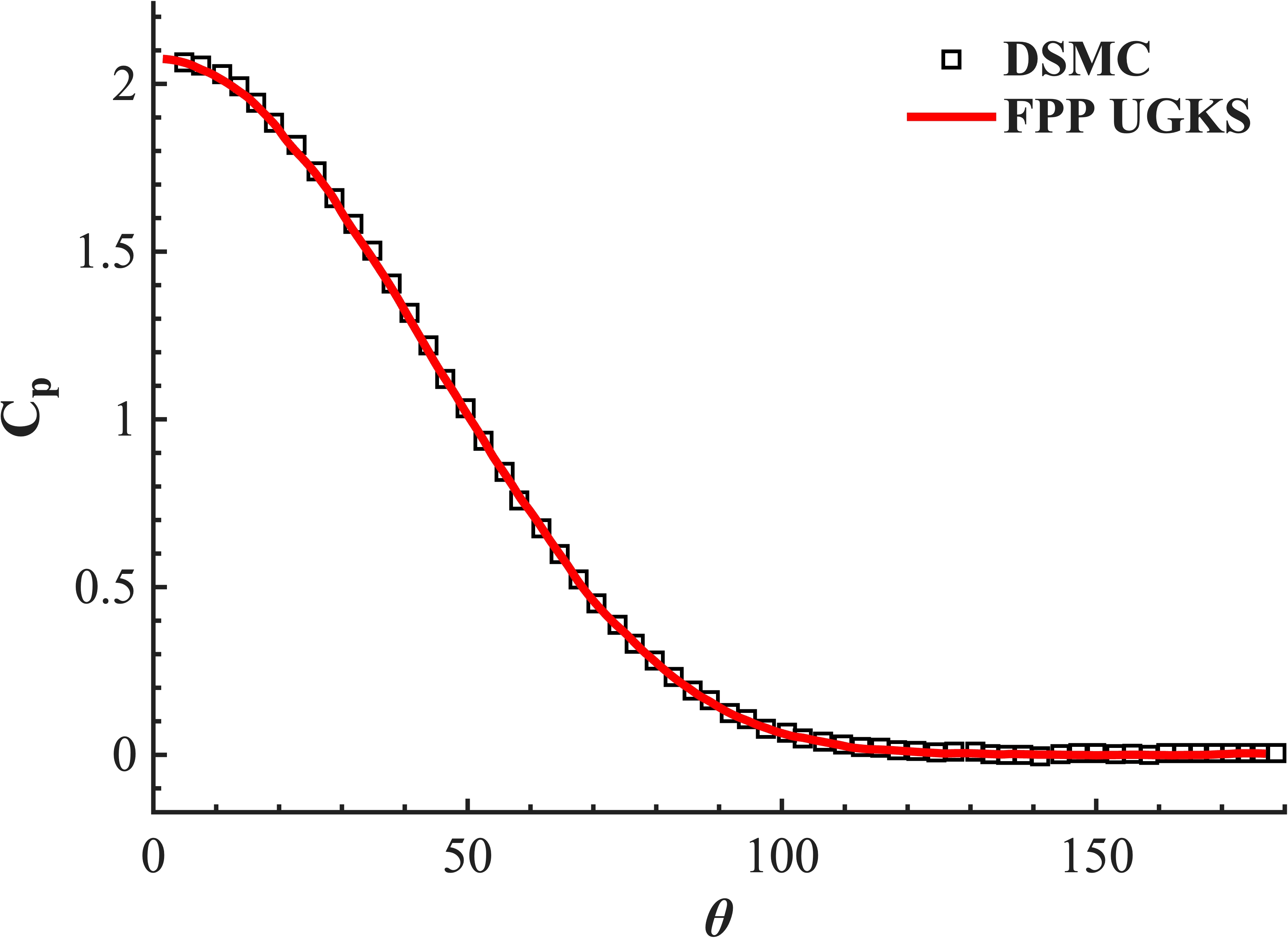}\\
			(b) $\mathrm{Kn}=0.121$
		\end{minipage}
		\caption{Pressure coefficient $C_p$ on the sphere surface. Symbols denote DSMC reference data from Ref.~\cite{Zhang2025}, and solid lines denote the present FPP-UGKS results.}
		\label{fig:sphere_cp}
	\end{figure}

\begin{figure}[htbp]
		\centering
		\begin{minipage}{0.48\linewidth}
			\centering
			\includegraphics[width=\linewidth]{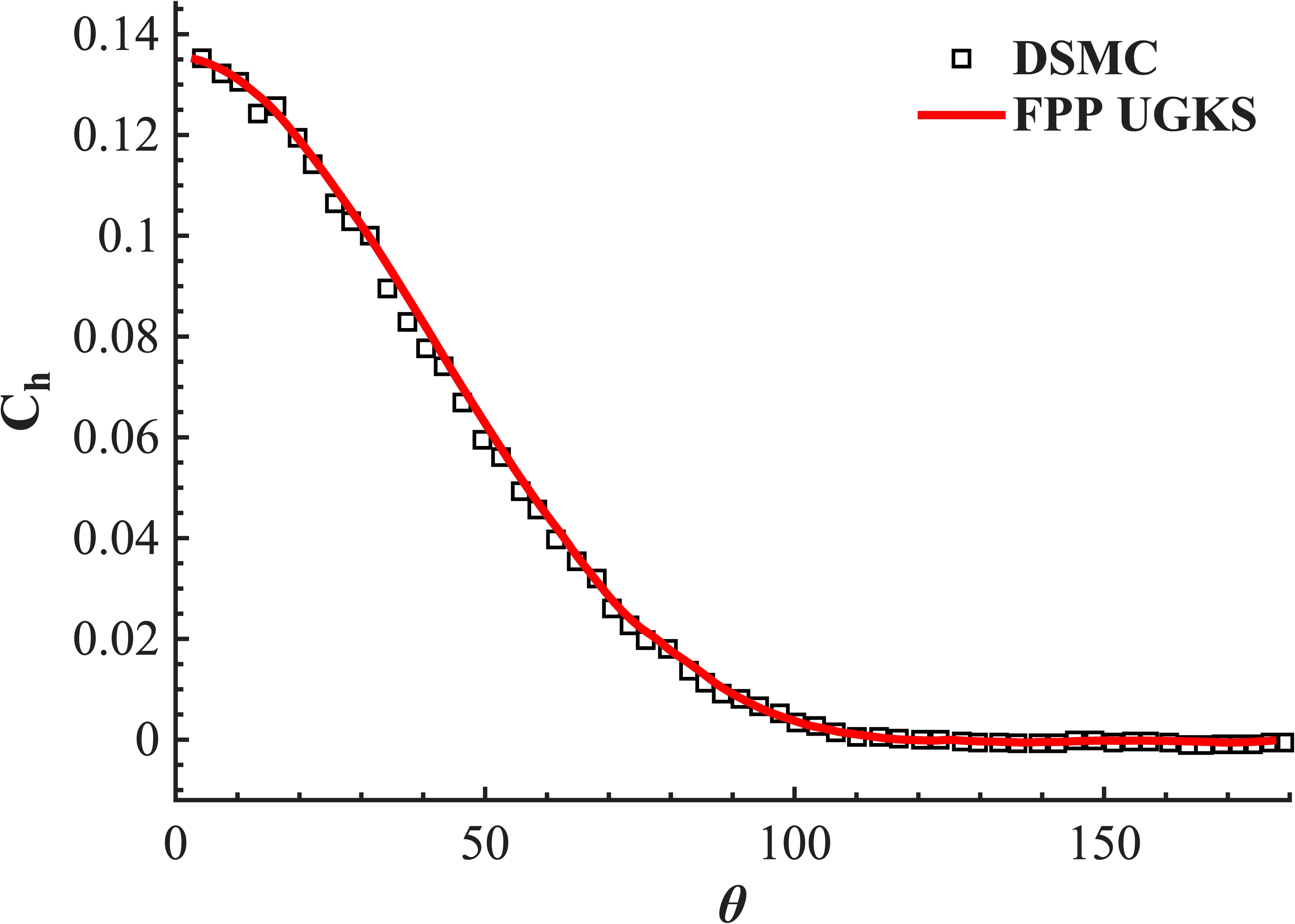}\\
			(a) $\mathrm{Kn}=0.031$
		\end{minipage}
		\hfill
		\begin{minipage}{0.48\linewidth}
			\centering
			\includegraphics[width=\linewidth]{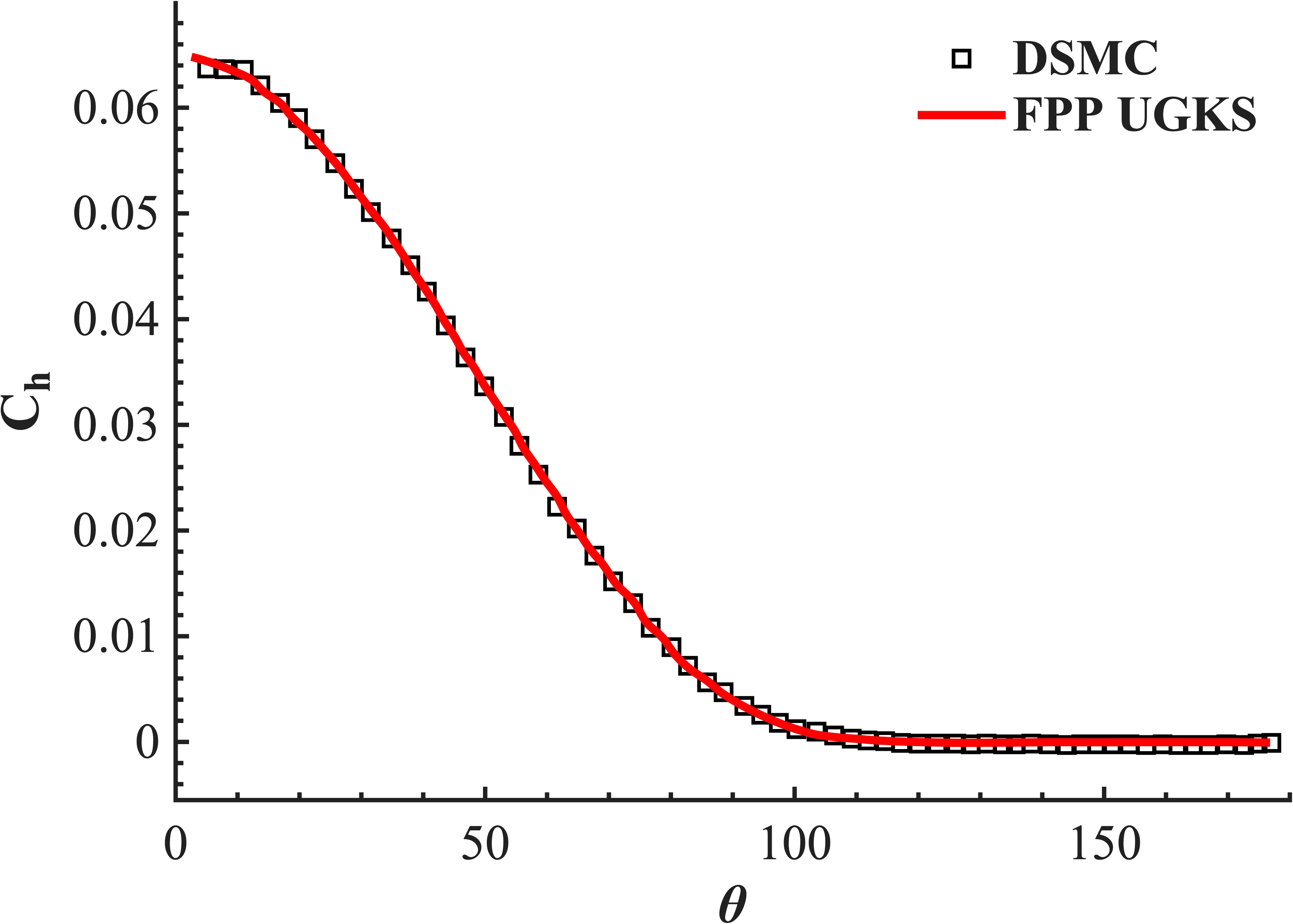}\\
			(b) $\mathrm{Kn}=0.121$
		\end{minipage}
		\caption{Heat-transfer coefficient $C_h$ on the sphere surface. Symbols denote DSMC reference data from Ref.~\cite{Zhang2025}, and solid lines denote the present FPP-UGKS results.}
		\label{fig:sphere_ch}
	\end{figure}

\begin{figure}[htbp]
		\centering
		\begin{minipage}{0.48\linewidth}
			\centering
			\includegraphics[width=\linewidth]{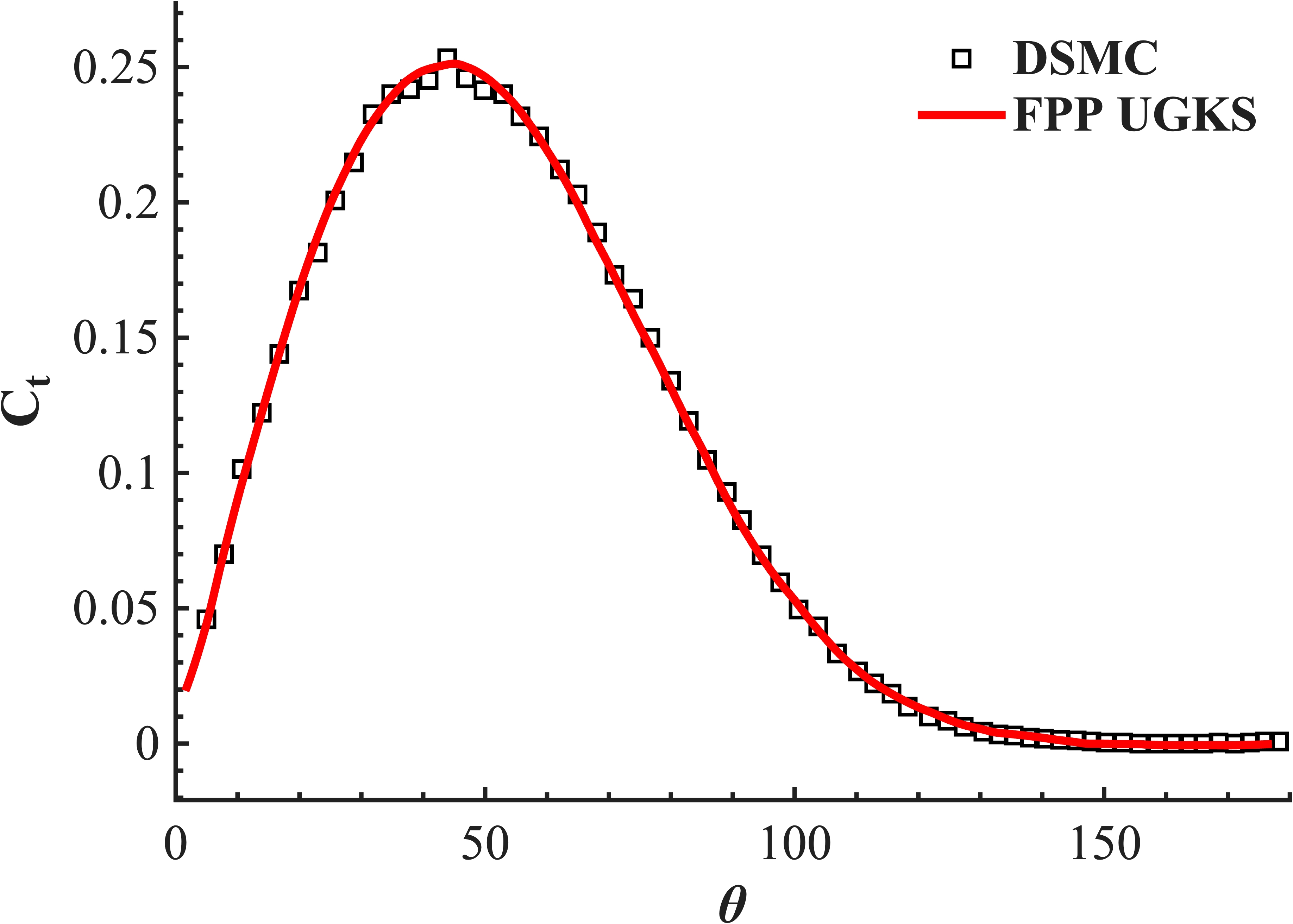}\\
			(a) $\mathrm{Kn}=0.031$
		\end{minipage}
		\hfill
		\begin{minipage}{0.48\linewidth}
			\centering
			\includegraphics[width=\linewidth]{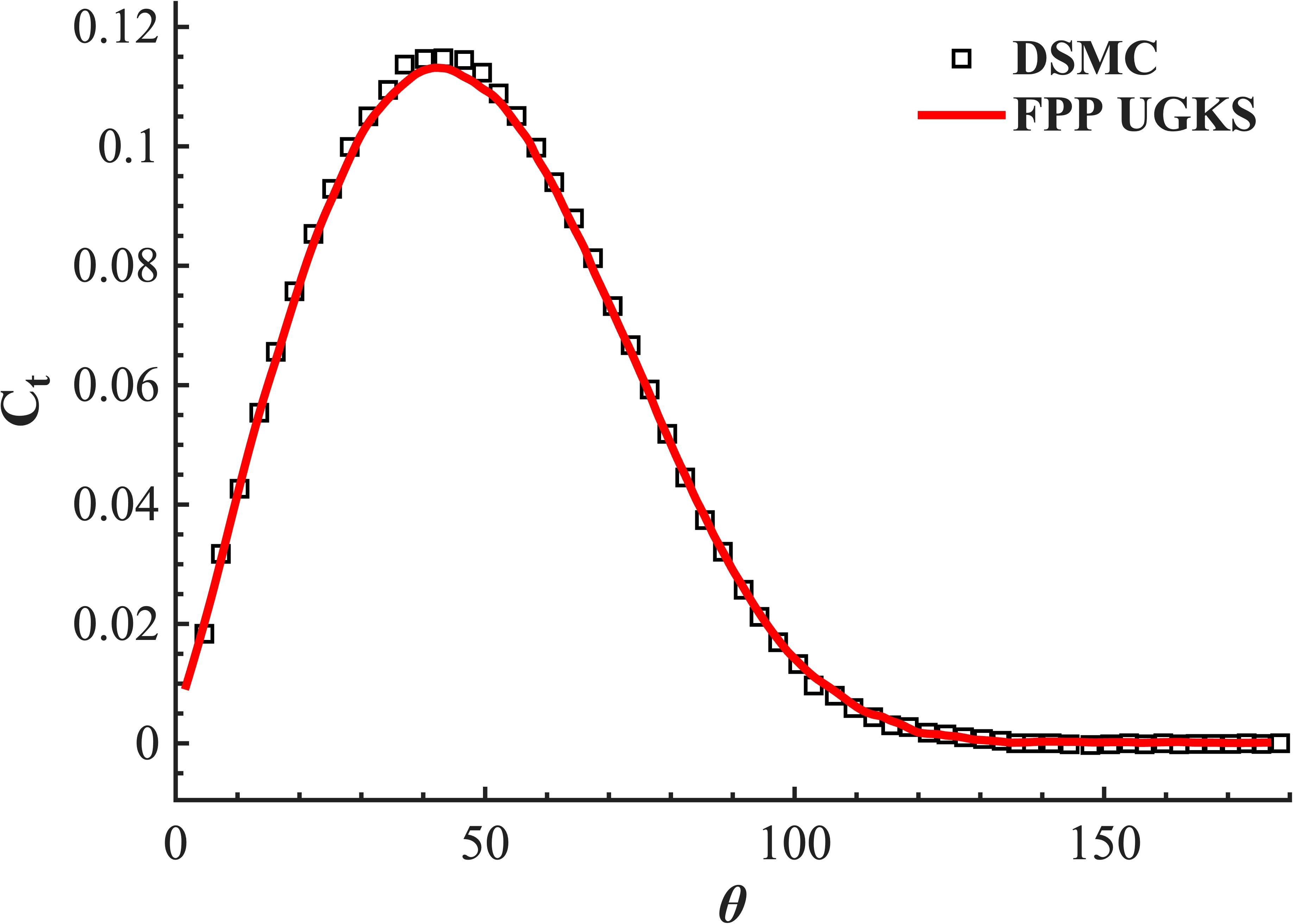}\\
			(b) $\mathrm{Kn}=0.121$
		\end{minipage}
		\caption{Tangential-stress coefficient $C_t$ on the sphere surface. Symbols denote DSMC reference data from Ref.~\cite{Zhang2025}, and solid lines denote the present FPP-UGKS results.}
		\label{fig:sphere_ct}
	\end{figure}

	The surface coefficient distributions are compared with DSMC reference data~\cite{Zhang2025} in Figs.~\ref{fig:sphere_cp}--\ref{fig:sphere_ct}. The pressure coefficient $C_p$ decreases monotonically from the stagnation point toward the downstream side, while the heat-transfer coefficient $C_h$ and tangential-stress coefficient $C_t$ show peak values away from the stagnation point. For both Knudsen numbers, the present FPP-UGKS results follow the DSMC data closely over the full polar-angle range. This agreement indicates that the velocity-block-pipelined formulation preserves the capability of UGKS to resolve rarefied hypersonic flow features while using the proposed phase-space parallel execution strategy.
	
	\FloatBarrier
	
	\subsection{Hypersonic flow around the X38-like vehicle}
	\label{subsec:x38_vehicle}
	
	The next validation case considers hypersonic flow around an X38-like reentry vehicle, which provides a more demanding external-flow configuration with a complex three-dimensional body shape. The freestream and wall conditions follow the near-continuum argon case reported by Li et al.~\cite{Li2021X38KineticComparison}. The freestream Mach number is $M_\infty=8.0$, the angle of attack is $\alpha=20^\circ$, the Knudsen number is $\mathrm{Kn}=0.00275$, and the Reynolds number is $\mathrm{Re}=5937$. The freestream density, temperature, and velocity are $\rho_\infty=1.11\times10^{-4}~\mathrm{kg/m^3}$, $T_\infty=56~\mathrm{K}$, and $U_\infty=1115.31~\mathrm{m/s}$, respectively. The wall temperature is fixed at $T_w=300~\mathrm{K}$. The scaled model has a characteristic length of $0.28~\mathrm{m}$ and a reference area of $0.012~\mathrm{m^2}$. In the present computation, the physical space is discretized by an unstructured mesh with 913784 cells, and the velocity space contains 19563 discrete velocity points.
	
\begin{figure}[htbp]
		\centering
		\begin{minipage}{0.32\linewidth}
			\centering
			\includegraphics[width=\linewidth]{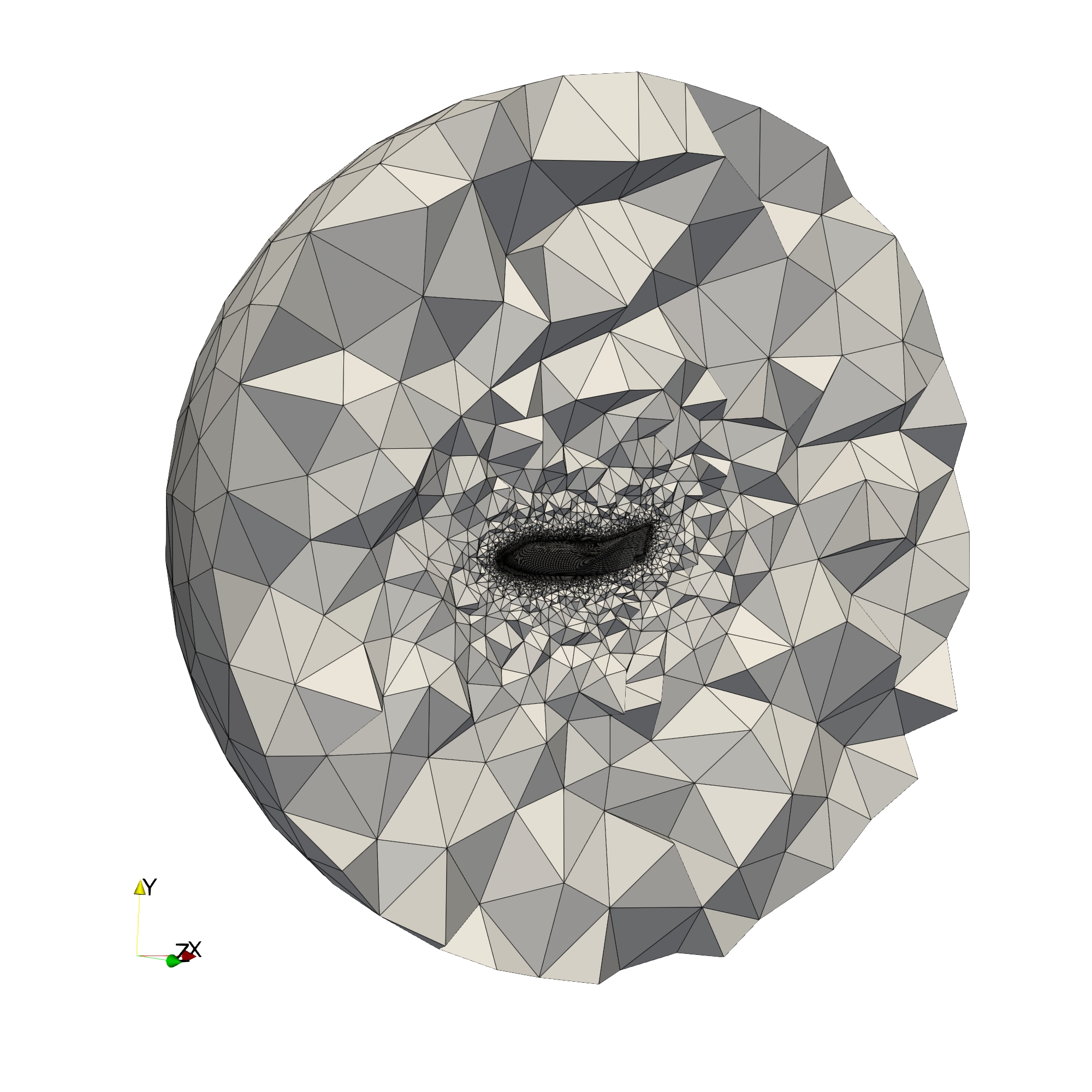}\\
			(a) Physical-space mesh
		\end{minipage}
		\hfill
		\begin{minipage}{0.32\linewidth}
			\centering
			\includegraphics[width=\linewidth]{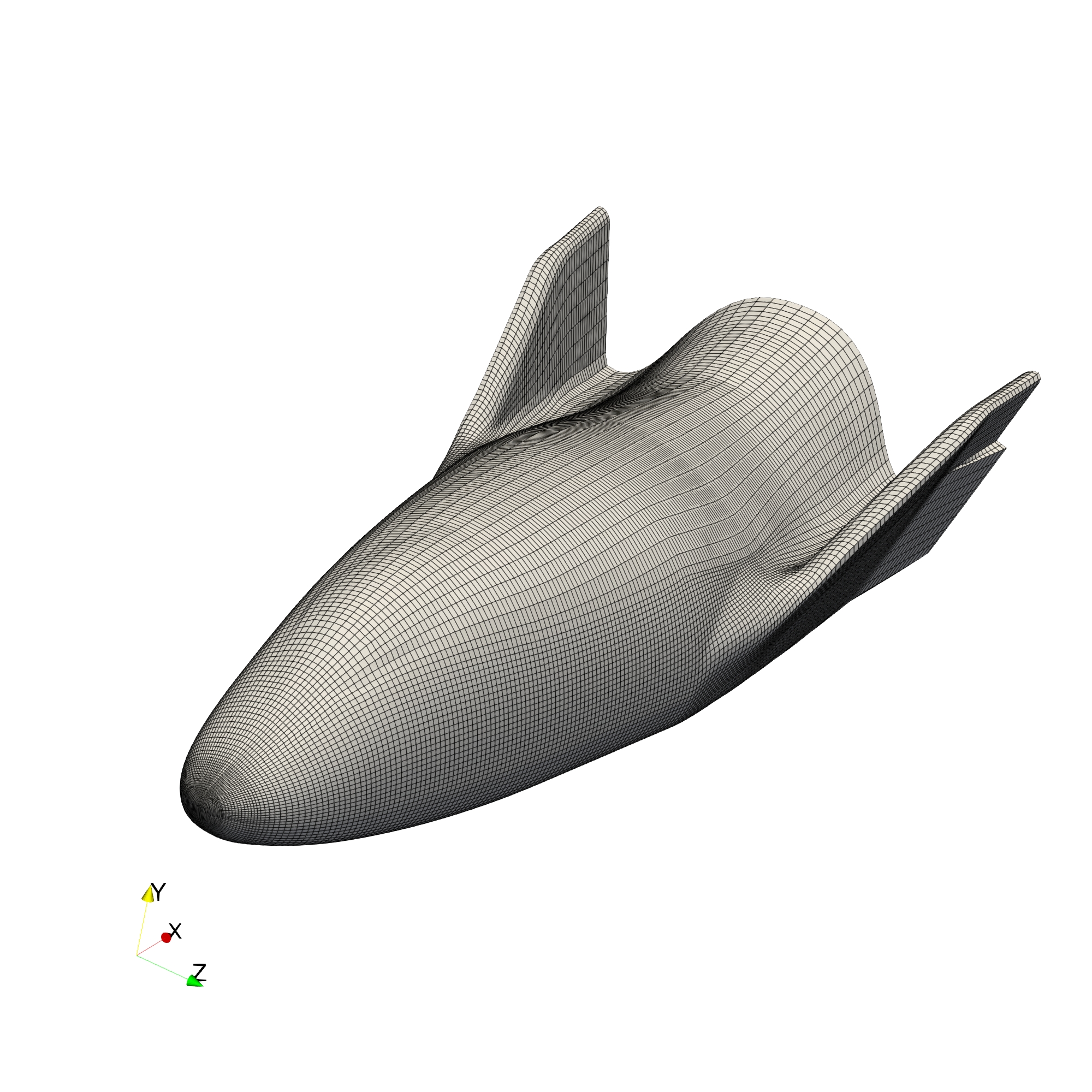}\\
			(b) Surface mesh
		\end{minipage}
		\hfill
		\begin{minipage}{0.32\linewidth}
			\centering
			\includegraphics[width=\linewidth]{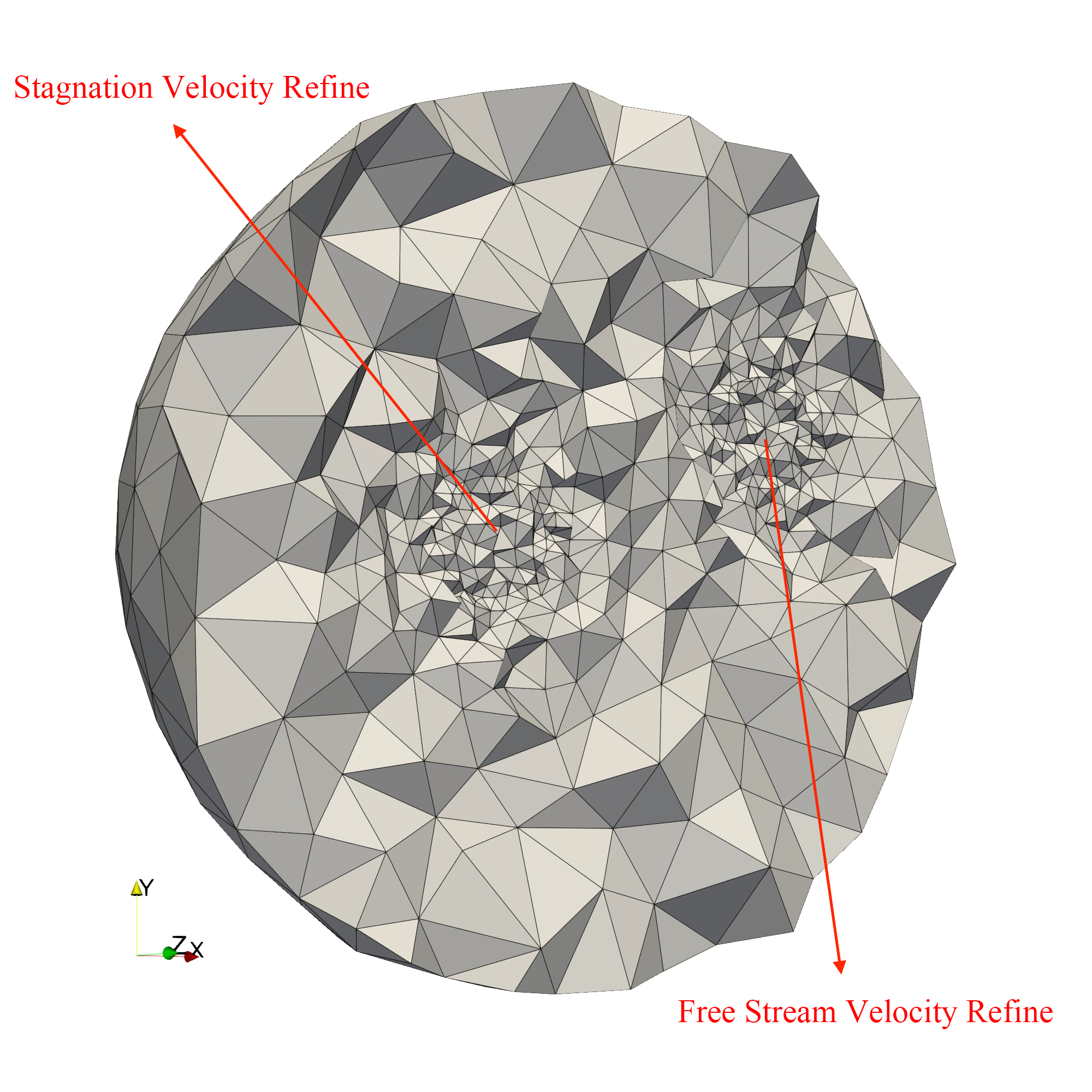}\\
			(c) Velocity-space mesh
		\end{minipage}
		\caption{Physical-space and velocity-space meshes for the X38-like vehicle calculation.}
		\label{fig:x38_meshes}
	\end{figure}

		The physical-space and velocity-space discretizations are shown in Fig.~\ref{fig:x38_meshes}. The physical mesh is refined around the forebody, the control-surface region, and the near-wall flow around the vehicle, while the surface mesh resolves the rounded nose, body curvature, and vertical fins. The discrete velocity mesh is refined in both the freestream-velocity and stagnation-velocity regions, so that the incoming molecular beam and the compressed distribution near the forebody are represented with higher velocity-space resolution.
		
		This case also provides a direct memory-footprint indication of the benefit of velocity-block pipelining and phase-space decomposition. On 64 GPUs, one complete double-precision full-velocity copy of the two reduced distributions $h$ and $b$ for this mesh would require about $2N_cN_v\times8/64 \approx 4.47~\mathrm{GB}$ per GPU. A conventional physical-domain-parallel UGKS implementation that keeps two full distribution states, three full-velocity microscopic-gradient arrays, and a full-velocity face-flux cache would require roughly $(2+3+2)\times4.47 \approx 31~\mathrm{GB}$ per GPU, where the face number is estimated by $N_f\approx2N_c$ for the unstructured mesh. This exceeds the 16~GB memory capacity of one Tesla V100-SXM2 GPU. In the present FPP-UGKS implementation, the measured memory footprint of the X38-like vehicle calculation is about 8.5~GB per GPU on average, showing that the same UGKS discretization can be executed within the available device memory.
		
\begin{figure}[htbp]
		\centering
		\includegraphics[width=0.95\linewidth]{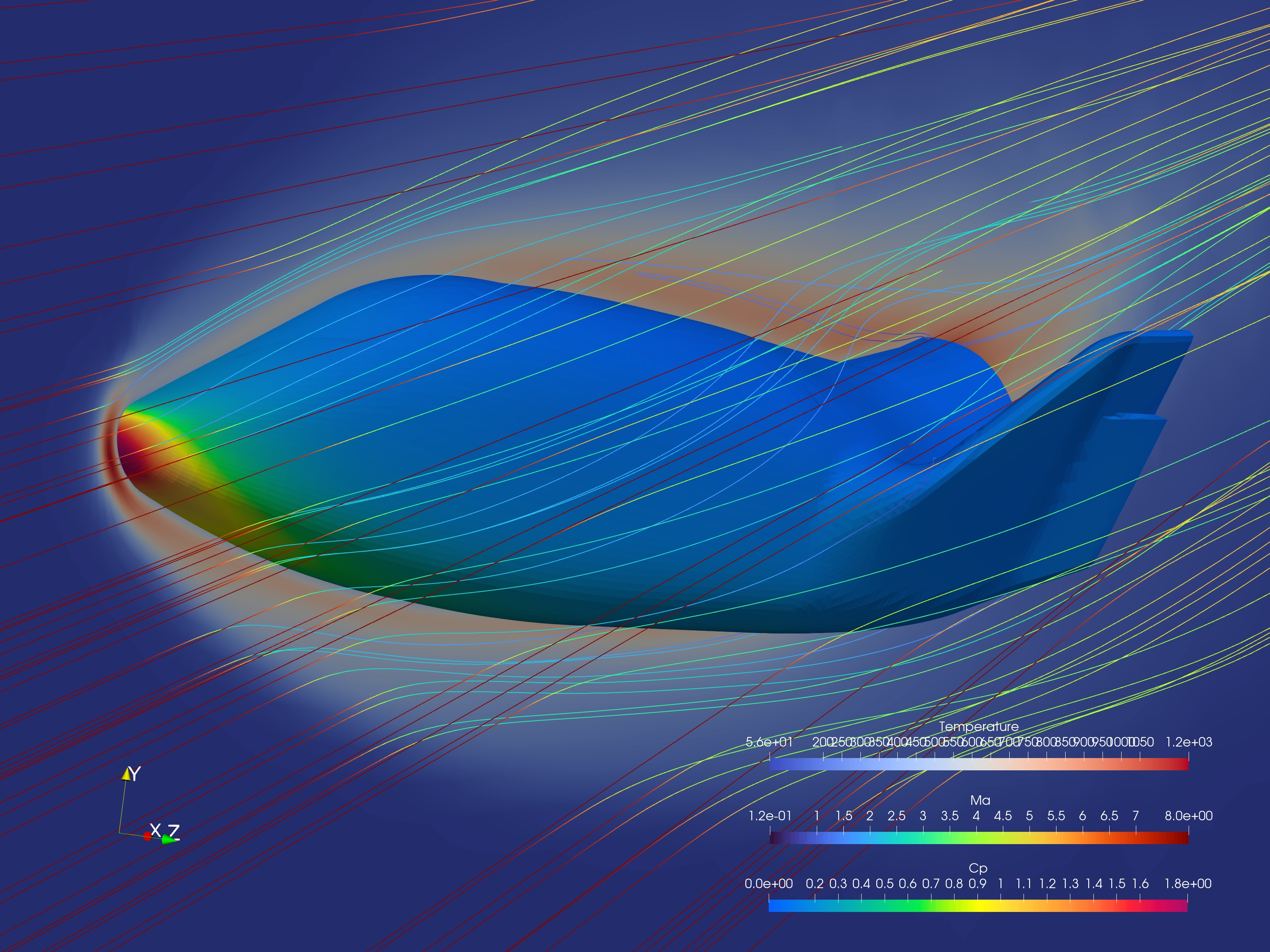}
		\caption{Computed temperature, Mach-number, and pressure distributions around the X38-like vehicle at $M_\infty=8.0$, $\alpha=20^\circ$, and $\mathrm{Kn}=0.00275$.}
		\label{fig:x38_flowfield}
	\end{figure}

	Figure~\ref{fig:x38_flowfield} shows the computed flow field around the X38-like vehicle. A strong bow shock forms ahead of the blunt forebody, and the highest temperature and pressure occur near the windward stagnation region. Downstream of the forebody, the flow accelerates around the body and fins, producing a strongly three-dimensional nonequilibrium structure. This case therefore exercises both the unstructured physical-space treatment and the locally refined velocity-space discretization in a configuration that is more representative of practical hypersonic reentry applications.
	
\begin{figure}[htbp]
		\centering
		\begin{minipage}{0.32\linewidth}
			\centering
			\includegraphics[width=\linewidth]{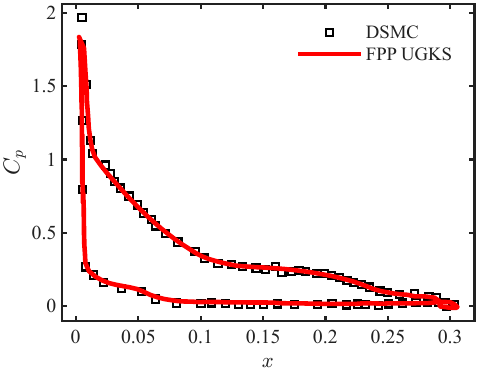}\\
			(a) $C_p$
		\end{minipage}
		\hfill
		\begin{minipage}{0.32\linewidth}
			\centering
			\includegraphics[width=\linewidth]{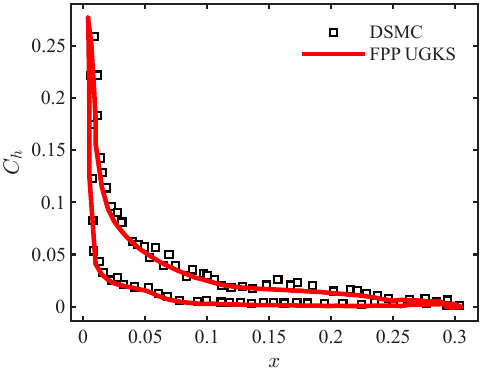}\\
			(b) $C_h$
		\end{minipage}
		\hfill
		\begin{minipage}{0.32\linewidth}
			\centering
			\includegraphics[width=\linewidth]{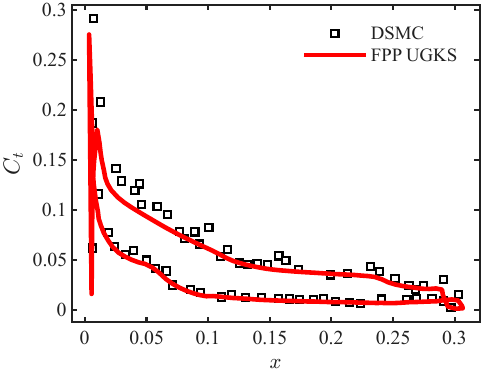}\\
			(c) $C_t$
		\end{minipage}
		\caption{Surface coefficient distributions on the X38-like vehicle. Symbols denote DSMC reference data from Ref.~\cite{Long2024ImplicitAdaptiveUGKS}, and solid lines denote the present FPP-UGKS results.}
		\label{fig:x38_surface_coefficients}
	\end{figure}

	The surface pressure, heat-transfer, and tangential-stress coefficients are compared with DSMC reference data~\cite{Long2024ImplicitAdaptiveUGKS} in Fig.~\ref{fig:x38_surface_coefficients}. The present FPP-UGKS results capture the high pressure and heat-transfer levels near the forebody and the rapid reduction of the surface quantities along the downstream body. The upper- and lower-surface branches also follow the DSMC trends over the main body region. The remaining differences are mainly located near regions where the reference DSMC data exhibit visible statistical scatter, especially for heat transfer and tangential stress at this low Knudsen number. Overall, the comparison confirms that the proposed implementation preserves the accuracy of UGKS for a complex hypersonic rarefied-flow configuration.
	
	For the X38-like vehicle calculation, the x38 Nsight Systems profiling run is summarized in Table~\ref{tab:x38_nsys_main_summary} and Appendix~\ref{app:x38_nsys_profiling}. The summary reports the GPU kernel timing breakdown, staged host-device transfers, MPI message volume, and an analytical memory-footprint estimate for the profiled rank. Appendix~\ref{app:x38_nsys_profiling} also includes the representative timeline for the same profiling run.
	
	\begin{table}[H]
	\centering
	\footnotesize
	\setlength{\tabcolsep}{4pt}
	\caption{Compact rank-0 summary for the x38 profiling case over 50 time steps. Timing and transfer values are from Nsight Systems; the gradient-cache footprint is an analytical estimate from the same run configuration.}
	\label{tab:x38_nsys_main_summary}
	\begin{tabular}{lp{0.62\linewidth}}
	\hline
	Metric & Rank-0 value \\
	\hline
	Top three GPU kernels & 89.9\% of accumulated GPU kernel time \\
	CUDA staged copies & 68.27~GB H2D and 68.28~GB D2H \\
	Point-to-point MPI sends & 62.85~GB payload \\
	Resident microscopic gradient cache & 13.93~GiB full-cache estimate reduced to 0.615~GiB triple-buffered cache, a 22.7x reduction \\
	\hline
	\end{tabular}
	\end{table}
	
	\FloatBarrier
	
		\subsection{Large-scale capability demonstration around the Orion-like capsule}
		\label{subsec:orion_capsule}
		
		After the validation cases above, hypersonic flow around an Orion-like capsule is used as a large-scale capability demonstration of the present implementation. The calculation is performed on the heterogeneous accelerator platform described in Section~\ref{subsec:speedup_test}. The freestream Mach number, Knudsen number, and velocity-space discretization are the same as those used in the X38-like vehicle case, namely $M_\infty=8.0$, $\mathrm{Kn}=0.00275$, and 19563 discrete velocity points. The physical space is discretized by an unstructured mesh with approximately 3.4 million cells, which is about 3.7 times larger than the X38-like vehicle mesh. Using $N_{\mathrm{dof}}=2N_cN_v$, where $N_c$ is the number of physical-space cells and $N_v$ is the number of discrete velocity points, the Orion-like capsule calculation contains approximately $2\times 3.4\times10^6\times19563 \approx 1.33\times10^{11}$ phase-space degrees of freedom.
	
\begin{figure}[htbp]
		\centering
		\begin{minipage}{0.48\linewidth}
			\centering
			\includegraphics[width=\linewidth]{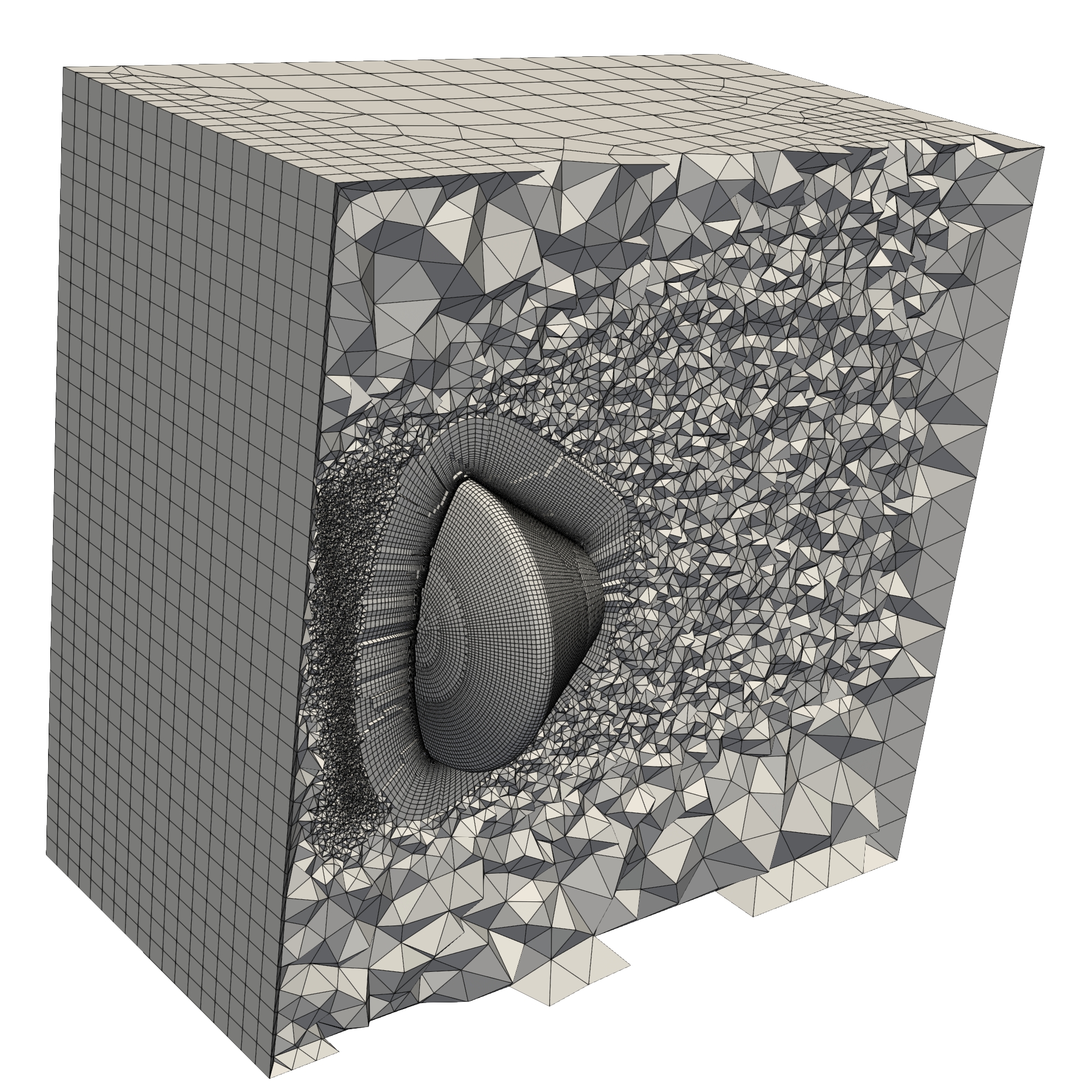}\\
			(a) Physical-space mesh
		\end{minipage}
		\hfill
		\begin{minipage}{0.48\linewidth}
			\centering
			\includegraphics[width=\linewidth]{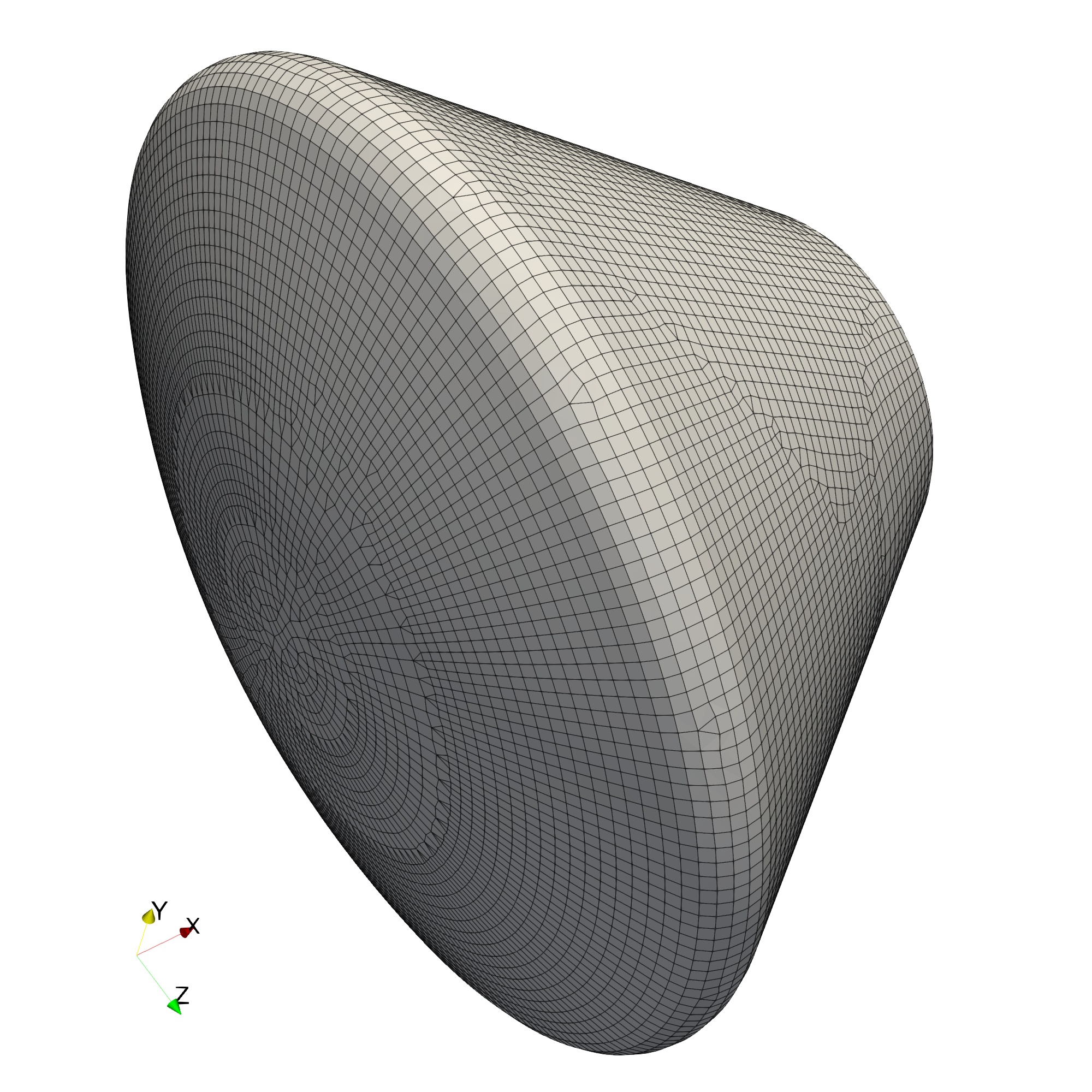}\\
			(b) Surface mesh
		\end{minipage}
		\caption{Physical-space and surface meshes for the Orion-like capsule calculation.}
		\label{fig:orion_meshes}
	\end{figure}

	The physical-space mesh and surface mesh are shown in Fig.~\ref{fig:orion_meshes}. The mesh is refined near the capsule forebody and shoulder region to resolve the detached bow shock and near-wall gradients, while the far-field region is coarsened to control the total cell count. The calculation was performed using 4096 GPGPU accelerators from the scalability-test platform, with one MPI process bound to each accelerator. The complete computation required about 2~h of total wall-clock time, demonstrating that the proposed full phase-space decomposition can make multi-million-cell deterministic UGKS simulations practical on large accelerator systems.
	
\begin{figure}[htbp]
		\centering
		\includegraphics[width=0.95\linewidth]{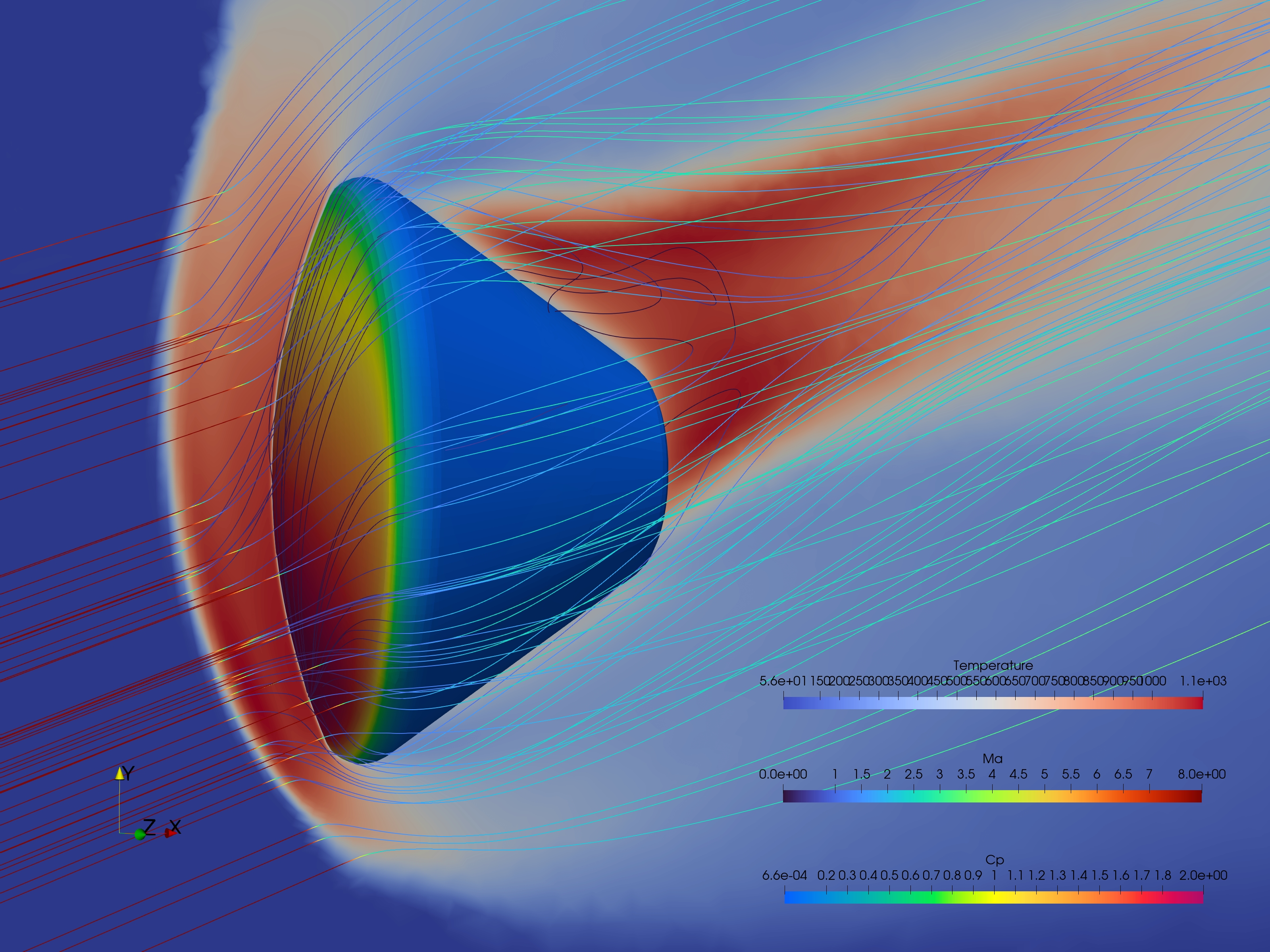}
		\caption{Computed temperature, pressure, Mach-number, and streamline distributions around the Orion-like capsule at $M_\infty=8.0$ and $\mathrm{Kn}=0.00275$.}
		\label{fig:orion_flowfield}
	\end{figure}

		Figure~\ref{fig:orion_flowfield} shows the computed flow field around the Orion-like capsule. A detached shock is formed in front of the blunt heat-shield surface, and strong compression and heating occur between the shock and the forebody. The streamline pattern illustrates the three-dimensional deflection and acceleration of the flow around the capsule shoulder. This large-scale demonstration shows that the proposed velocity-block-pipelined full phase-space decomposition can be applied to realistic blunt-body hypersonic rarefied-flow configurations at substantially larger physical-space resolution.
	
	\FloatBarrier
	
	\section{Conclusion}\label{sec:conclusion}
	
	A GPGPU-oriented FPP-UGKS with velocity-block pipelining has been developed for large unstructured kinetic simulations. The formulation retains the original UGKS flux construction and the two-stage temporal discretization, but reorganizes microscopic evolution around fixed-size velocity blocks. Within each block, nonequilibrium flux evaluation, local moment accumulation, and the first-stage microscopic update are executed with data reuse at the interface and cell levels, making the method more suitable for many-core accelerator execution.
	
	The key algorithmic change is the combination of velocity-block sweeping with a full phase-space MPI decomposition. Global ranks are split into physical-space and velocity-space communicators, so each rank owns only a physical subdomain and a subset of velocity blocks. Physical-halo communication is confined to ranks with the same velocity subset, whereas heat fluxes, wall moments, interface coefficients, and face fluxes are completed through reductions or broadcasts over the velocity communicator. The triple-buffered pipeline further overlaps reconstruction, halo exchange, nonequilibrium flux evaluation, and intermediate microscopic updates. As a result, persistent microscopic storage scales with the local velocity subset, and temporary gradients and fluxes are limited to the active velocity-block working set.
	
		The numerical tests confirm that the proposed implementation preserves the accuracy of the underlying UGKS while improving its memory behavior and parallel scalability. The validation cases cover both internal and external rarefied-flow configurations, including canonical cavity flows, supersonic flow around a sphere, and hypersonic flow around the X38-like vehicle. Across these validation cases, the method retains the ability of UGKS to capture nonequilibrium transport, detached shocks, aerodynamic heating, and surface-force distributions. The Orion-like capsule calculation is used separately as a large-scale capability demonstration, showing that full phase-space decomposition extends the feasible problem size without changing the physical flux model. Using 4,096 GPGPU accelerators, this demonstration represents, to the best of our knowledge, the largest reported parallel computation in terms of GPU count for DVM-type deterministic kinetic methods applied to transitional-flow simulation.
	
	These results show that velocity-block pipelining and full phase-space decomposition provide an effective route toward large-scale deterministic UGKS simulations on modern heterogeneous accelerator platforms. The present study also indicates that performance depends on a balance between local microscopic workload reduction and the cost of velocity-communicator reductions. Future work will therefore focus on automatic selection of the velocity-block size and velocity partition, stronger use of device-aware communication, and extensions to more general gas models, adaptive velocity spaces, and implicit or steady-state acceleration strategies.
	
	\section*{Acknowledgements}
	
The current research is supported by the GHfund A  (ghfund202407012738), the National Natural Science Foundation of China (12302378, 92371201, and 92371107), the National Numerical Windtunnel Project, the Funding of National Key Laboratory of Computational Physics, the Natural Science Basic Research Plan in Shaanxi Province of China (2025SYS-SYSZD-070)
	
	\clearpage
	\appendix
	\section{Raw scalability timing data}
	\label{app:raw_scalability_timings}
	
	This appendix lists the raw wall-clock runtimes used in Figs.~\ref{fig:scalability_bv32}--\ref{fig:scalability_bv256}. Each entry is the runtime in seconds for 100 time steps. The columns $t_1$, $t_{\mathrm{node}}$, $t_4$, and $t_8$ correspond to $P_v=1$, $P_v=N_{\mathrm{node}}$, $P_v=4$, and $P_v=8$, respectively.
	
	\begin{center}
		\begin{minipage}{\linewidth}
			\noindent\textbf{Table A.1}\par
			\noindent Raw strong- and weak-scaling timings for $B_v=32$.\par\medskip
			\centering
			\scriptsize
			\setlength{\tabcolsep}{3.5pt}
			\begin{tabular}{lrrrrrrrr}
				\toprule
				Mode & Nodes & $n_p$ & $N$ & $N_c$ & $t_1$ & $t_{\mathrm{node}}$ & $t_4$ & $t_8$ \\
				\midrule
				Strong & 1  & 8   & 80  & 512000  & 492 & 492 & 674 & 673 \\
				Strong & 2  & 16  & 80  & 512000  & 258 & 320 & 338 & 345 \\
				Strong & 4  & 32  & 80  & 512000  & 127 & 177 & 176 & 182 \\
				Strong & 8  & 64  & 80  & 512000  & 85  & 96  & 93  & 96  \\
				Strong & 16 & 128 & 80  & 512000  & 62  & 88  & 64  & 55  \\
				Strong & 32 & 256 & 80  & 512000  & 46  & 60  & 38  & 34  \\
				Strong & 64 & 512 & 80  & 512000  & 31  & 43  & 26  & 19  \\
				\midrule
				Weak   & 1  & 8   & 30  & 27000   & 30  & 30 & 26  & 28  \\
				Weak   & 2  & 16  & 38  & 54872   & 33  & 30 & 27  & 29  \\
				Weak   & 4  & 32  & 48  & 110592  & 41  & 27 & 27  & 29  \\
				Weak   & 8  & 64  & 60  & 216000  & 50  & 30 & 28  & 30  \\
				Weak   & 16 & 128 & 76  & 438976  & 57  & 41 & 32  & 30  \\
				Weak   & 32 & 256 & 96  & 884736  & 62  & 47 & 35  & 31  \\
				Weak   & 64 & 512 & 120 & 1728000 & 65  & 56 & 38  & 31  \\
				\bottomrule
			\end{tabular}
		\end{minipage}
	\end{center}
	
	\begin{center}
		\begin{minipage}{\linewidth}
			\noindent\textbf{Table A.2}\par
			\noindent Raw strong- and weak-scaling timings for $B_v=64$.\par\medskip
			\centering
			\scriptsize
			\setlength{\tabcolsep}{3.5pt}
			\begin{tabular}{lrrrrrrrr}
				\toprule
				Mode & Nodes & $n_p$ & $N$ & $N_c$ & $t_1$ & $t_{\mathrm{node}}$ & $t_4$ & $t_8$ \\
				\midrule
				Strong & 1  & 8   & 80  & 512000  & 488 & 488 & 620 & 635 \\
				Strong & 2  & 16  & 80  & 512000  & 256 & 284 & 311 & 326 \\
				Strong & 4  & 32  & 80  & 512000  & 126 & 162 & 162 & 172 \\
				Strong & 8  & 64  & 80  & 512000  & 85  & 92  & 86  & 91  \\
				Strong & 16 & 128 & 80  & 512000  & 61  & 72  & 59  & 52  \\
				Strong & 32 & 256 & 80  & 512000  & 45  & 53  & 35  & 32  \\
				Strong & 64 & 512 & 80  & 512000  & 31  & 38  & 24  & 18  \\
				\midrule
				Weak   & 1  & 8   & 30  & 27000   & 30  & 30 & 27  & 30  \\
				Weak   & 2  & 16  & 38  & 54872   & 33  & 30 & 29  & 31  \\
				Weak   & 4  & 32  & 48  & 110592  & 41  & 29 & 29  & 31  \\
				Weak   & 8  & 64  & 60  & 216000  & 50  & 32 & 30  & 32  \\
				Weak   & 16 & 128 & 76  & 438976  & 57  & 41 & 33  & 32  \\
				Weak   & 32 & 256 & 96  & 884736  & 62  & 51 & 36  & 32  \\
				Weak   & 64 & 512 & 120 & 1728000 & 65  & 67 & 40  & 33  \\
				\bottomrule
			\end{tabular}
		\end{minipage}
	\end{center}
	
	\begin{center}
		\begin{minipage}{\linewidth}
			\noindent\textbf{Table A.3}\par
			\noindent Raw strong- and weak-scaling timings for $B_v=128$.\par\medskip
			\centering
			\scriptsize
			\setlength{\tabcolsep}{3.5pt}
			\begin{tabular}{lrrrrrrrr}
				\toprule
				Mode & Nodes & $n_p$ & $N$ & $N_c$ & $t_1$ & $t_{\mathrm{node}}$ & $t_4$ & $t_8$ \\
				\midrule
				Strong & 1  & 8   & 80  & 512000  & 484 & 484 & 570 & 600 \\
				Strong & 2  & 16  & 80  & 512000  & 254 & 252 & 286 & 308 \\
				Strong & 4  & 32  & 80  & 512000  & 125 & 148 & 149 & 162 \\
				Strong & 8  & 64  & 80  & 512000  & 84  & 85  & 79  & 86  \\
				Strong & 16 & 128 & 80  & 512000  & 61  & 59  & 54  & 49  \\
				Strong & 32 & 256 & 80  & 512000  & 45  & 47  & 32  & 30  \\
				Strong & 64 & 512 & 80  & 512000  & 31  & 34  & 22  & 17  \\
				\midrule
				Weak   & 1  & 8   & 30  & 27000   & 30  & 30 & 28  & 32  \\
				Weak   & 2  & 16  & 38  & 54872   & 33  & 30 & 30  & 33  \\
				Weak   & 4  & 32  & 48  & 110592  & 41  & 30 & 30  & 33  \\
				Weak   & 8  & 64  & 60  & 216000  & 50  & 34 & 31  & 34  \\
				Weak   & 16 & 128 & 76  & 438976  & 57  & 41 & 35  & 35  \\
				Weak   & 32 & 256 & 96  & 884736  & 62  & 54 & 38  & 35  \\
				Weak   & 64 & 512 & 120 & 1728000 & 65  & 72 & 42  & 36  \\
				\bottomrule
			\end{tabular}
		\end{minipage}
	\end{center}
	
	\begin{center}
		\begin{minipage}{\linewidth}
			\noindent\textbf{Table A.4}\par
			\noindent Raw strong- and weak-scaling timings for $B_v=256$.\par\medskip
			\centering
			\scriptsize
			\setlength{\tabcolsep}{3.5pt}
			\begin{tabular}{lrrrrrrrr}
				\toprule
				Mode & Nodes & $n_p$ & $N$ & $N_c$ & $t_1$ & $t_{\mathrm{node}}$ & $t_4$ & $t_8$ \\
				\midrule
				Strong & 1  & 8   & 80  & 512000  & 486 & 486 & 529 & 601 \\
				Strong & 2  & 16  & 80  & 512000  & 252 & 254 & 266 & 304 \\
				Strong & 4  & 32  & 80  & 512000  & 125 & 137 & 137 & 153 \\
				Strong & 8  & 64  & 80  & 512000  & 82  & 80  & 72  & 80  \\
				Strong & 16 & 128 & 80  & 512000  & 59  & 48  & 48  & 44  \\
				Strong & 32 & 256 & 80  & 512000  & 42  & 35  & 27  & 28  \\
				Strong & 64 & 512 & 80  & 512000  & 37  & 29  & 21  & 18  \\
				\midrule
				Weak   & 1  & 8   & 30  & 27000   & 30  & 30 & 26  & 31  \\
				Weak   & 2  & 16  & 38  & 54872   & 35  & 30 & 28  & 32  \\
				Weak   & 4  & 32  & 48  & 110592  & 40  & 29 & 29  & 35  \\
				Weak   & 8  & 64  & 60  & 216000  & 46  & 35 & 35  & 35  \\
				Weak   & 16 & 128 & 76  & 438976  & 55  & 42 & 39  & 36  \\
				Weak   & 32 & 256 & 96  & 884736  & 63  & 62 & 43  & 38  \\
				Weak   & 64 & 512 & 120 & 1728000 & 72  & 94 & 44  & 40  \\
				\bottomrule
			\end{tabular}
		\end{minipage}
	\end{center}
	
	\clearpage
	\section{Nsight Systems profiling of the X38 case}
	\label{app:x38_nsys_profiling}
	\label{app:x38_nsight_timeline}
	
	The x38 case was additionally profiled with Nsight Systems using the CUDA profiler API range already placed around the solver time loop. The profiling window covered 50 UGKS time steps. The current subsection reports rank-0 per-rank timing, transfer, and message-volume measurements from that capture, together with an analytical memory-footprint estimate for the same profiled rank.
	
	\begin{center}
		\begin{minipage}{\linewidth}
			\noindent\textbf{Table B.1}\par
			\noindent Nsight Systems GPU kernel summary for the x38 case on MPI rank 0 over 50 time steps.\par\medskip
			\centering
			\scriptsize
			\setlength{\tabcolsep}{3.5pt}
			\begin{tabular}{lrrrr}
				\toprule
				Kernel & Calls & Total time (s) & Time/step (ms) & Kernel time (\%) \\
				\midrule
				Non-equilibrium interior flux & 3400 & 40.661 & 813.2 & 44.9 \\
				DVS least-squares reconstruction & 3500 & 24.611 & 492.2 & 27.2 \\
				Update $f$ tilde & 3400 & 16.108 & 322.2 & 17.8 \\
				Conservative-variable update & 50 & 3.626 & 72.5 & 4.0 \\
				Unpack $f$ and gradients & 3450 & 1.724 & 34.5 & 1.9 \\
				Non-equilibrium parallel-face flux & 3400 & 1.440 & 28.8 & 1.6 \\
				Pack $f$ and gradients & 3450 & 1.321 & 26.4 & 1.5 \\
				\bottomrule
			\end{tabular}
		\end{minipage}
	\end{center}
	
	The non-equilibrium interior flux, DVS least-squares reconstruction, and distribution-function update together account for 89.9\% of the accumulated GPU kernel time on the profiled rank, identifying these kernels as the primary compute-side bottlenecks visible to Nsight Systems.
	
	\begin{center}
		\begin{minipage}{\linewidth}
			\noindent\textbf{Table B.2}\par
			\noindent Host-device transfer and MPI message volume measured by Nsight Systems for the x38 case on MPI rank 0 over 50 time steps. For MPI rows, the time column is the Nsight Systems MPI API event duration.\par\medskip
			\centering
			\scriptsize
			\setlength{\tabcolsep}{3.5pt}
			\begin{tabular}{lrrrr}
				\toprule
				Category & Count & Total volume (GB) & Volume/step (GB) & Total time (s) \\
				\midrule
				CUDA H2D memcpy & 4200 & 68.274 & 1.365 & 20.412 \\
				CUDA D2H memcpy & 4225 & 68.277 & 1.366 & 16.770 \\
				MPI Isend payload & 17750 & 62.851 & 1.257 & 0.042 \\
				MPI Irecv payload & 17750 & 62.851 & 1.257 & 0.019 \\
				MPI Bcast send volume & 450 & 36.738 & 0.735 & 3.537 \\
				MPI Bcast receive volume & 450 & 4.592 & 0.092 & 3.537 \\
				MPI Allreduce send volume & 205 & 5.815 & 0.116 & 2.010 \\
				MPI Allreduce receive volume & 205 & 5.815 & 0.116 & 2.010 \\
				\bottomrule
			\end{tabular}
		\end{minipage}
	\end{center}
	
	The point-to-point MPI payload is consistent with the halo exchange model of the implementation: the dominant term is the repeated exchange of $f$ and microscopic gradients for each velocity block. The measured point-to-point send volume is 1.257~GB per step on rank 0, while the corresponding staged CUDA transfer volume is 1.365~GB/step in each direction.
	
	\begin{center}
		\begin{minipage}{\linewidth}
			\noindent\textbf{Table B.3}\par
			\noindent Analytical GPU memory footprint estimate for major UGKS arrays on the profiled x38 rank, derived from the rank-0 problem sizes and array dimensions in the profiling run.\par\medskip
			\centering
			\scriptsize
			\setlength{\tabcolsep}{3.5pt}
			\begin{tabular}{lrr}
				\toprule
				Array group & Size (GB) & Size (GiB) \\
				\midrule
				Distribution function $f$ & 5.207 & 4.849 \\
				Boundary distribution cache & 0.122 & 0.114 \\
				Triple microscopic gradient cache & 0.660 & 0.615 \\
				Microscopic flux/update cache $f_{dt}$ & 0.168 & 0.156 \\
				Face macro/coefficient cache & 0.089 & 0.083 \\
				MPI pre buffers, send+recv & 0.019 & 0.018 \\
				MPI gradient buffers, send+recv & 0.002 & 0.002 \\
				Triple MPI $f$-gradient buffers, send+recv & 0.108 & 0.101 \\
				\midrule
				Major microscopic/cache subtotal & 6.246 & 5.817 \\
				\bottomrule
			\end{tabular}
		\end{minipage}
	\end{center}
	
	For rank 0, a full resident microscopic gradient cache over all 2,176 local velocities would require 13.93~GiB, whereas the triple-buffered velocity-block cache requires 0.615~GiB. Thus the resident microscopic gradient cache is reduced to 4.4\% of the full-cache footprint, corresponding to a 22.7x reduction for this array group. Table B.3 should be interpreted as an analytical accounting of the dominant microscopic and communication-buffer arrays on the profiled rank, not as the complete device-resident memory reported by the runtime. The listed arrays account for 5.817~GiB, while the measured X38-like vehicle memory footprint of about 8.5~GB per GPU reported in Section~\ref{subsec:x38_vehicle} also includes auxiliary macroscopic arrays, mesh and connectivity storage, boundary and reduction workspaces, allocator padding, and runtime/library overhead. The comparison therefore shows both the main source of memory reduction relative to a full resident gradient cache and the consistency between the major-array estimate and the measured total footprint.
	
	\clearpage
	\begin{landscape}
		Figure B.1 shows a representative NVIDIA Nsight Systems timeline for the X38-like vehicle calculation. The device timeline contains repeated velocity-block cycles consisting of memory transfers, least-squares reconstruction, nonequilibrium interface-flux evaluation, and microscopic distribution update kernels. On the host side, MPI API events and CUDA stream synchronization calls appear between consecutive block cycles.
		
		\begin{center}
			\includegraphics[width=0.96\linewidth]{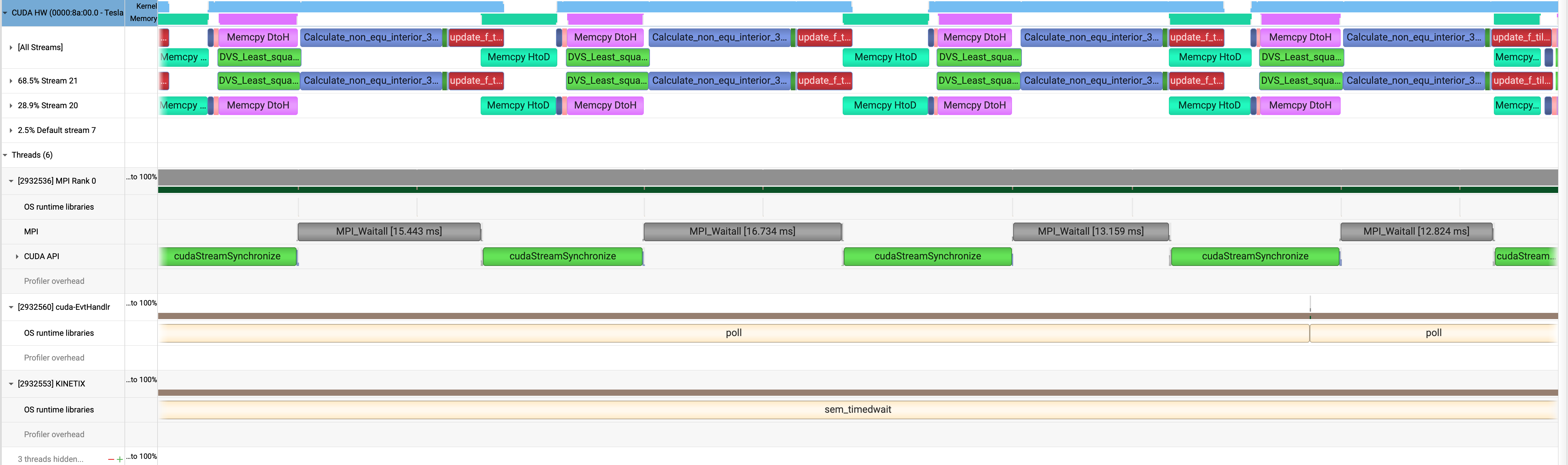}
			
			\noindent\textbf{Figure B.1}\quad Representative NVIDIA Nsight Systems timeline for the X38-like vehicle calculation, illustrating the repeated velocity-block execution pattern and the interaction between GPU kernels, memory transfers, MPI waiting intervals, and CUDA stream synchronization.
		\end{center}
	\end{landscape}
		
	\clearpage
	
	\bibliographystyle{unsrt}
	
	\bibliography{cas-refs}
	
	
	
\end{document}